\documentclass[journal,twoclumn]{IEEEtran}
\usepackage{multicol}
\usepackage{etoolbox}
\makeatletter
\patchcmd{\@makecaption}
  {\scshape}
  {}
  {}
  {}
\makeatletter
\patchcmd{\@makecaption}
  {\\}
  {.\ }
  {}
  {}
\makeatother

\usepackage{amsfonts}
\usepackage{amsmath}
\usepackage{mathrsfs}
\usepackage{mathtools}
\usepackage{amsfonts}
\usepackage{amssymb}
\usepackage{graphicx}
\usepackage{epsfig}
\usepackage{psfrag}{}
\usepackage{array}
\usepackage{cases}
\usepackage{eufrak}
\usepackage{cite,graphicx,amssymb,color}
\usepackage{algorithmic}
\usepackage{algorithm}
\usepackage{setspace}
\usepackage{bm}
\usepackage{multirow}
\usepackage{threeparttable}
\usepackage{array}
\usepackage{makecell}
\usepackage{soul}
\usepackage{float}
\usepackage{booktabs}
\usepackage{diagbox}
\usepackage{caption}
\usepackage{booktabs}
\usepackage{breqn}
\usepackage{subfig}

\usepackage{xfrac} % nice fractions

\usepackage{nicematrix}
\usepackage{tikz}
\newfloat{figtab}{htb}{fgtb}
\makeatletter
  \newcommand\figcaption{\def\@captype{figure}\caption}
  \newcommand\tabcaption{\def\@captype{table}\caption}
\makeatother

\DeclareMathOperator*{\argmax}{argmax}

\newtheorem{Thm}{Theorem}
\newtheorem{Lem}{Lemma}

\newtheorem{Prob}{Problem}

\newtheorem{Asump}{Assumption}

\newtheorem{Proof}{Proof}

\newcommand{\bigzero}{\mbox{\normalfont\Large\bfseries 0}}

\NiceMatrixOptions
{
    custom-line ={command= H, tikz= dashed, width= 1mm}, % horizontal, dashed
    custom-line = {letter= I, tikz= dashed, width= 1mm}, % vertical, dashed
}

\IEEEoverridecommandlockouts
\begin{document}
\title{Enhancing Neural Adaptive Wireless Video Streaming via Lower-Layer Information Exposure and Online Tuning (Technical Report)}
\author{\IEEEauthorblockN{Lingzhi Zhao, Ying Cui, \textit{Member, IEEE}, Yuhang Jia, Yunfei Zhang, Klara Nahrstedt, \textit{Fellow, IEEE, ACM} }
\thanks{L. Zhao and K. Nahrstedt are with University of Illinois Urbana-Champaign, USA. Y. Cui is with IoT
Thrust, the Hong Kong University of Science and Technology (Guangzhou), China, and also with The Hong Kong University of Science and
Technology, Hong Kong, SAR, China, Y. Jia and Y. Zhang are with Tencent Technology. \textcolor{black}{This paper was submitted in part to IEEE ICC
2024\cite{ICC24}.}}
}
%Y. Han and Y. Zhang are with Tencent Technology Co., Ltd, Shenzhen, 518054, China.

\maketitle

\begin{abstract}
Deep reinforcement learning (DRL) demonstrates its promising potential in the realm of adaptive video streaming and has recently received increasing attention. However, existing DRL-based methods for adaptive video streaming use only application (APP) layer information, adopt heuristic training methods, and train generalized neural networks with pre-collected data. This paper aims to boost the quality of experience (QoE) of adaptive wireless video streaming by using lower-layer information, deriving a rigorous training method, and adopting online tuning with real-time data. First, we formulate a more comprehensive and accurate adaptive wireless video streaming problem as an infinite stage discounted Markov decision process (MDP) problem by additionally incorporating past and lower-layer information, allowing a flexible tradeoff between QoE and costs for obtaining system information and solving the problem. In the offline scenario (only with pre-collected data), we propose an enhanced asynchronous advantage actor-critic (eA3C) method by jointly optimizing the parameters of parameterized policy and value function. Specifically, we build an eA3C network consisting of a policy network and a value network that can utilize cross-layer, past, and current information and jointly train the eA3C network using pre-collected samples. 
In the online scenario (with additional real-time data), we propose two \textcolor{black}{continual} learning-based online tuning methods for designing better policies for a specific user with different QoE and training time tradeoffs. 
In particular, they incorporate different numbers of tunable components into the respective offline trained eA3C networks and adjust the tunable components based on collected real-time samples. 
Finally, experimental results show that the proposed offline policy can improve the QoE by $6.8\% \sim 14.4\%$ compared to the state-of-arts in the offline scenario, and the proposed online policies can further achieve $6\% \sim 28\%$ gains in QoE over the proposed offline policy in the online scenario.
\end{abstract}


\section{Introduction}
%Video on demand (VoD) allows users to select and watch video content of their choices on their own devices (e.g., TV or computer). 
Video has become the most widely used application on the Internet, witnessing a substantial surge in traffic volume in recent years\cite{report}. Video on demand (VoD) service, which accounts for 29\% of the traffic carried over the Internet\cite{report}, enables users to, on demand, select and view video content, leveraging the video streaming technology via which video is delivered and consumed continuously from a source.   
%In the past few years, it has been observed that video streaming has accounted for 55\% of the traffic carried over the Internet and it continues to steadily increase every year.
During video streaming, network fluctuations can easily cause rebuffering at viewers, severely \textcolor{black}{degrading} their viewing experiences. To address this issue, 
adaptive video streaming\cite{JSAC14,sigcomm14,TMM16,TMM19,CL14,TON20,sigcomm15,sigcomm16,TCCN17,mobihoc19,TON21,TMM20,icnp18,acmmm19,sigcomm17,JSAC20,NSDI20} has been proposed to adapt the video chunk bitrate to the dynamic network condition and user's buffer occupancy. In particular, adaptive video streaming can be formulated as a dynamic programming (DP) problem where a (bitrate adaptation) policy that maps the system state to the video chunk bitrate is optimized to maximize the video quality\textcolor{black}{\cite{JSAC14,sigcomm14,TMM16,TMM19,CL14,TON20,sigcomm15,sigcomm16,TCCN17,mobihoc19,TON21,TMM20,icnp18,acmmm19,sigcomm17,JSAC20,NSDI20}} or to minimize the video quality variation\cite{JSAC14,TMM16,TMM19,CL14,TON20,sigcomm15,NSDI20,sigcomm16,mobihoc19,icnp18,sigcomm17,TON21,TMM20,acmmm19,JSAC20}, rebuffering time\cite{JSAC14,sigcomm14,TMM16,TMM19,CL14,TON20,sigcomm15,NSDI20,sigcomm16,TCCN17,mobihoc19,icnp18,sigcomm17,TMM20,acmmm19,JSAC20}, and startup delay\cite{sigcomm15,sigcomm16}. 

%Adaptive video streaming can be modelled as an optimal control problem of a dynamic system in which the system chooses a sequence of policies over time to select the bitrate of the video in order to maximize the user' s quality of experience (QoE).

\begin{figure*}[t]
%\vspace*{-1.2cm}
\begin{center}
   %\subfloat[\small{Average QoE versus number of layers}]
   %{\resizebox{8cm}{!}{\includegraphics{Offline_QoE_layer.eps}}}
   %\subfloat[\small{Average QoE versus number of neurons}]
   %{\resizebox{8cm}{!}{\includegraphics{Offline_QoE_neuron.eps}}}
      \subfloat[$R_{X,C}(\tau)$ on Set-1]
   {\resizebox{3.6cm}{!}{\includegraphics{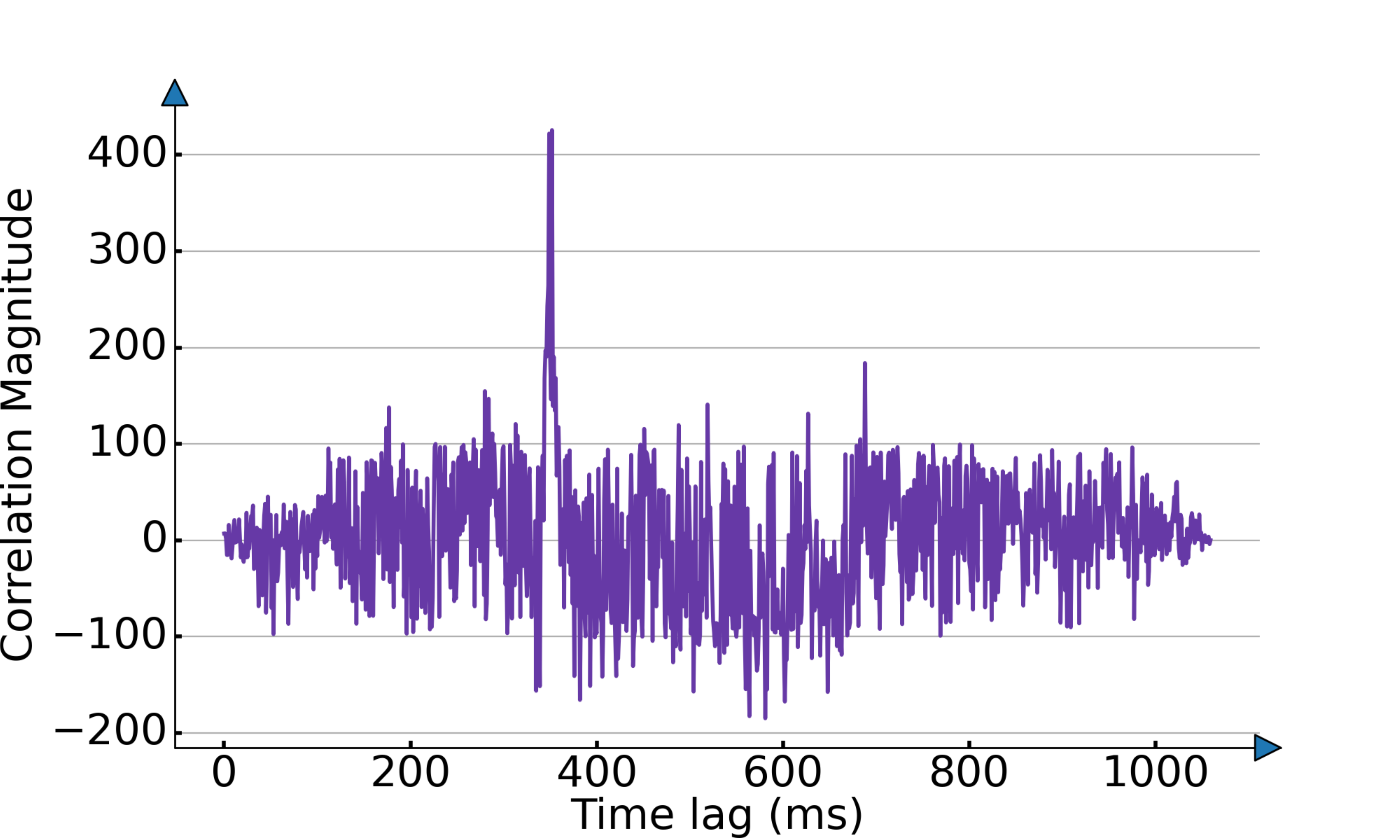}}}
   \subfloat[$R_{X,C}(\tau)$ on Set-2]
   {\resizebox{3.6cm}{!}{\includegraphics{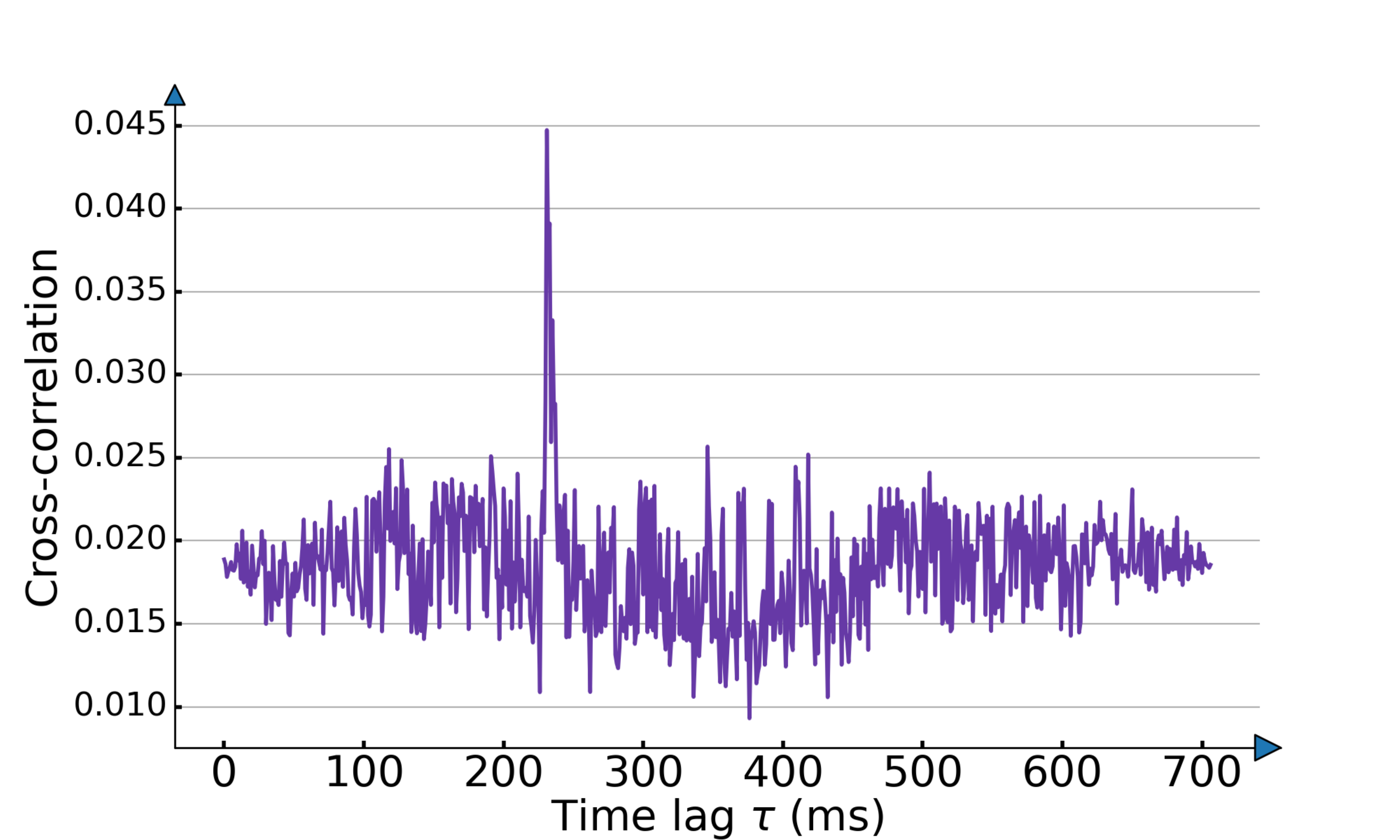}}}
   \subfloat[Distributions]
   {\resizebox{3.6cm}{!}{\includegraphics{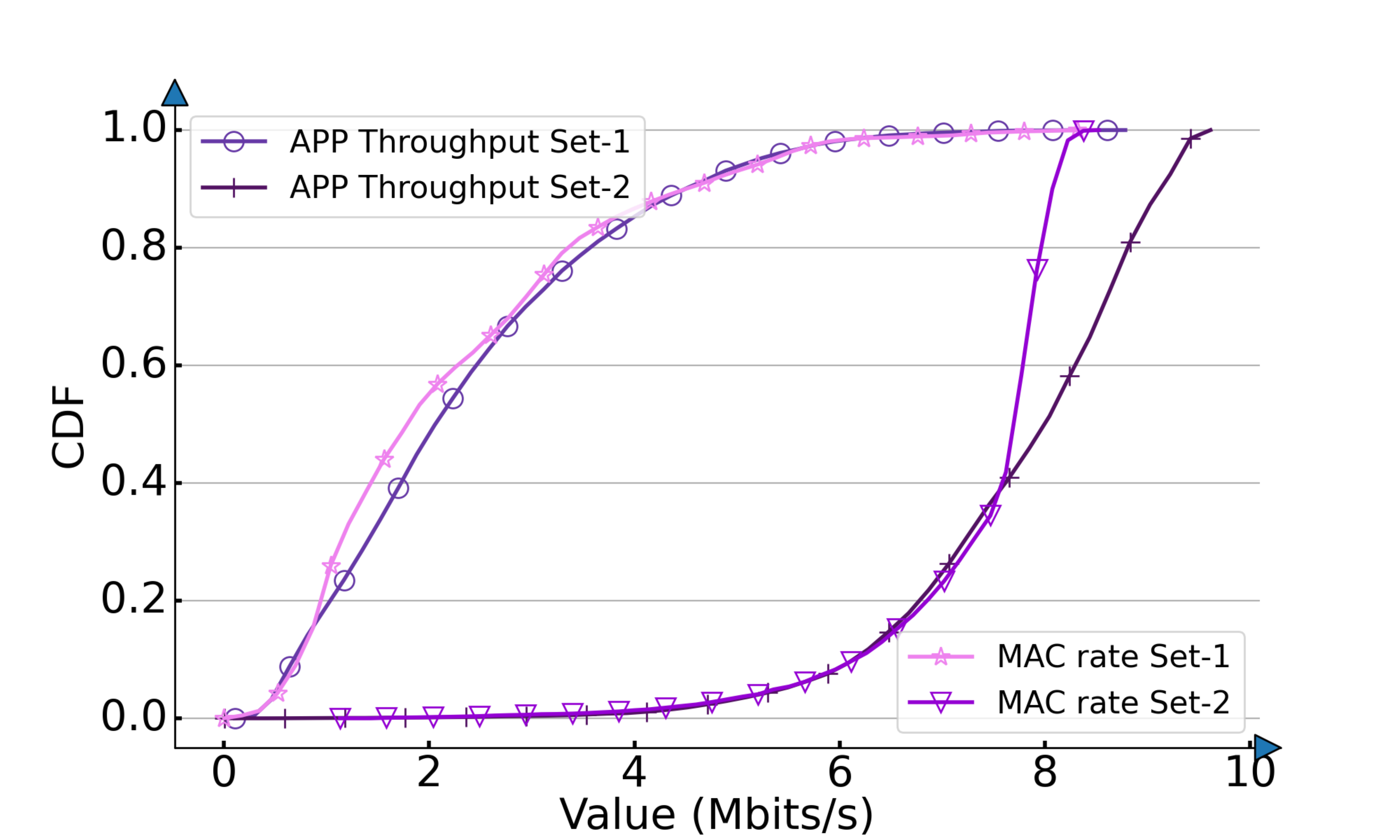}}}
   \subfloat[$R_{C,C}(\tau)$ on Set-1]
   {\resizebox{3.6cm}{!}{\includegraphics{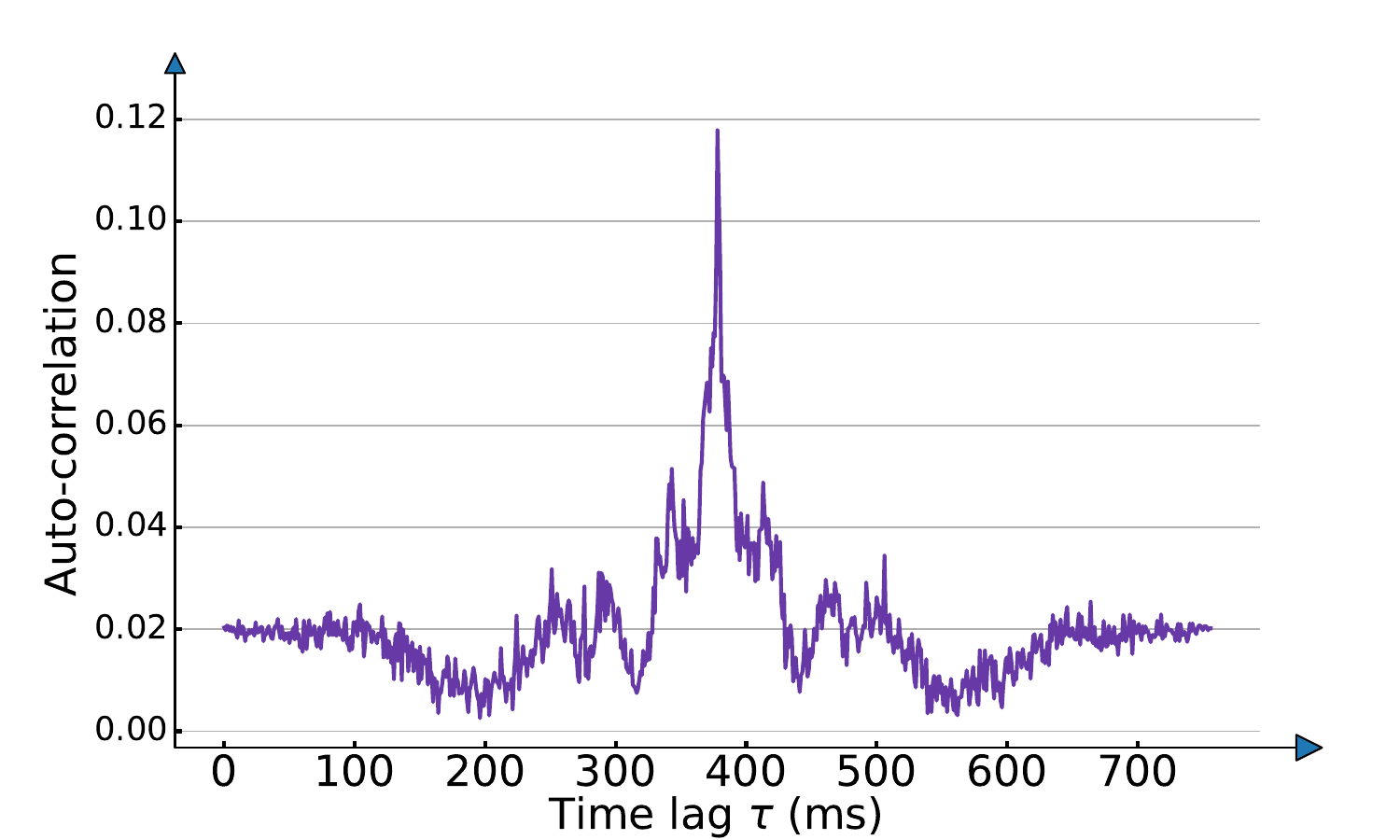}}}
      \subfloat[$R_{C,C}(\tau)$ on Set-2]
   {\resizebox{3.6cm}{!}{\includegraphics{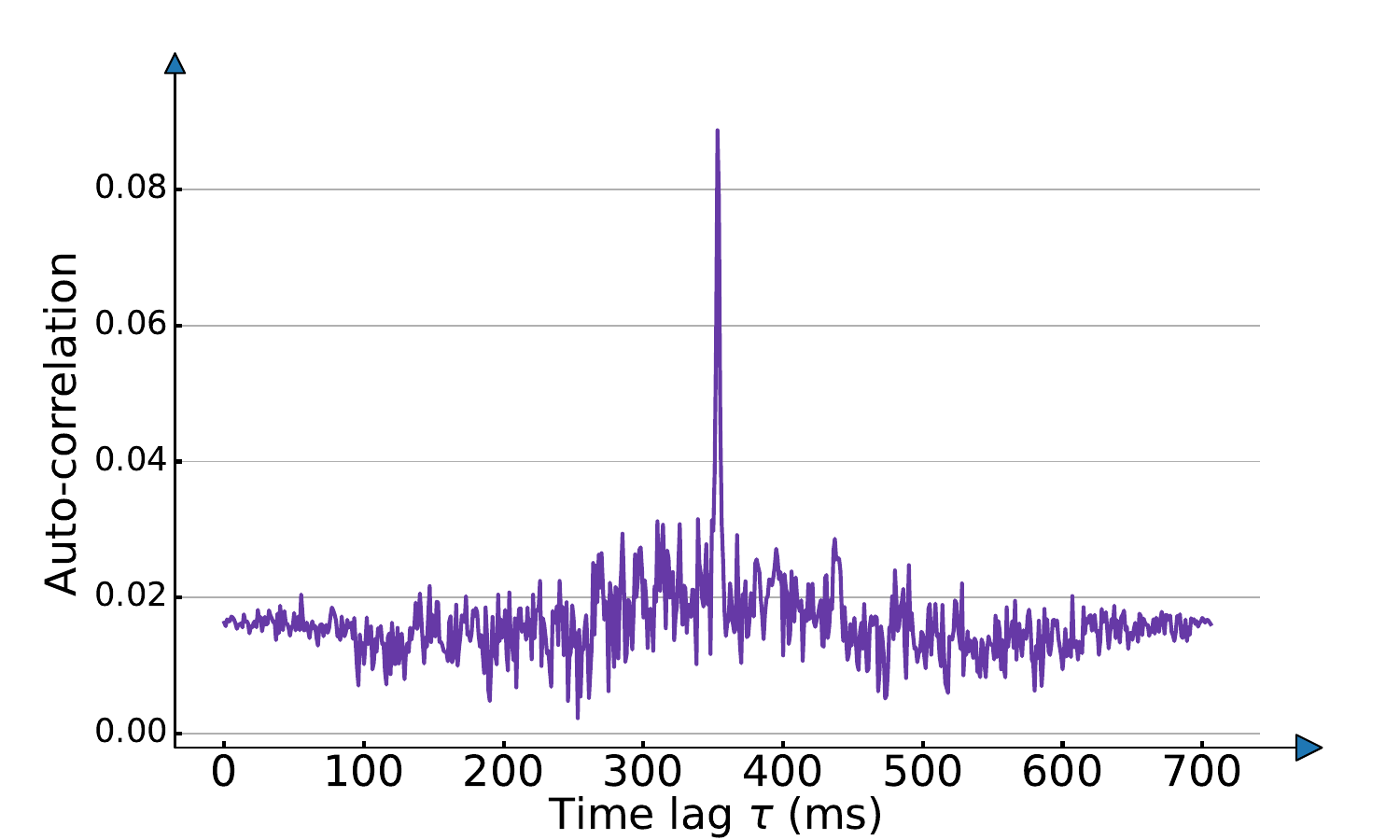}}}
 %\subfloat[\small{CDF of rebuffering time}]
 %{\resizebox{5.9cm}{!}{\includegraphics{pic/multi_rebuffering.eps}}}
 %\vspace*{-0.07cm}
 \end{center}
 %\vspace*{-0.37cm}
   \caption{Relationship between APP layer throughput sequence $\{C_{n}:n=1,\ldots,\}$ and MAC rate sequence $\{X_{n} : n = 1,\ldots,\}$ on two collected datasets\cite{report} (see Section~\ref{sec:sim} for details). The (sample) cross-correlation between $X$ and $C$ and auto-correlation of $C$ are defined as $R_{X,C}(\tau) \triangleq\frac{1}{N-\tau}\sum^{N-\tau}_{n=1}X_{n}C_{n+\tau}$, $R_{C,C}(\tau) \triangleq\frac{1}{N-\tau}\sum^{N-\tau}_{n=1}C_{n}C_{n+\tau}$, respectively, where $\tau \in \{0,1,\ldots,N\}$ is time lag, and $N$ is the number of samples.}
   \label{fig:cor}
%\vspace*{-0.60cm}
\end{figure*}

There are mainly four methods to solve the \textcolor{black}{DP} problems for adaptive video streaming\textcolor{black}{\cite{sigcomm14,TON20,TMM16,TMM19,JSAC14,sigcomm15,NSDI20,sigcomm16,CL14}}. Firstly, in \cite{sigcomm14}, a heuristic (buffer-based) policy is proposed to select the bitrate of each chunk according to only the instantaneous user's buffer occupancy without considering the instantaneous network condition. Secondly, \cite{TON20} presents BOLA that determines the bitrate of each chunk based on the instantaneous user's buffer occupancy by optimizing the one-step Lyapunov drift (expected change of the Lyapunov function values at two consecutive downloaded video chunks). \textcolor{black}{Both the heuristic\cite{sigcomm14} and Lyapunov optimization-based\cite{TON20} policies are myopic and can be far from optimal, as they cannot take into account the instant bitrate adaptation's long-term (beyond one step) influence.}
Thirdly, in \cite{TMM16,TMM19,JSAC14}, the network state evolution is assumed to follow an Markov chain with known transition probabilities, and DP techniques such as value iteration\cite[pp. 84]{dp_2} and policy iteration\cite[pp. 97]{dp_2} are adopted to \textcolor{black}{solve the DP problems}. However, DP may suffer from the curse of modeling as network statistics in practice are usually difficult to obtain. Finally, in \cite{sigcomm15,NSDI20,sigcomm16,CL14}, the network statistics are assumed unknown, and reinforcement learning (RL)\cite{rl} is applied to resolve the DP problems. The polices obtained by DP and RL can \textcolor{black}{take into account} the long-term influence. However, both have high computational complexities due to the curse of dimensionality.
%The authors in \cite{TMM16} and \cite{TMM19} assume that the statistics of the network throughput are known while the authors in \cite{sigcomm14,sigcomm15,NSDI20,sigcomm16} assume that the statistics of the network throughput are unknown. The proposed algorithms\cite{sigcomm14,TMM16,TMM19,sigcomm15,NSDI20,sigcomm16} have low computational complexity but do not have performance guarantee. 

Deep reinforcement learning (DRL)\cite{a3c,alphago,rl} has been applied to adaptive video streaming to solve the curse of dimensionality issue for RL\cite{TCCN17,mobihoc19,icnp18,sigcomm17,JSAC20,mm22,TON21,acmmm19,TMM20}. 
%Deep RL is an area of machine learning concerned with how intelligent agents take actions in an environment in order to maximize the target reward. 
There are two main DRL algorithms, i.e., deep Q-learning (DQ)\cite{TCCN17,mobihoc19,doubleDQN} and asynchronous advantage actor-critic (A3C)\cite{icnp18,sigcomm17,JSAC20,TON21,acmmm19,TMM20,mm22}. \textcolor{black}{Specifically, DQ  represents Q-factors using a deep Q-network (DQN), adopts a separate copy of the DQN, and trains the DQN with the aid of its copy using deep learning (DL) and Q-learning\cite{dp,doubleDQN}. The trained DQN produces the approximate Q-factors, and 
%The DNN produces approximate Q-factor. 
a policy is then obtained by maximizing the approximate Q-factors with respect to all bitrates at each network state. Moreover,
A3C represents a randomized policy and a value function using a policy neural network and a value neural network, respectively, and trains these two neural networks alternatively. The trained policy neural network directly produces the \textcolor{black}{optimized} randomized policy.} A3C naturally balances exploration and exploitation through its \textcolor{black}{randomized policy design} and is shown in \cite{icnp18,sigcomm17,JSAC20} to achieve better performance than DQ in adaptive video streaming. 
%A3C is designed to take advantage of parallel computation by training the neural network across multiple threads, while DQN is typically run on a single thread. This makes A3C more computationally efficient and improves the policy exploration. A3C applies policy gradient method\cite{a3c} to solve an expected value function maximization problem. Specifically, the policy gradient method approximates the policy and value function with policy and value parameters, respectively, separates the problem into two optimization problems and solve them alternatively. To implement the policy gradient method, A3C alternatively trains two neural networks, i.e., policy network for bitrate selection and value network for evaluating the policy. 
%which is based on the policy gradient method\cite{a3c}, is the state-of-art for adaptive video streaming. This algorithm alternatively trains two neural networks, i.e., policy network which is responsible for selecting bitrate and value network which evaluates the policy produced by the policy network.

%Besides, some work focuses on adaptive live video streaming\cite{TON21,TMM20,acmmm19}. Live streaming is streaming media simultaneously recorded and broadcast in real-time over the internet. Live streaming has additional metrics such as latency and target buffer comapring to the VoD. In general, live streaming and VoD are very similar and it is relatively easy to extend the VoD to live streaming. So this paper mainly focuses on adaptive video streaming for VoD.

\textit{Limitations and Motivations:} there are three major limitations in the existing work on
adaptive video streaming\cite{JSAC14,sigcomm14,TMM16,TMM19,CL14,TON20,sigcomm15,sigcomm16,TCCN17,mobihoc19,TON21,TMM20,icnp18,acmmm19,sigcomm17,JSAC20,NSDI20}. Firstly, most of the existing policies\cite{JSAC14,sigcomm14,TMM16,TMM19,CL14,TON20,sigcomm15,sigcomm16,TCCN17,mobihoc19,TON21,TMM20,icnp18,acmmm19,sigcomm17,JSAC20,NSDI20} \textit{rely only on application (APP) layer information} (e.g., APP layer throughput, buffer occupancy, and bitrate selection). Since the APP layer relies on all lower layers to complete its process and lower-layers can respond more rapidly to the changes in the environment, information from the lower layers may also be helpful for adaptive video streaming\cite{NG,mobicom15}.
%Although some work\cite{TWC20,TCSVT19,globecom19} tries to jointly optimize bitrate and physical (PHY) layer communication resource, such resource cannot usually be adjusted in a real system. 
Fig.~\ref{fig:cor} (a)-(b) show that the MAC rate can respond $200-300$ milliseconds faster to the environment changes than the APP layer throughput. \textcolor{black}{This indicates that lower-layer information exposure may help predict the future changes in the application layer and advance its adaptations as suggested in the work item AIS 5G of 3GG\cite{NG_2,NG_1}.}
Secondly, A3C\cite{a3c,icnp18,sigcomm17,JSAC20} breaks the original optimization problem for the policy and value parameters into two separate but related problems and trains the policy and value networks alternatively \textcolor{black}{by solving the optimization problems for the value and policy parameters alternatively}. \textit{The training method is rather heuristic} and hardly yields any convergence or performance guarantee.
%\textcolor{black}{Some work \cite{JSAC20,mm22} aiming to adapt the offline trained networks to the real-time environment does not adjust the neural network structures, thereby failing to achieve satisfactory QoE for the specific user. }
%Since video streaming is a long-lasting application, the real-time samples of a user can be gradually collected during the video streaming process and used to produce a better policy tailored for a specific user.
%Recently, \cite{mm22,JSAC20} propose to predict the throughput in the online scenario for  are rather heuristic.  
Last but not least, A3C for adaptive video streaming\cite{TCCN17,mobihoc19,icnp18,sigcomm17} usually train deep neural networks offline based on the pre-collected samples reflecting a general environment. Offline trained DQ and A3C networks suffer from \textit{potential model mismatches} and yield performance degradation when applied to a particular user since the real-time environment for the specific user, \textcolor{black}{determined by user's mobility, distance to the transmitter, geographical locations (urban or rural areas), etc.}, may differ from the general offline environment as shown in Fig.~\ref{fig:cor}.
%the policy gradient method used in \cite{a3c,icnp18,sigcomm17} separates the original expected value function maximization problem\cite{a3c} into two separate optimization problems and solves them alternatively. The separation and optimization methods are rather heuristic and yield less performance guarantee since they fail to fully capture the interactions between the policy and value parameters.

%The neural network framework\cite{a3c,sigcomm17} for implementing the policy gradient method separately trains the actor and critic networks. 

In this paper, we would like to address the above limitations. In particular, we boost the performance (users' QoE) of adaptive wireless video streaming by incorporating lower-layer information in the problem formulation, obtaining an enhanced offline policy via improving the training method for A3C, and fine-tuning the offline policy based on the online collected samples. Our main
contributions are summarized below.
\begin{itemize}
  \item We model the \textit{impacts of lower-layer information} (e.g., downlink media access control (MAC) rate, \textcolor{black}{number of} occupied physical resource blocks (PRBs), and modulation and coding scheme (MCS) \textcolor{black}{index}) together with the APP layer information on adaptive wireless video streaming. Note that such information is usually not Markovian. By additionally incorporating past and lower-layer information, we formulate a more comprehensive and accurate adaptive wireless video streaming problem as an infinite stage discounted MDP problem, offering possibilities for \textcolor{black}{enhancing} QoE. The proposed formulation allows a flexible tradeoff \textcolor{black}{between} QoE and costs for obtaining system information \textcolor{black}{and solving the problem}. 
  \item %We parameterize the policy and formulate an expected value function maximization problem under the policy. Using some optimization techniques, we equivalently transform the problem into an unconstrained problem which incorporates the Bellman equation constraint. 
 We propose an \textit{enhanced A3C method} for the offline scenario \textcolor{black}{(only with pre-collected data)}, improving the state-of-art A3C method\cite{a3c}.
  %by adopting a parameterized random policy and value function and jointly optimizing the policy and value parameters. 
  %The enhanced policy gradient method has better performance and convergence guarantee comparing to the previous policy gradient method\cite{a3c}. 
  Specifically, we build an A3C network that consists of a policy neural network and a value neural network \textcolor{black}{to represent a parameterized policy and a parameterized value function, respectively} and take cross-layer information and past and current information as input. Moreover, we jointly train the \textcolor{black}{policy and value neural networks (i.e., jointly optimize the policy and value parameters)} using pre-collected samples. \textcolor{black}{The joint optimization can capture intrinsic relationship between policy and value parameters and thus} achieve fast convergence speed and better convergent performance. 
 %to produce a bitrate adaptation policy with good generalization ability. The experimental results show that the enhanced policy gradient can reduce the training time by 4.6\%-6.7\% comparing to previous policy gradient method\cite{a3c}. 
  \item  We propose two \textit{\textcolor{black}{continual} learning-based online tuning methods} for the online scenario \textcolor{black}{(with additional real-time data)} to design better \textcolor{black}{policies} for a specific user leveraging the offline trained A3C network and \textcolor{black}{online} collected  samples. The first online tuning method builds a new policy network based on the policy network of the offline trained eA3C network, adopts the value network of the offline trained eA3C network, and trains the tunable components of its policy network \textcolor{black}{(i.e., optimizes the newly added parameters of the online policy)}. The optimization of the tunable \textcolor{black}{components of the new} policy network can be viewed as a constrained policy improvement given the value function under the offline policy.
  The second online tuning method builds new policy and value networks based on the policy and value networks of the offline trained eA3C network, respectively, and jointly trains the tunable components of its policy and value networks \textcolor{black}{(i.e., jointly optimizes the newly added parameters of the online policy and the value function)}. The joint optimization of the tunable components of the policy and value networks can be viewed as an online tuning version \textcolor{black}{of the proposed eA3C method}. 
  %\textcolor{black}{The network structures corresponding to the two online tuning methods are set flexible to capture the temporal features of the system state.} 
  These two \textcolor{black}{continual} learning-based online tuning methods produce online polices with better \textcolor{black}{QoE} for the specific user than the policy of the offline trained eA3C network. The second online tuning method \textcolor{black}{achieves higher QoE with a longer} training time \textcolor{black}{than the first online tuning method}.
  \item We collect a bunch of datasets that contain the APP layer throughput and some lower-layer \textcolor{black}{quantities} and evaluate the proposed and existing methods. The experimental results firstly show that the proposed offline policy can improve the QoE by $6.8\% \sim 14.4\%$ compared to the state-of-arts in the offline scenario, which reveals the importance of utilizing lower-layer, past and current information and the advantage of the enhanced policy gradient method. The experimental results secondly show that the proposed online policies can achieve $6\% \sim 28\%$ gains in QoE compared to the proposed offline policy in the online scenario, indicating the advantage of online tuning for video streaming which is a long-lasting application.
\end{itemize}

\begin{table}[t]
\caption{KEY NOTATION AND ABBREVIATIONS.}
\begin{center}
\textcolor{black}{
\resizebox{9cm}{!}{
\begin{tabular}{|c|c|} 
\hline 
Notation & Description \\ \hline  
$B$ & buffer size (in seconds)   \\ \hline
$T$ & length of each chunk (in seconds)\\ \hline
$N~(\mathcal{N})$ & number (set) of chunks\\ \hline
$D~(\mathcal{D})$ &  number (set) of quality levels \\ \hline
$M~(\mathcal{M})$ & number (set) of lower-layer quantities   \\ \hline
$R_{n}~(\mathcal{R})$ &  bitrate (action) of $n$-th chunk (bitrate set) \\ \hline
$\mathbf{C}_{n},B_{n}$ &  APP layer throughput, buffer occupancy at stage $n$  \\ \hline
$\overline{\mathbf{X}}_{n}$ & lower-layer quantities at stage $n$   \\ \hline
%$B_{n}$ & buffer occupancy at stage $n$  \\ \hline
$Y_{n}$ & bitrate of $(n-1)$-th chunk \\\hline
$\mathbf{S}_{n}\triangleq (B_{n},\mathbf{C}_{n},\overline{\mathbf{X}}_{n},Y_{n}) $ & system state at stage $n$ \\\hline
$\pi(\cdot;\bm{\Theta}_{\pi})$ & parameterized offline policy \\ \hline
$V^{\pi}(\cdot;\bm{\Theta}_{v})$ & parameterized value function under offline policy $\pi(\cdot;\bm{\Theta}_{\pi})$ \\\hline
$\hat{\pi}(\cdot;\hat{\bm{\Theta}}_{\pi_{\text{OTP}}},\bm{\Theta}^{\star}_{\pi})$ & parameterized online policy of eA3C-OTP \\ \hline
$\hat{\pi}(\cdot;\hat{\bm{\Theta}}_{\pi_{\text{OTPV}}},\bm{\Theta}^{\star}_{\pi})$ & parameterized online policy of eA3C-OTPV \\ \hline
$V^{\pi_{\text{OTPV}}}(\cdot;\bm{\Theta}^{\star}_{v},\hat{\bm{\Theta}}_{v_{\text{OTPV}}})$ & parameterized value function under online policy $\hat{\pi}(\cdot;\hat{\bm{\Theta}}^{\star}_{\pi_{\text{OTPV}}},\bm{\Theta}^{\star}_{\pi})$ \\\hline
$\bm{\Theta}_{\pi}~(\bm{\Theta}_{v})$ & eA3C policy (value) network parameters \\\hline
$\hat{\bm{\Theta}}_{\pi_{\text{OTP}}}$ & eA3C-OTP policy network tunable parameters \\\hline
$\hat{\bm{\Theta}}_{\pi_{\text{OTPV}}}~(\hat{\bm{\Theta}}_{v_{\text{OTPV}}})$ & eA3C-OTPV policy (value) network tunable parameters \\\hline
\textcolor{black}{Abbreviations} & Description \\ \hline  
MDP & Markov decision process\\ \hline
APP & Application\\\hline
PRB & Physical layer resource block \\\hline
MAC & Media access control \\\hline
MCS & Modulation and coding scheme \\\hline
(eA3C) A3C & (Enhanced) asynchronous advantage actor-critic \\\hline
eA3C-OTP & eA3C with online tuning for policy network \\\hline
eA3C-OTPV & eA3C with online tuning for policy and value networks \\\hline
DP & Dynamic programming \\\hline
DRL & Deep reinforcement learning\\\hline
DQ & Deep Q-learning\\\hline
MPC & Model predictive control\\\hline
RMSP & Root mean squared propagation\\\hline
\end{tabular}
}} \label{table0}
\end{center}
\end{table}

\textcolor{black}{
\textbf{Notation:} We represent scalar constants by non-boldface letters (e.g., $x$), sets by calligraphic letters
(e.g., $\mathcal{X}$ ), and sets of sets by boldface calligraphic letters (e.g., $\boldsymbol{\mathcal{X}}$). $*$ denotes the convolution operation. $\text{vec}(\cdot)$ denotes vectorization. $\mathbf{0}_{M\times M}$ denotes the $M\times M$ zero matrix. $\mathbb{N}$ denotes the set of natural numbers.
}
%\subsection{Related Work}

%\subsubsection{\textcolor{black}{continual} Learning for Deep Reinforcement Learning}
%A number of previous papers have used \textcolor{black}{continual} learning to tackle the challengings faced by the deep reinforcement learning where the knowledge is \textcolor{black}{continual}ed from the external expertise to facilitate the efficiency and effectiveness of the another reinforcement learning process. Out of a lot of \textcolor{black}{continual} approaches, representation \textcolor{black}{continual} is one of the most useful approaches. Specifically, \cite{tflearning,pathnet} propose a progressive neural network which enables knowledge \textcolor{black}{continual} across multiple reinforcement learning tasks in a progressive way. However, the numbers of hidden layers and neurons of the progressive networks in  \cite{tflearning,pathnet} are usually fixed which may lead to poor performance for some learning tasks. 
%Our motivation stems primarily from the fact that the structure of the progressive network can be tuned for different tasks. Thus, we adjust the the numbers of hidden layers and neurons of the progressive network corresponding to the environment in our system and make the network more flexible. 

\section{System Model}
In this section, we introduce the system model for adaptive wireless video streaming 
%of one video from a base station (BS) to a user 
using information from the APP layer and lower layers. \textcolor{black}{The key notation and abbreviations used in this paper is listed in Table~\ref{table0}.}
\subsection{System Model}
\begin{figure*}[t]
\begin{center}
 {\resizebox{16cm}{!}{\includegraphics{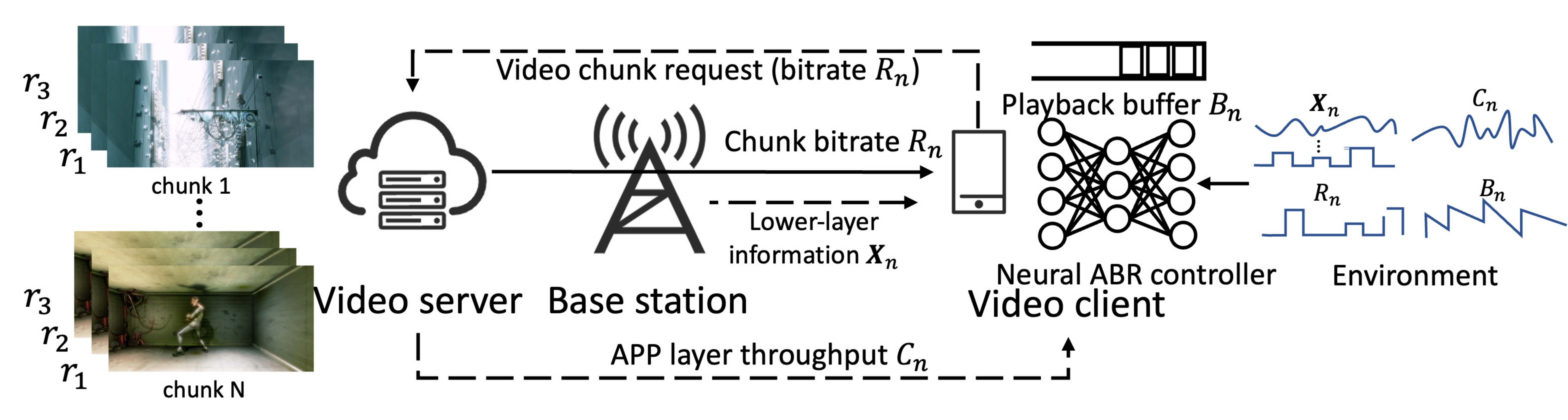}}}
 %\vspace*{-0.57cm}
\end{center}
   \caption{System model. $D = 3, \mathcal{D} = \{1,2,3\}, \mathcal{R} = \{r_{1},r_{2},r_{3}\}$.}
   \label{fig:system}
%\vspace*{-0.60cm}
\end{figure*}

As illustrated in Fig.~\ref{fig:system}, we consider adaptive wireless video streaming.
%of one video from a BS to a user.
\textcolor{black}{On the server side}, the video is segmented into $N$ chunks. Let $\mathcal{N}\triangleq \{1,\ldots,N\}$ denote the set of video chunks. The length of each chunk is $T$ (in second). The \textcolor{black}{typical} value of $T$ is $1\sim 4$.
%Considering user heterogeneity
%(e.g., in cellular usage costs, display resolutions of devices,
%channel conditions, etc.), 
To well adapt to the network conditions, each chunk is pre-encoded into $D$
representations corresponding to $D$ quality levels using High
Efficiency Video Coding (HEVC), as in Dynamic Adaptive
Streaming over HTTP (DASH). Let $\mathcal{D} \triangleq \{1,\ldots, D\}$ denote the set of $D$ quality levels. 
%For all $d\in\mathcal{D}$, the $d$-th representation of each chunk corresponds to the $d$-th lowest quality. 
For ease of exposition, assume that the bitrates of the chunks with the same quality level are identical\cite{JSAC14,sigcomm14}.\footnote{Various chunks of the same quality may have different bitrates due to their distinct spatial redundancies. The variation of bitrates corresponding to the same quality is usually small in practice and hence is ignored for tractability \cite{JSAC14,sigcomm14}.} \textcolor{black}{For all $d\in\mathcal{D}$,} the bitrate of the $d$-th representation of a chunk is denoted by $r_{d}$ (in bits/s). Note that $r_{d},d\in\mathcal{D}$ satisfy $r_{1} < \cdots < r_{D}$. \textcolor{black}{Let $\mathcal{R}\triangleq \{r_{1},\ldots,r_{D}\}$ denote the set of bitrates corresponding to the $D$ quality levels.}
The $N$ chunks will be sequentially transmitted to the user (downloaded by the user). For each chunk, only one of the $D$ representations will be transmitted to the user. Let $R_{n}$ (in bits/s) denote the \textcolor{black}{(APP layer)} bitrate of (the representation of) the $n$-th chunk that is transmitted to the user, where
\begin{align}
R_{n} \in \mathcal{R},~n\in\mathcal{N}.\label{eq:bitrate_selection}
\end{align}

Let $C_{n}\in\mathcal{C}$ represent the average \textcolor{black}{APP layer throughput} experienced during the download process for chunk $n\in\mathcal{N}$, \textcolor{black}{where $\mathcal{C}$ denotes the \textcolor{black}{APP layer throughput} space which is assumed to be a finite set for simplicity.} We model $C_{n},n\in\mathcal{N}$ as random variables, since the APP layer throughput usually varies over time (\textcolor{black}{at the scale of seconds}) due to the change in large-scale fading (including path loss and shadowing) between the base station and the user caused by the movement of the user, the change in the communication resources allocated to the user, etc.\footnote{\textcolor{black}{$C_{n}$ reflects the average of small-scale fading over $1\sim 4$ seconds.}} Then, the download time (in seconds) for chunk $n$ is given by $\frac{R_{n}T}{C_{n}}$. As the APP layer relies on all lower-layers to complete its process, and lower-layers are able to respond more rapidly to the changes in the environment as shown in Fig.~\ref{fig:cor} (a) and (b), information from the lower-layers, \textcolor{black}{such as MAC rate, PRB number, and MCS index,} may also be helpful for adaptive wireless video streaming. Therefore, \textcolor{black}{it is important to} explicitly model lower-layer information\cite{NG,mobicom15,TMC06}. Specifically, let $M$ and $\mathcal{M}\triangleq \{1,\ldots, M\}$ denote the number and set of lower-layer \textcolor{black}{quantities}, respectively. For all $n\in\mathcal{N}$ and $m\in\mathcal{M}$, let $X_{n,m}$ denote the value of the $m$-th lower-layer quantity (\textcolor{black}{e.g., MAC rate, PRB number or MCS index}) when downloading chunk $n$, referred to as the $m$-th lower layer information. Similarly, we model $X_{n,m},n\in\mathcal{N},m\in\mathcal{M}$ as random variables to reflect the possible changes \textcolor{black}{over time}. Denote $\mathbf{X}_{n} \triangleq (X_{n,m})_{m\in\mathcal{M}} \in \boldsymbol{\mathcal{X}}$ as the overal lower-layer information when downloading chunk $n$ where $\boldsymbol{\mathcal{X}}$ denotes the overall lower-layer information space which is assumed to be a finite set for simplicity. Mathematically, $C_{n},n\in\mathcal{N}$ and $\mathbf{X}_{n}, n\in\mathcal{N}$ are characterized by probability distributions $\Pr[C_{n}|C_{n-1},C_{n-2},\ldots, C_{1},\mathbf{X}_{n},\mathbf{X}_{n-1},\ldots,\mathbf{X}_{1}]$ and $\Pr[\mathbf{X}_{n}|\mathbf{X}_{n-1},\ldots,\mathbf{X}_{1}]$, respectively.\footnote{\textcolor{black}{We assume that the lower-layer information can be obtained by the user\cite{NG,mobicom15,TMC06}}.} It is worth noting that most existing work fails to utilize the lower-layer information in adaptive wireless video streaming, leading to \textcolor{black}{nonnegligible} performance loss. The appealing advantages of properly modeling and utilizing lower-layer information will be evidenced in Section~\ref{sec:sim} by extensive experiments.

The $N$ video chunks are downloaded into a playback buffer \textcolor{black}{at the user side}, which contains downloaded but as yet unviewed video chunks. Let $B>0$ denote the buffer size (in seconds) which depends on the policy of the service provider and storage limitation on the user. A typical buffer may hold a few tens of seconds of video chunks. Let $B_{n}\in [0,B]$ denote the \textcolor{black}{(APP layer)} buffer occupancy (in seconds), i.e.,
the play time of the video chunks left in the buffer, when the user starts to download chunk $n\in\mathcal{N}$. Thus, the buffer occupancy evolves according to\footnote{As in \cite{sigcomm15,sigcomm17}, we assume that the user immediately starts to download chunk $n+1$ as soon as chunk $n$ is downloaded. The only exception is that when the buffer is full, the user waits for the buffer to reduce to a certain level, resulting in the suspension of downloading chunk $n+1$.}
\begin{align}
&B_{n+1} = \left(\left(B_{n} - \frac{R_{n}T}{C_{n}}\right)^{+} + T, B \right)^{+}, ~n=1,\ldots, N-1,\label{eq:buffer_evolve}
\end{align}
where $(x)^{+} \triangleq \max\{x,0\}$. \textcolor{black}{Let $\mathcal{B}$ denote the buffer occupancy space. Since $\mathcal{R}$ and $\mathcal{C}$ are finite sets, $\frac{R_{n}}{C_{n}}$ falls in a finite set, implying that $\mathcal{B}$ is a finite set.} Note that $B_{n},n\in\mathcal{N}$ are also random due to \textcolor{black}{the randomness of} $C_{n},n\in\mathcal{N}$. The buffer update equation in \eqref{eq:buffer_evolve} indicates that the buffer occupancy increases by $T$ seconds after each chunk is downloaded and decreases as the user watches the video. While downloading chunk $n$, the buffer occupancy keeps on decreasing if $B_{n} \geq \frac{R_{n}T}{C_{n}}$, as the user is continuously watching the video. The buffer is empty, leading to rebuffering of time $\left(\frac{R_{n}T}{C_{n}} - B_{n} \right)^{+}$, \textcolor{black}{if $B_{n} < \frac{R_{n}T}{C_{n}}$.}

%Note that in the practical system, the number of information is pretty large and it is unrealistic to convey all information between the network and application. So in the following, we will formulate the adaptive video streaming as a Partially Observable Markov Decision Process (POMDP) problem by jointly considering download rate at the application layer and network information and identify what information can be sent from networks to applications.
%as well as the frequency of information delivering.
\subsection{Performance Metrics}
We refer to the stage where chunk $n$ is being downloaded as stage $n$, for all $n\in \mathcal{N}$. The system evolves over $N$ stages. 
%Recalling that the system information states are random, we would like to optimize the bitrates of the $N$ chunks to maximize the average QoE of the user over the $N$ stages. 
We consider three metrics for QoE \textcolor{black}{of video streaming}, i.e., video quality, video quality variation, and rebuffering time. Specifically, let $U(R)$ denote the utility for a chunk with the bitrate $R$. Here, $U(\cdot)$ can be any nonnegative, nondecreasing, and concave function \textcolor{black}{that is bounded on the finite set $\mathcal{D}$}, and $U(0) = 0$.\footnote{Logarithmic functions satisfy the requirements on $U(\cdot)$. Besides, numerical results show that Peak Signal-to-Noise Ratio (PSNR) and Structural Similarity Index Measure (SSIM) also satisfy the requirements on $U(\cdot)$\cite{PSNR,SSIM}.} Its monotonicity can capture the notion that perceptual quality increases with bitrate. Its concavity can capture the notion that the increase rate of perceptual quality decreases with bitrate. We define the \textcolor{black}{user's} QoE at stage $n$
%under a given stationary unichain policy $\pi$ 
as:
\begin{align}
g(C_{n},B_{n},R_{n},R_{n-1}) \triangleq & ~   U(R_{n}) - \alpha|U(R_{n}) - U(R_{n-1})|\nonumber\\
  &~~~~~~~ - \beta\left(\frac{R_{n}T}{C_{n}} - B_{n} \right)^{+},\label{eq:per_stage_reward}
\end{align}
where $\alpha>0$ and $\beta>0$ are the associated weights for the video quality variation (\textcolor{black}{i.e., quality smoothness}) $|U(R_{n}) - U(R_{n-1})|$\cite{sigcomm17,sigcomm15} and rebuffering time (\textcolor{black}{i.e., stall time}) $\left(\frac{R_{n}T}{C_{n}} - B_{n} \right)^{+}$\cite{sigcomm17,sigcomm15}, respectively. \textcolor{black}{The weight  $\alpha$ and $\beta$ can be thought of as quantifying the desire to make the video quality variation and rebuffering time small, respectively, and provide the trade-offs among three metrics.}

\section{QoE Maximization Problem Formulation}
\textcolor{black}{Note that a video usually consists of a large number of chunks (e.g., a 40-minute video contains 600-2400 chunks). Thus, in the rest of the paper, we consider the extreme scenario where $N\rightarrow\infty$, and $\mathcal{N}$ turns to the set of natural numbers, denoted by $\mathbb{N}$, for tractability.} % in Section~\ref{sec:MDP}.
In this section, we formulate the adaptive wireless streaming problem as an infinite stage discounted MDP under certain conditions. Specifically, to ensure a stationary MDP, we first adopt the following assumption.
\textcolor{black}{
\begin{Asump}[\textcolor{black}{Stationary Markov Chains of Order $k$}]\label{asmp:1}
%For all , the random APP layer throughput $C_{n}$ and lower-layer information $\mathbf{X}_{n}$ are characterized by 
For all $n\in\mathbb{N}$ and for some $k\in\mathbb{N}^{+}$, the probability distributions $\Pr[C_{n}|C_{n-1},C_{n-2},\ldots, C_{1},\mathbf{X}_{n},\mathbf{X}_{n-1},\ldots,\mathbf{X}_{1}]$ and $\Pr[\mathbf{X}_{n}|\mathbf{X}_{n-1},\ldots,\mathbf{X}_{1}]$ satisfy:
\begin{dmath*}
	\Pr[C_{n}|C_{n-1},C_{n-2},\ldots, C_{1},\mathbf{X}_{n},\mathbf{X}_{n-1},\ldots,\mathbf{X}_{1}] \nonumber\\
	= \Pr[C_{n}|C_{n-1},C_{n-2},\ldots,C_{n-k},\mathbf{X}_{n},\mathbf{X}_{n-1},\ldots,\mathbf{X}_{n-k}],\nonumber\\
	\Pr[\mathbf{X}_{n}|\mathbf{X}_{n-1},\ldots,\mathbf{X}_{1}] = \Pr[\mathbf{X}_{n}|\mathbf{X}_{n-1},\mathbf{X}_{n-2},\ldots,\mathbf{X}_{n-k}],\nonumber
\end{dmath*} and are independent of stage $n$.
\end{Asump}
}
\textcolor{black}{Under Assumption~\ref{asmp:1}, 
$\{(\mathbf{X}_{n},\mathbf{X}_{n-1},\ldots,\mathbf{X}_{n-k+1}):n\in\mathbb{N}\}$ and $\{(C_{n},C_{n-1},\ldots,C_{n-k+1},\mathbf{X}_{n},\mathbf{X}_{n-1},\ldots,\mathbf{X}_{n-k+1}):n\in\mathbb{N}\}$ \textcolor{black}{are stationary Markov chains of order $k$}. }Furthermore, we define the system state at stage $n$, denoted by $\mathbf{S}_{n}$, as follows. Given Assumption~\ref{asmp:1}, we include $\mathbf{C}_{n} \triangleq (C_{n},C_{n-1},\ldots,C_{n-k+1})$ and $\overline{\mathbf{X}}_{n} \triangleq (\mathbf{X}_{n},\mathbf{X}_{n-1},\ldots,\mathbf{X}_{n-k+1})$ in $\mathbf{S}_{n}$ by using state augmentation\cite[pp. 38]{dp}. Besides, noting that for all $n\in\mathbb{N},$ $g(C_{n},B_{n},R_{n},R_{n-1})$ in \eqref{eq:per_stage_reward} depends not only on $(C_{n},B_{n}, R_{n})$ but also on $R_{n-1}$,\footnote{\textcolor{black}{The video quality and video quality variation are related to $R_{n}$ and $R_{n-1}$, the rebuffering time depends on $C_{n}$ and $B_{n}$.}} we include $R_{n-1}$ in $\mathbf{S}_{n}$ by using action augmentation\cite[pp. 38]{dp}.
%Thus, Problem~\ref{prob:QoE_max} cannot be solved directly by the standard algorithms for MDP problems. 
Specifically, we introduce auxiliary variable $Y_{n}$ at stage $n$, and let 
\begin{align}
Y_{n} = R_{n-1},~n\in\mathbb{N}.\label{eq:state_augmentation}
\end{align}
Eventually, we have $\mathbf{S}_{n}\triangleq (B_{n},\mathbf{C}_{n},\overline{\mathbf{X}}_{n},Y_{n}) \in \bm{\mathcal{S}}$, where $\bm{\mathcal{S}}=\mathcal{B}\times\mathcal{C}^{k}\times\boldsymbol{\mathcal{X}}^{k}\times \mathcal{R}$ denotes the system state space which is a finite state space due to the \textcolor{black}{finiteness} of $\mathcal{B}$, $\mathcal{C}$, $\boldsymbol{\mathcal{X}}$, and $\mathcal{R}$. Define $\mathbf{s}_{n} \triangleq (b_{n},\mathbf{c}_{n},\overline{\mathbf{x}}_{n},y_{n}),n\in\mathbb{N}$. Therefore, by Assumption~\ref{asmp:1} and the definition of the system state, we can show the following result on the transition probabilities of the system. 
\begin{Lem}[Transition Probabilities]\label{theorem}
For all $n\in\mathbb{N},$
%and $\mathbf{s}_{i}\in\boldsymbol{\mathcal{S}}, r_{i}\in\mathcal{D}, i=1,\ldots,n+1,$
\begin{align}
&\Pr[\mathbf{S}_{n+1} = \mathbf{s}_{n+1}|\mathbf{S}_{n}= \mathbf{s}_{n},\ldots,\mathbf{S}_{1} = \mathbf{s}_{1}, R_{n} = r_{n},\ldots R_{1} = r_{1}]\nonumber\\
& = \Pr[\mathbf{S}_{n+1} = \mathbf{s}_{n+1}|\mathbf{S}_{n}= \mathbf{s}_{n}, R_{n} = r_{n}] \nonumber\\
& = \Pr[\mathbf{C}_{n+1} = \mathbf{c}_{n+1} | \overline{\mathbf{X}}_{n+1}= \overline{\mathbf{x}}_{n+1}, \mathbf{C}_{n} = \mathbf{c}_{n},\overline{\mathbf{X}}_{n} = \overline{\mathbf{x}}_{n}] \nonumber\\
&\times \Pr[\overline{\mathbf{X}}_{n+1} = \overline{\mathbf{x}}_{n+1} |\overline{\mathbf{X}}_{n} =  \overline{\mathbf{x}}_{n}] \nonumber\\
&\times \mathbb{I}\left[b_{n+1} = \left(\left(b_{n}-\frac{r_{n}T}{c_{n}}\right)^{+} +T,B\right)^{+} \right] \times \mathbb{I}\left[y_{n+1} = r_{n}\right],
\end{align}
which is independent of $n$. Here, $\mathbb{I}[\cdot]$ denotes the indicator function.
\end{Lem}
\begin{Proof}
Please refer to Appendix A.
\end{Proof}

Define $\mathbf{s}' = (b',\mathbf{c}',\overline{\mathbf{x}}',y')$ and $\mathbf{s} = (b,\mathbf{c},\overline{\mathbf{x}},y)$. Based on Lemma~\ref{theorem}, for all $n\in\mathbb{N}$, $\mathbf{s},\mathbf{s}' \in\boldsymbol{\mathcal{S}}$ and $r\in\mathcal{R}$, we denote $\Pr[\mathbf{S}_{n+1} = \mathbf{s}'|\mathbf{S}_{n}= \mathbf{s}, R_{n} = r]$ by $p_{\mathbf{s},\mathbf{s}'}(r)$, referred to as the transition probability from the system state $\mathbf{s}$ to a successor system state $\mathbf{s}'$ for a given bitrate $r$. 

We refer to the bitrate at stage $n$, $R_{n}$, as the action at stage $n$ and term $\mathcal{R}$ as the action space.
Consider a stationary \textcolor{black}{randomized (bitrate adaptation) policy, denoted by $\pi$, which is a function that maps the system state $\mathbf{s}$ into a probability distribution $\pi(\mathbf{s},r),r\in\mathcal{R}$. Note that for all $\mathbf{s}\in\boldsymbol{\mathcal{S}}$, $\pi(\mathbf{s},\cdot)$ satisfies $\pi(\mathbf{s},r) \geq 0,r\in\mathcal{R}$ and $\sum_{r\in\mathcal{R}}\pi(\mathbf{s},r) =1$.} Under Assumption~\ref{asmp:1} and the stationary randomized policy $\pi$, we have a homogeneous Markov chain $\{\mathbf{S}_{n}:n\in\mathbb{N}\}$ with transition probabilities $\sum_{r\in\mathcal{R}}\pi(\mathbf{s},r)p_{\mathbf{s},\mathbf{s}'}(r),\mathbf{s}',\mathbf{s}\in\boldsymbol{\mathcal{S}}$.
%By Assumption~\ref{asmp:1} and under the stationary policy $\pi$, the system states $\mathbf{S}_{n},n\in\mathbb{N}$ are random variables with distributions defined by $\Pr[\mathbf{S}_{n+1} = \mathbf{s}_{n+1}|\mathbf{S}_{n} = \mathbf{s}_{n}, \pi(\mathbf{S}_{n}) = r_{n}] =p_{\mathbf{s},\mathbf{s}'}(\pi(\mathbf{s}))$.
Given the notion of the system state $\mathbf{S}_{n}$, we rewrite $g(C_{n},B_{n},R_{n},R_{n-1})$ as $g(\mathbf{S}_{n},R_{n}) = U(R_{n}) - \alpha|U(R_{n}) - U(Y_{n})| - \beta\left(\frac{R_{n}T}{C_{n}} - B_{n} \right)^{+}$, referred to as the reward at stage $n$.
Thus, given an initial state $\mathbf{s}$, the value function i.e., expected discounted QoE, under the stationary randomized policy $\pi$ is given by
\begin{align}
%q(\pi) \triangleq u(\pi) - \alpha_{v} v(\pi) - \beta_{h}h(\pi) = 
V^{\pi}(\mathbf{s}) = \limsup_{N\rightarrow\infty} \mathbb{E}\left[ \sum^{N}_{n=1} \gamma^{n-1}\sum_{r\in\mathcal{R}}\pi(\mathbf{S}_{n},r)g(\mathbf{S}_{n},r)\right],\label{eq:QoE}
\end{align}
where $\mathbf{S}_{1} = \mathbf{s}$, the system states $\mathbf{S}_{n},n=1,2,3,\ldots$ evolve according to $\sum_{r\in\mathcal{R}}\pi(\mathbf{s},r)p_{\mathbf{s},\mathbf{s}'}(r),\mathbf{s}',\mathbf{s}\in\boldsymbol{\mathcal{S}},$ and the expectation is taken over $\mathbf{S}_{n},n=2,3,\ldots$. 

We aim to optimize the stationary randomized policy $\pi$ to maximize the \textcolor{black}{expected discounted QoE} given in \eqref{eq:QoE}.
%Note that $q_{n}(\mathbf{s}_{n},r_{n},r_{n-1}),n\in\mathcal{N}$ in Problem~\ref{prob:QoE_max} depends not only on the preceding state $\mathbf{s}_{n}$ and action $r_{n}$ but also on earlier action $r_{n-1}$, which violates the assummption of the basic MDP problem. Thus, Problem~\ref{prob:QoE_max} cannot be solved directly by the standard algorithms for MDP problems.
The problem is readily formulated as follows.

\begin{Prob}[QoE Maximization]\label{prob:QoE_max_MDP}
\begin{align}
V(\mathbf{s}) = \max_{\pi}~\limsup_{N\rightarrow\infty} \mathbb{E}\left[ \sum^{N}_{n=1} \gamma^{n-1}\sum_{r\in\mathcal{R}}\pi(\mathbf{S}_{n},r)g(\mathbf{S}_{n},r)\right],\nonumber
\end{align}
where $\mathbf{S}_{1} = \mathbf{s}$, and $V(\mathbf{s})$ represents the optimal value function. Let $\pi^{\star}$ denote an optimal stationary randomized policy.
\end{Prob}

Based on the above illustration, we can conclude that Problem~\ref{prob:QoE_max_MDP} is a stochastic discounted MDP over \textcolor{black}{an infinite number of stages}.
Note that for all $n\in\mathbb{N}$, $g(\mathbf{S}_{n},R_{n})$ is bounded since $\mathbf{S}_{n}$ and $R_{n}$ lie in finite spaces $\boldsymbol{\mathcal{S}}$ and $\mathcal{R}$, respectively, and $U(\cdot)$ is bounded on $\mathcal{R}$. \textcolor{black}{By \cite[Proposition 1.2.3]{dp_2} and \cite[Proposition 1.2.5]{dp_2}, we have the following result.}
%Note that the overall QoE per stage $q(\mathbf{s}_{n},r_{n})$ satisfies $q(\mathbf{s}_{n},r_{n}) \leq U(R_{D})$,
%According to \cite[Theorem 6.2.6]{ddp}, for unichain infinite horizon discounted MDPs with finite state space and action space, there always exists a deterministic stationary policy that is optimal. Note that these requirements are satisfied by the MDP considered in our work, it is sufficient to focus on the deterministic stationary policy space. 
%\footnote{We restrict our attention to the stationary deterministic policy. If there exists an optimal stationary deterministic policy, there exists an optimal stationary randomized policy\cite[pp. 10]{dp}.}
\textcolor{black}{
\begin{Lem}[Optimal Value Function and Optimal Stationary Randomized Policy] 
The optimal value function $V(\cdot)$ satisfies\cite[Proposition 1.2.3]{dp_2}:
\begin{align}
V(\mathbf{s}) =  \max_{\pi} ~  V^{\pi}(\mathbf{s}), \quad \mathbf{s}\in\boldsymbol{\mathcal{S}},\label{eq:bellman_equ}
\end{align}
where $V^{\pi}(\cdot)$ is the value function under the stationary randomized policy $\pi$ and satisfies:
\begin{align}
	V^{\pi}(\mathbf{s}) &= \sum_{r\in\mathcal{R}}\pi(\mathbf{s},r)\left(g(\mathbf{s},r) + \gamma\sum_{\mathbf{s}'\in\boldsymbol{\mathcal{S}}}p_{\mathbf{s}',\mathbf{s}}(r) V^{\pi}(\mathbf{s}') \right) , \nonumber \\
&~~~~~~~~~~~~~~~~~~~~~~~~~~~~~~~~~~~~~~~~~~~~~~\mathbf{s}\in\boldsymbol{\mathcal{S}}.\label{eq:bellman_equ_policy}
\end{align}
Furthermore, an optimal stationary randomized policy $\pi^{\star}$ is given by\cite[Proposition 1.2.5]{dp_2}:
\begin{align}
\pi^{\star}(\mathbf{s},r) = \left\{ \begin{array}{ll}
\frac{1}{|\mathcal{R}^{\star}(\mathbf{s})|}, & \textrm{if $r\in \mathcal{R}^{\star}(\mathbf{s}) $}\\
0, & \textrm{otherwise}
\end{array} \right.
\quad \mathbf{s}\in\boldsymbol{\mathcal{S}},\label{eq:bellman_policy}
\end{align}
where $\mathcal{R}^{\star}(\mathbf{s}) \triangleq \argmax_{r\in\mathcal{R}} ~g(\mathbf{s},r)+ \gamma \sum_{\mathbf{s}'\in\boldsymbol{\mathcal{S}}}p_{\mathbf{s}',\mathbf{s}}(r) V(\mathbf{s}')$.
\end{Lem}
}

Standard DP algorithms such as value iteration, policy iteration (both applicable for the case where the transition probabilities are known), and Q-learning\cite[pp. 495]{dp_2} (applicable for the case where the transition probabilities are unknown) can be used to obtain an optimal stationary policy by solving the Bellman equation given in \eqref{eq:bellman_equ} and \eqref{eq:bellman_equ_policy}. However, their computational complexities are prohibitively high due to the extremely large state spaces. Additionally, Assumption~\ref{asmp:1} may not be satisfied in practical systems, \textcolor{black}{implying that Problem~\ref{prob:QoE_max_MDP} may not be viewed as a stochastic discounted MDP. Thus, those standard DP may not provide satisfactory QoE} in practice. \textcolor{black}{DRL, leveraging DL and approximate DP, is put forward as a solution to address the above limitations. A3C, a widely used DRL algorithm, demonstrates promising potential in the realm of adaptive video streaming. In Section~\ref{sec:A3C} and Section~\ref{sec_online_MDP}, we \textcolor{black}{present} novel methods to enhance the performance of A3C in the offline scenario (only with pre-collected data) and online scenario (with additional real-time data), respectively.}

\section{Enhanced A3C Method in the Offline Scenario}\label{sec:A3C}
\textcolor{black}{
%do not rely on Assumption~\ref{asmp:1} and 
%propose an actor-critic deep reinforcement learning (DRL) algorithm to deal with  . 
%DRL generally adopts value and/or policy approximation via deep neural networks.
%A variety of different algorithms could be used to train the learning agent in the abstract RL framework (e.g., DQN [29], REINFORCE [44], etc.). 
%because (1) to the best of our knowledge, it is the state-of-art and it has been successfully applied to many other concrete learning problems; and (2) in the video streaming application, the asynchronous parallel training framework supports online training in which many users concurrently send their experience feedback to the agent. 
%Actor-critic DRL algorithm\cite{a3c} is a widely used DRL method that is based on policy gradient method\cite[pp. 620]{dp_2}. 
The state-of-art A3C method\cite{a3c} breaks the original optimization problem for the policy and value parameters into two separate but related problems and alternatively trains the policy and value networks (i.e., solve the optimization problems for the value and policy parameters). Such training method is rather heuristic and hardly yields any convergence or performance guarantee.
In this section, we \textcolor{black}{consider the offline scenario only with pre-collected data from many users and} \textcolor{black}{present} an enhanced A3C (eA3C) method \textcolor{black}{which jointly trains policy and value networks. eA3C captures intrinsic relationship between policy and value parameters and thus reduces training time and improves QoE.} }

\subsection{Joint Optimization Formulation}\label{sec:off_formulation}
%We propose a  policy gradient method to solve the problem in \eqref{eq:bellman_equ}. 
%We parameterize the stationary policy $\pi$ and value function $V(\cdot)$ by the parameters $\bm{\Theta}_{\pi}\in\mathbb{R}^{n_{\pi}}$ and $\bm{\Theta}_{v}\in\mathbb{R}^{n_{v}}$, respectively.
%and reward function $V(\cdot)$ ,  and $\bm{\Theta}_{v}\in\mathbb{R}^{n_{v}}$, respectively.
\textcolor{black}{We approximate an arbitrary stationary randomized policy $\pi$ by a parametric form $\pi(\cdot;\bm{\Theta}_{\pi})$, with policy parameters $\bm{\Theta}_{\pi}\in\mathbb{R}^{n_{\pi}}$, and optimize $\bm{\Theta}_{\pi}$ instead of $\pi$ for tractability as in \cite{a3c,sigcomm17}.}
%Instead of allowing an arbitrary randomized policy, we restrict to a randomized stationary policy $\pi$, the distribution of which has a given parametric form\cite{a3c,sigcomm17}:\footnote{Note that in the rest of the paper, we refer $\pi(\cdot)$ to as a parameterized policy instead of an arbitrary policy with a slight abuse of notation.}
%\begin{align}
%\Pr[\pi(\mathbf{s}) = r] = \pi(\mathbf{s},r;\bm{\Theta}_{\pi}),~r\in\mathcal{D}, \label{eq:randomized_policy}
%\end{align}
%where $\pi(\cdot;\bm{\Theta}_{\pi})$ is a parameterized function, and \textcolor{black}{$\bm{\Theta}_{\pi}\in\mathbb{R}^{n_{\pi}}$ is the policy parameter}. 
%For a given $\bm{\Theta}_{\pi}$,
%$\pi(\mathbf{s},r;\bm{\Theta}_{\pi})$ represents the probability that the action $r$ is applied when the state is $\mathbf{s}$. Note that for all $\mathbf{s}\in\boldsymbol{\mathcal{S}}$, $\pi(\mathbf{s},\cdot;\bm{\Theta}_{\pi})$ satisfies $\pi(\mathbf{s},r;\bm{\Theta}_{\pi}) \geq 0,r\in\mathcal{D}$ and $\sum_{r\in\mathcal{D}}\pi(\mathbf{s},r;\bm{\Theta}_{\pi}) =1$. 
Specifically, we  optimize $\bm{\Theta}_{\pi}$ to maximize the expected value function (which is equal to the expected discounted QoE)\cite{rl,a3c}:
%For ease of presentation, define $J(\bm{\Theta}_{\pi},\mathbf{s}_{0}) =  V^{\pi}(\mathbf{s}_{0})$. Then, we 
%We consider a reward function under the randomized stationary policy $\pi$ of a given parametric form $V^{\pi}(\mathbf{s};\bm{\Theta}_{v})$. 
%which represents the reward at state $\mathbf{s}$. 
\begin{align}
\max_{\bm{\Theta}_{\pi}}&\quad \sum_{\mathbf{s}\in\boldsymbol{\mathcal{S}}}
\eta^{\pi}(\mathbf{s}) V^{\pi}(\mathbf{s}),\label{prob:offline_policy_gradient_formulation}
\end{align}
\iffalse
\begin{align}
\max_{\bm{\Theta}_{\pi},(V^{\pi}(\mathbf{s}))_{\mathbf{s}\in\boldsymbol{\mathcal{S}}}}&\quad \sum_{\mathbf{s}\in\boldsymbol{\mathcal{S}}}
\eta^{\pi}(\mathbf{s}) V^{\pi}(\mathbf{s}),\label{prob:offline_policy_gradient_formulation}\\
	&\mathrm{s.t.}\quad \eqref{eq:bellman_equ_policy},\nonumber
\end{align}
\fi
where $\eta^{\pi}(\cdot)$ is the stationary probability distribution of the system state under policy $\pi$, and $V^{\pi}(\cdot)$ is the solution of the Bellman equation in \eqref{eq:bellman_equ_policy} under policy $\pi$. Solving the problem in \eqref{prob:offline_policy_gradient_formulation} is challenging, since $V^{\pi}(\cdot)$ is a solution to an extremely large number of equations given by \eqref{eq:bellman_equ_policy} and cannot be obtained analytically.
%the number of the constraints in \eqref{eq:bellman_equ_policy}, $|\boldsymbol{\mathcal{S}}|$, is extremely large. Second, the number of optimization variables, $|\boldsymbol{\mathcal{S}}| + n_{\pi}$, is prohibitively large. 
To address the challenge, we equivalently transform the problem in \eqref{prob:offline_policy_gradient_formulation} \textcolor{black}{into the problem in \eqref{prob:policy_gradient_equal}, as shown at the top of the next page, with $(V^{\pi}(\mathbf{s}))_{\mathbf{s}\in\boldsymbol{\mathcal{S}}}$ as additional optimization variables,}
\begin{figure*}
\begin{align}
\max_{\bm{\Theta}_{\pi},(V^{\pi}(\mathbf{s}))_{\mathbf{s}\in\boldsymbol{\mathcal{S}}}}&\quad \sum_{\mathbf{s}\in\boldsymbol{\mathcal{S}}}
\eta^{\pi}(\mathbf{s})  \sum_{r\in\mathcal{R}} \pi(\mathbf{s},r;\bm{\Theta}_{\pi})
 \left(g(\mathbf{s},r) + \gamma \sum_{\mathbf{s}'\in\boldsymbol{\mathcal{S}}}p_{\mathbf{s},\mathbf{s}'}(r)V^{\pi}(\mathbf{s}') \right)\nonumber\\
 	&-\xi \sum_{\mathbf{s}\in\boldsymbol{\mathcal{S}}}
\eta^{\pi}(\mathbf{s})  
	\left(\sum_{r\in\mathcal{R}} \pi(\mathbf{s},r;\bm{\Theta}_{\pi}) \left( g(\mathbf{s},r) + \gamma \sum_{\mathbf{s}'\in\boldsymbol{\mathcal{S}}}p_{\mathbf{s},\mathbf{s}'}(r)V^{\pi}(\mathbf{s}')\right) - V^{\pi}(\mathbf{s})\right)^{2},\label{prob:policy_gradient_equal}
\end{align}
\hrulefill
\end{figure*}
where $\xi >0$. The equivalence is shown by the following lemma.
\textcolor{black}{\begin{Lem}[Relationship between the Problems in \eqref{prob:offline_policy_gradient_formulation} and \eqref{prob:policy_gradient_equal}]\label{lemma_2}
Consider a sequence $\{\xi^{k}\}$ with $0 < \xi^{k} <\xi^{k+1}$ for all $k = 0,1,2,\ldots$, and $\xi^{k} \rightarrow \infty$. Then every limit point of the sequence $\{\bm{\Theta}^{k}_{\pi}\}$ is a globally optimal point of the problem in \eqref{prob:offline_policy_gradient_formulation}, where $\bm{\Theta}^{k}_{\pi}$ represents a globally optimal point of the problem in \eqref{prob:policy_gradient_equal} with $\xi = \xi^{k}$.
\end{Lem}
}
\begin{Proof}
Please refer to Appendix B.
\end{Proof}

\textcolor{black}{Based on Lemma~\ref{lemma_2}, we can solve the problem in \eqref{prob:policy_gradient_equal} with a sufficiently large $\xi$ for tractability.} Note that the problem in \eqref{prob:policy_gradient_equal} is still challenging since the number of optimization variables, $|\boldsymbol{\mathcal{S}}| + n_{\pi}$, is prohibitively large. As in \cite{a3c,sigcomm17}, we approximate the value function under policy $\pi$, $V^{\pi}(\cdot)$, by a parametric form $V^{\pi}(\cdot;\bm{\Theta}_{v})$, with value parameters $\bm{\Theta}_{v}\in\mathbb{R}^{n_{v}}$. \textcolor{black}{Then, we consider the joint optimization of the policy and value parameters given by the problem in \eqref{prob:policy_gradient_equal_approxi}, as shown at the top of the next page.} Note that it is a simplified version of the problem in \eqref{prob:policy_gradient_equal}.
\begin{figure*}
\begin{align}
\max_{\bm{\Theta}_{\pi},\bm{\Theta}_{v}}&\quad \sum_{\mathbf{s}\in\boldsymbol{\mathcal{S}}}
\eta^{\pi}(\mathbf{s})  \sum_{r\in\mathcal{R}} \pi(\mathbf{s},r;\bm{\Theta}_{\pi})
 \left(g(\mathbf{s},r) + \gamma \sum_{\mathbf{s}'\in\boldsymbol{\mathcal{S}}}p_{\mathbf{s},\mathbf{s}'}(r)V^{\pi}(\mathbf{s}';\bm{\Theta}_{v}) \right)\nonumber\\
 	&-\xi \sum_{\mathbf{s}\in\boldsymbol{\mathcal{S}}}
\eta^{\pi}(\mathbf{s})  
	\left(\sum_{r\in\mathcal{R}} \pi(\mathbf{s},r;\bm{\Theta}_{\pi})\left( g(\mathbf{s},r) + \gamma \sum_{\mathbf{s}'\in\boldsymbol{\mathcal{S}}}p_{\mathbf{s},\mathbf{s}'}(r)V^{\pi}(\mathbf{s}';\bm{\Theta}_{v})\right) - V^{\pi}(\mathbf{s};\bm{\Theta}_{v})\right)^{2}.\label{prob:policy_gradient_equal_approxi}
\end{align}
\hrulefill
\end{figure*}
%Let $(\bm{\Theta}^{\star}_{v},\bm{\Theta}^{\star}_{\pi})$ denote an optimal solution of the problem in \eqref{prob:final_policy_gradient_formulation}. Then, $V^{\pi}(\cdot;\bm{\Theta}^{\star}_{v})$ can be regarded as an approximate solution of the equation~\eqref{eq:bellman_equ_policy} in norm $\|\cdot\|_{2}$.
and the number of the optimization variables reduces from $|\boldsymbol{\mathcal{S}}| + n_{\pi}$ to $n_{v} + n_{\pi}$. The problem in \eqref{prob:policy_gradient_equal_approxi} is generally a non-convex stochastic problem with unknown stationary probability distribution $\eta^{\pi}(\cdot)$ and transition probabilities $p_{\mathbf{s},\mathbf{s}'}(r),\mathbf{s},\mathbf{s}'\in\boldsymbol{\mathcal{S}},r\in\mathcal{D}$. \textcolor{black}{We can obtain a stationary point of the problem in \eqref{prob:policy_gradient_equal_approxi} using standard stochastic gradient methods\cite{siam}.}\footnote{One can use standard alternating optimization method to tackle the problem \eqref{prob:policy_gradient_equal_approxi}. \textcolor{black}{However, it does not guarantee to converge to a stationary point, because the optimization of $\bm{\Theta}_{\pi}$ (for fixed $\bm{\Theta}_{v}$) and the optimization of $\bm{\Theta}_{v}$ (for fixed $\bm{\Theta}_{\pi}$) are non-convex and usually cannot be solved optimally\cite{cvx}.}}

\begin{figure}[t]
\begin{center}
 {\resizebox{8cm}{!}{\includegraphics{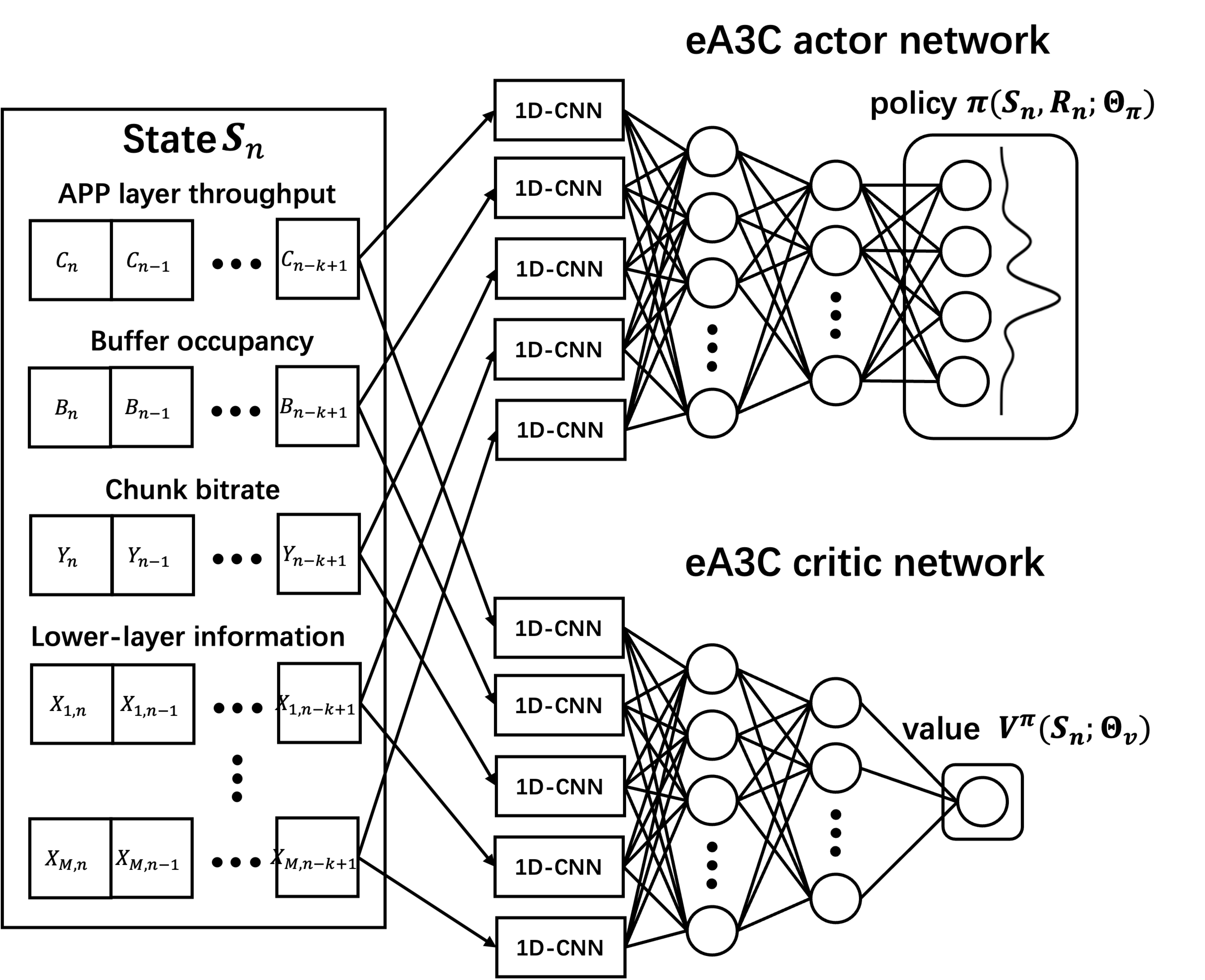}}}
 %\vspace*{-0.57cm}
\end{center}
   \caption{Structure of the eA3C network}
   \label{fig:A3C}
%\vspace*{-0.60cm}
\end{figure}

%In the following, we use neural networks to obtain the approximate policy $\pi_{\bm{\Theta}^{m}}(\cdot)$ and the approximate value function $\tilde{V}(\cdot)$. 
\subsection{Neural Network Architecture} We present an actor-critic neural network corresponding to the eA3C method, namely eA3C network. The eA3C network consists of two neural networks, namely actor network (policy network) and critic network (value network), as shown in Figure~\ref{fig:A3C}.
%to obtain the parameters $\bm{\Theta}_{\pi}\in\mathbb{R}^{n_{\pi}}$ and $\bm{\Theta}_{v}\in\mathbb{R}^{n_{v}}$ according to \eqref{eq:policy_parameter_update} and \eqref{eq:value_parameter_update}, respectively.
Specifically, the actor network has weights $\bm{\Theta}_{\pi}\in\mathbb{R}^{n_{\pi}}$ and produces $\pi(\mathbf{s},r;\bm{\Theta}_{\pi})$, whereas the critic network has weights $\bm{\Theta}_{v}\in\mathbb{R}^{n_{v}}$ and produces $V^{\pi}(\mathbf{s};\bm{\Theta}_{v})$.
The inputs of the two neural networks are the states over most recent $k$ chunks, $\mathbf{S}_{n}$. For ease of implementation, the inputs are rearranged as a $(M+3)\times k$ matrix, $[\mathbf{C}^{T}_{n}~\mathbf{B}^{T}_{n}~\mathbf{Y}^{T}_{n}~\mathbf{X}^{T}_{1,n}~\ldots~\mathbf{X}^{T}_{M,n}]$.
%\begin{displaymath}
%\left[ \begin{array}{c}
%\mathbf{C}^{T}_{n} \\
%\mathbf{B}^{T}_{n} \\
%\mathbf{Y}^{T}_{n}\\
%\mathbf{X}^{T}_{1,n}\\
%\vdots \\
%\mathbf{X}^{T}_{M,n}
%\end{array}
%\right] \in\boldsymbol{\mathcal{S}}^{k}, 
%\end{displaymath}
Here, $\mathbf{C}_{n} \triangleq (C_{n}, C_{n-1}, \ldots ,C_{n-k+1})\in\mathcal{C}^{k}$ denotes the most recent $k$ APP layer throughputs by stage $n$; $\mathbf{B}_{n} \triangleq (B_{n}, B_{n-1}, \ldots ,B_{n-k+1})\in\mathcal{B}^{k}$ denotes the most recent $k$ buffer occupancies by stage $n$; $\mathbf{Y}_{n} \triangleq (Y_{n}, Y_{n-1}, \ldots ,Y_{n-k+1})\in\mathcal{D}^{k}$ denotes the most recent $k$ selected bitrates by stage $n$; for all $m=1,\ldots,M$, $\mathbf{X}_{m,n} \triangleq (X_{m,n}, X_{m,n-1}, \ldots ,X_{m,n-k+1})\in\boldsymbol{\mathcal{X}}^{k}_{m}$ denotes the most recent $k$ $m$-th lower-layer information by stage $n$. \textcolor{black}{The notation of the weights and layer outputs of the eA3C network is summarized in Table~\ref{table_offline}.}

\begin{table}[t]
\caption{Notation of the weights and layer outputs of the eA3C network ($j = \pi $ or $v$). For ease of presentation, we let $\pi$ network and $v$ network represent actor network and critic network, respectively.}
\begin{center}
\resizebox{9cm}{!}{
\begin{tabular}{|c|c|} 
\hline 
Notation & Description \\ \hline  
$\mathbf{w}^{(i)}_{j,0}\in\mathbb{R}^{n_{j,0}}$ &  weight vector of $i$-th 1D CNN in input layer of $j$ network, $i = 1,\ldots, M+3$ \\ \hline
$\mathbf{W}_{j, 1} \in \mathbb{R}^{n_{j,1}\times (M+3)(k-n_{j,0}+1)}$ &  weight matrix of first hidden layer of $j$ network\\ \hline
$\mathbf{W}_{j,l} \in\mathbb{R}^{n_{j,l}\times n_{j,l-1}}$ &  weight matrix of $l$-th layers of $j$ network, $l = 2,\ldots, L_{j}-1$\\ \hline
$\mathbf{w}_{v,L_{v}}\in\mathbb{R}^{n_{v,L_{v}-1}}$ &  weight vector of output layer of critic network\\ \hline
$\mathbf{W}_{\pi,L_{\pi}}\in\mathbb{R}^{n_{\pi,L_{\pi}} \times n_{\pi,L_{\pi}-1}}$ &  weight vector of output layer of actor network\\ \hline
$\mathbf{h}_{j,l}\in\mathbb{R}^{n_{j,l}}$ &  output of $l$-th layer of $j$ network, $l = 0,1,\ldots,L_{j}-1$ \\ \hline
$h_{v,L_{v}} \in\mathbb{R}$ & output of $L_{v}$-th layer of critic network \\ \hline
$\mathbf{h}_{\pi,L_{\pi}} \in\mathbb{R}^{n_{\pi,L_{\pi}}}$ & output of $L_{\pi}$-th layer of actor network \\ \hline
\end{tabular}
} 
\label{table_offline}
\end{center}
\end{table}

\noindent\textbf{eA3C actor network:} The eA3C actor network has one input layer, $L_{\pi}-1$ hidden layers, and one output layer, indexed by $0,1,\ldots, L_{\pi}-1,L_{\pi}$, respectively. The input layer consists of $M+3$ one-dimensional convolutional neural networks (1D CNNs) which are used to extract the temporal features of the $M+3$ vectors, $\mathbf{C}_{n},\mathbf{B}_{n},\mathbf{Y}_{n},\mathbf{X}_{1,n},\ldots \mathbf{X}_{M,n}$, respectively. The lengths of the $M+3$ 1D CNNs' filters are the same, denoted by $n_{\pi,0}$. The $l$-th hidden layer is a fully connected layer with $n_{\pi,l}$ neurons that utilize the Rectified Linear Unit (ReLU) activation function. The output layer is a fully connected layer with $n_{\pi,L_{a}}$ neurons that utilize the softmax activation function. 
  %For all $l=0,1,\ldots,L$, let $\mathbf{h}_{\pi,l}$ denote the output of $l$-th layer. Then, we have:
%\begin{align}
%&\mathbf{h}_{\pi,0} 
%&\in \mathbb{R}^{(M+3) \times (k-g+1)}\nonumber\\
%= (\mathbf{w}^{(1)}_{\pi,0} * \mathbf{C}_{n},\mathbf{w}^{(2)}_{\pi,0} *\mathbf{B}_{n}, \mathbf{w}^{(3)}_{\pi,0} *\mathbf{Y}_{n}, \mathbf{w}^{(4)}_{\pi,0} * \mathbf{X}_{1,n},  \mathbf{w}^{(5)}_{\pi,0} *\mathbf{X}_{2,n}, \ldots,  \mathbf{w}^{(M+3)}_{\pi,0} *\mathbf{X}_{M,n} )^{T} ,\nonumber\\
%&\mathbf{h}_{\pi,l} = \textbf{ReLU}(\mathbf{W}_{\pi,l}\mathbf{h}_{\pi,l-1}),~l= 1,\ldots,L-1,\nonumber\\
%&\mathbf{h}_{\pi,L} = \textbf{Softmax}(\mathbf{W}_{\pi,L}\mathbf{h}_{\pi,L-1}),\nonumber
%&\mathbf{h}_{\pi,L} = \pi(\mathbf{S}_{n},f_{i};\bm{\Theta}_{\pi,n}) =  \frac{\mathbf{w}^{T}_{\pi,L,i}\mathbf{h}_{L-1}}{\sum_{i=1}^{n_{\pi,L}}\mathbf{w}^{T}_{\pi,L,i}\mathbf{h}_{L-1}},~i=1,\ldots,n_{\pi,L}, \nonumber
%\end{align}
%Let $\mathbf{w}^{(i)}_{\pi,0}\in\mathbb{R}^{n_{\pi,0}}$ denote the weight vector of the $i$-th 1D CNN for all $i = 1,\ldots, M+3$, let $\mathbf{W}_{\pi, 1} \in \mathbb{R}^{n_{\pi,1}\times (M+3)(k-n_{\pi,0}+1)}$ denote the weight matrix of the first hidden layer, and let $\mathbf{W}_{\pi,l} \in\mathbb{R}^{n_{\pi,l}\times n_{\pi,l-1}}$ denote the weight matrix of the $l$-th layers for all $l = 2,\ldots, L_{a}$. 
We form a column vector $\bm{\Theta}_{\pi} = (\mathbf{w}^{(1)}_{\pi,0},\mathbf{w}^{(2)}_{\pi,0},\ldots, \mathbf{w}^{(M+3)}_{\pi,0}, \text{vec}(\mathbf{W}_{\pi,1}),\ldots, \text{vec}(\mathbf{W}_{\pi,L_{\pi}}))\in\mathbb{R}^{n_{\pi}}$, where $n_{\pi} = (M+3)n_{\pi,0} + n_{\pi,1}(M+3)(k-n_{\pi,0} +1) + \sum^{L_{\pi}}_{l=2}n_{\pi,l}n_{\pi,l-1}$, and $\text{vec}(\cdot)$ denotes vectorization. 
%For all $l=0,1,\ldots,L_{a}$, let $\mathbf{h}_{\pi,l}$ denote the output of the $l$-th layer. 
Then, we have:
\begin{align}
\begin{split}
&\mathbf{h}_{\pi,0} 
= (\mathbf{w}^{(1)}_{\pi,0} * \mathbf{C}_{n},\mathbf{w}^{(2)}_{\pi,0} *\mathbf{B}_{n}, \mathbf{w}^{(3)}_{\pi,0} *\mathbf{Y}_{n}, \mathbf{w}^{(4)}_{\pi,0} * \mathbf{X}_{1,n},  \\
&~~~~~~~~~~ \mathbf{w}^{(5)}_{\pi,0} *\mathbf{X}_{2,n}, \ldots,  \mathbf{w}^{(M+3)}_{\pi,0} *\mathbf{X}_{M,n} ) ,\\
&\mathbf{h}_{\pi,l} = \textbf{ReLU}(\mathbf{W}_{\pi,l}\mathbf{h}_{\pi,l-1}),~l= 1,\ldots,L_{\pi}-1,\\
&\mathbf{h}_{\pi,L_{\pi}} = \textbf{Softmax}(\mathbf{W}_{\pi,L_{\pi}}\mathbf{h}_{\pi,L_{\pi}-1}),
%&\mathbf{h}_{\pi,L} = \pi(\mathbf{S}_{n},f_{i};\bm{\Theta}_{\pi,n}) =  \frac{\mathbf{w}^{T}_{\pi,L,i}\mathbf{h}_{L-1}}{\sum_{i=1}^{n_{\pi,L}}\mathbf{w}^{T}_{\pi,L,i}\mathbf{h}_{L-1}},~i=1,\ldots,n_{\pi,L}, \nonumber
\end{split}
\label{eq:output_actor}
\end{align}
where $*$ denotes the convolution operation. The output of the network $\mathbf{h}_{\pi,L_{\pi}}$ is the distribution of policy $\pi$, i.e., $(\pi(\mathbf{s},r;\bm{\Theta}_{\pi}))_{r\in\mathcal{D}}$.

\noindent\textbf{eA3C critic network:} The eA3C critic network has the same neural network structure as the eA3C actor network except that its output layer is a linear neural network without any activation function. Similarly, the critic network has one input layer, $L_{v}-1$ hidden layers, and one output layer, indexed by $0,1,\ldots, L_{v}-1,L_{v}$, respectively. The length of the filter of each 1D CNN and the number of neurons in the $l$-th layer are $n_{v,0}$ and $n_{v,l}$, respectively.
%Let $\mathbf{w}^{(i)}_{v,0}\in\mathbb{R}^{n_{v,0}}$ denote the weight vector of the $i$-th 1D CNN for all $i= 1,2,\ldots, M+3$, let $\mathbf{W}_{v, 1} \in \mathbb{R}^{n_{v,1}\times (M+3)(k-n_{v,0}+1)}$ denote the weight matrix of the first hidden layer, let $\mathbf{W}_{v,l} \in\mathbb{R}^{n_{v,l}\times n_{v,l-1}}$ denote the weight matrix of the $l$-th layer for all $l=2,\ldots,L_{c}-1$, and let $\mathbf{w}_{v,L_{c}}\in\mathbb{R}^{n_{v,L_{c}-1}}$ denote the weight vector of the output layer. 
Similarly, we have $\bm{\Theta}_{v} = (\mathbf{w}^{(1)}_{v,0},\mathbf{w}^{(2)}_{v,0},\ldots, \mathbf{w}^{(M+3)}_{v,0}, \text{vec}(\mathbf{W}_{v,1}),\ldots, \mathbf{w}_{v,L_{v}})\in\mathbb{R}^{n_{v}}$, where $n_{v} = (M+3)n_{v,0} + n_{v,1}(M+3)(k-n_{v,0} +1) + \sum^{L_{v}}_{l=2}n_{v,l}n_{v,l-1}$. 
%For all $l=0,1,\ldots,L_{c}-1$, let $\mathbf{h}_{v,l}$ denote the output of the $l$-th layer. Let $h_{v,L_{c}}$ denote the output of the $L_{c}$-th layer. 
Then, we have:
\begin{align}
\begin{split}
&\mathbf{h}_{v,0} 
= (\mathbf{w}^{(1)}_{v,0} * \mathbf{C}_{n},\mathbf{w}^{(2)}_{v,0} *\mathbf{B}_{n}, \mathbf{w}^{(3)}_{v,0} *\mathbf{Y}_{n}, \mathbf{w}^{(4)}_{v,0} * \mathbf{X}_{1,n}, \\
&~~~~~~~~~~  \mathbf{w}^{(5)}_{v,0} *\mathbf{X}_{2,n}, \ldots,  \mathbf{w}^{(M+3)}_{v,0} *\mathbf{X}_{M,n} ) ,\\
&\mathbf{h}_{v,l} = \textbf{ReLU}(\mathbf{W}_{v,l}\mathbf{h}_{v,l-1}),~l= 1,\ldots,L_{v}-1,\\
&h_{v,L_{v}} = \mathbf{w}^{T}_{v,L_{v}}\mathbf{h}_{v,L_{v}-1},
%&\mathbf{h}_{\pi,L} = \pi(\mathbf{S}_{n},f_{i};\bm{\Theta}_{\pi,n}) =  \frac{\mathbf{w}^{T}_{\pi,L,i}\mathbf{h}_{L-1}}{\sum_{i=1}^{n_{\pi,L}}\mathbf{w}^{T}_{\pi,L,i}\mathbf{h}_{L-1}},~i=1,\ldots,n_{\pi,L}, \nonumber
\end{split}
\label{eq:out_value}
\end{align}
where the output of the network $h_{v,L_{v}}$ is the approximated value function under policy $\pi$, i.e., $V^{\pi}(\mathbf{s};\bm{\Theta}_{v})$.

%where $\mathbf{w}_{\pi,2,i}, i=1,\ldots,n_{\pi,1} \in\mathbb{R}^{(M+3)(k-n_{\pi,2}-1) \times 1}$. 
%where $\mathbf{w}_{\pi,l,i},i=1,\ldots,n_{\pi,l} \in\mathbb{R}^{n_{\pi,l-1}\times 1}$. 

\textcolor{black}{We consider the offline scenario. Let $\{(\mathbf{C}_{n},\mathbf{X}_{n}): n=1,2,\ldots\}$ denote the pre-collected data from a large number of users.} Given the sample $\mathbf{S}_{n}$ at stage $n$, we generate the action $R_{n}$ according to the policy distribution $\pi(\mathbf{S}_{n},r;\bm{\Theta}_{\pi,n}),r\in\mathcal{R}$, i.e., $\Pr[R_{n} = r] = \pi(\mathbf{S}_{n},r;\bm{\Theta}_{\pi,n})$ for all $r\in\mathcal{R}$, and obtain the sample $\mathbf{S}_{n+1}$ at stage $n+1$ as follows: obtain $\mathbf{C}_{n+1}$ and $\mathbf{X}_{n+1}$ from the pre-collected sequence $\{(\mathbf{C}_{n},\mathbf{X}_{n}): n=1,2,\ldots\}$; based on $C_{n}$, obtain $B_{n+1}$ and $Y_{n+1}$ according to \eqref{eq:buffer_evolve} and \eqref{eq:state_augmentation}, respectively. Consequently, we have the sample $\mathbf{S}_{n+1} = (\mathbf{C}_{n+1},\mathbf{X}_{n+1},\mathbf{B}_{n+1},\mathbf{Y}_{n+1})$. At each stage $n$, we use a batch of the latest $q$ samples and choose:
\begin{dmath}
 \sum^{n-1}_{i=n- q}\frac{1}{q}\left(\log\pi(\mathbf{S}_{i},R_{i};\bm{\Theta}_{\pi}) \left(g(\mathbf{S}_{i},R_{i}) + \gamma V^{\pi}(\mathbf{S}_{i+1};\bm{\Theta}_{v}) \right) \\
 - \xi  \left(g(\mathbf{S}_{i},R_{i}) + \gamma V^{\pi}(\mathbf{S}_{i+1};\bm{\Theta}_{v}) - V^{\pi}(\mathbf{S}_{i};\bm{\Theta}_{v})\right)^{2}\\
 + \beta \sum_{r\in\mathcal{R}}\pi(\mathbf{S}_{i},r;\bm{\Theta}_{\pi}) 
	\log\pi(\mathbf{S}_{i},r;\bm{\Theta}_{\pi})\right) \label{eq:loss_function}
	\end{dmath}
as the loss function for jointly training the eA3C actor and critic networks. Here, $V^{\pi}(\cdot;\bm{\Theta}_{v})$ is the output of the critic network, $\sum_{r\in\mathcal{R}} \pi(\mathbf{S}_{i},r;\bm{\Theta}_{\pi}) 
	\log \pi(\mathbf{S}_{i},r;\bm{\Theta}_{\pi})$ is the entropy regularization term, and $\beta$ is the associated weights for the entropy. The entropy regularization term is used to ensure adequate exploration of the action space for 
 discovering good policies. Large $\beta$ encourages policy exploration, whereas small $\beta$ encourages policy exploitation. We train eA3C network using the root mean squared propagation (RMSP) algorithm \cite{rmsp}.
\textcolor{black}{After training, the values of the weights of the actor network and critic network are denoted as $\bm{\Theta}^{\star}_{\pi}$ \textcolor{black}{($\mathbf{w}^{\star(i)}_{\pi,0},i=1,\ldots,M+3, \mathbf{W}^{\star}_{\pi,l},l=1,\ldots,L_{\pi}$)} and $\bm{\Theta}^{\star}_{v}$ \textcolor{black}{($\mathbf{w}^{\star(i)}_{v,0},i=1,\ldots,M+3, \mathbf{W}^{\star}_{v,l},l=1,\ldots,L_{v}-1, \mathbf{w}^{\star}_{v,L_{v}}$)}, respectively.} The output of the actor network $\pi(\mathbf{s},r;\bm{\Theta}^{\star}_{\pi}),r\in\mathcal{R}$ is the resulting policy $\pi$. The output of the critic network $V^{\pi}(\mathbf{s};\bm{\Theta}^{\star}_{v})$ serves as an approximation of the value function under the policy $\pi$.
%and can be used for reducing the computational complexity of traditional DP algorithms.
\subsection{Comparisons with the A3C Method}
First, we compare the optimization formulations and solution methods for the policy and value parameters. \textcolor{black}{The proposed joint optimization of policy and value parameters in \eqref{prob:policy_gradient_equal_approxi} comes from an equivalent transformation of the problem in \eqref{prob:offline_policy_gradient_formulation}. By contrast, in \cite{a3c,sigcomm17}, the problem in \eqref{prob:offline_policy_gradient_formulation} is separated into the two optimization problems for policy parameter $\bm{\Theta}_{\pi}$ and value parameter $\bm{\Theta}_{v}$, which are given in \eqref{prob:a3c_policy} and \eqref{prob:a3c_value}, respectively, \textcolor{black}{as shown at the top of the next page.}
\begin{figure*}
\begin{align}
	&\max_{\bm{\Theta}_{\pi}}\quad \sum_{\mathbf{s}\in\boldsymbol{\mathcal{S}}}
\eta^{\pi}(\mathbf{s})  \sum_{r\in\mathcal{R}} \pi(\mathbf{s},r;\bm{\Theta}_{\pi})
 \left(g(\mathbf{s},r) + \gamma \sum_{\mathbf{s}'\in\boldsymbol{\mathcal{S}}}p_{\mathbf{s},\mathbf{s}'}(r)V^{\pi}(\mathbf{s}';\bm{\Theta}_{v}) \right),\label{prob:a3c_policy}\\
 &\min_{\bm{\Theta}_{v}}\quad \sum_{\mathbf{s}\in\boldsymbol{\mathcal{S}}}
\eta^{\pi}(\mathbf{s})  
	\left( \sum_{r\in\mathcal{R}} \pi(\mathbf{s},r;\bm{\Theta}_{\pi})\left( g(\mathbf{s},r) + \gamma \sum_{\mathbf{s}'\in\boldsymbol{\mathcal{S}}}p_{\mathbf{s},\mathbf{s}'}(r)V^{\pi}(\mathbf{s}';\bm{\Theta}_{v}) \right) - V^{\pi}(\mathbf{s};\bm{\Theta}_{v})\right)^{2}.\label{prob:a3c_value}
\end{align}
\hrulefill
\end{figure*}
The two problems are solved alternatively. The separate optimization method is more heuristic and may yield lower QoE and longer computation time than the joint optimization method as illustrated below. Firstly, applying the standard alternating optimization method\cite{cvx} to the joint optimization problem in \eqref{prob:policy_gradient_equal_approxi} results in two optimization problems that are different from the problems in \eqref{prob:a3c_policy} and \eqref{prob:a3c_value} and captures more interactions between $\bm{\Theta}_{\pi}$ and $\bm{\Theta}_{v}$. Note that the objective function of the problem in \eqref{prob:a3c_policy} omits $\sum_{\mathbf{s}\in\boldsymbol{\mathcal{S}}}
\eta^{\pi}(\mathbf{s})  \sum_{r\in\mathcal{R}} \pi(\mathbf{s},r;\bm{\Theta}_{\pi})
	(g(\mathbf{s},r) $ 
$+\gamma \sum_{\mathbf{s}'\in\boldsymbol{\mathcal{S}}}p_{\mathbf{s},\mathbf{s}'}(r)V^{\pi}(\mathbf{s}';\bm{\Theta}_{v}) - V^{\pi}(\mathbf{s};\bm{\Theta}_{v}))^{2}$, thereby ignoring some influence of $\bm{\Theta}_{v}$. Besides, the objective function of the problem in \eqref{prob:a3c_value} omits $\sum_{\mathbf{s}\in\boldsymbol{\mathcal{S}}}
\eta^{\pi}(\mathbf{s})  \sum_{r\in\mathcal{R}} \pi(\mathbf{s},r;\bm{\Theta}_{\pi})
 (g(\mathbf{s},r)$ $+ \gamma \sum_{\mathbf{s}'\in\boldsymbol{\mathcal{S}}}p_{\mathbf{s},\mathbf{s}'}(r)V^{\pi}(\mathbf{s}';\bm{\Theta}_{v}) )$, thus ignoring some influence of $\bm{\Theta}_{\pi}$. 
Secondly, alternatively solving the non-convex problems in \eqref{prob:a3c_policy} and \eqref{prob:a3c_value} (suboptimally with stationary points) does not have any convergence guarantee to a stationary point of the problem in \eqref{prob:policy_gradient_equal_approxi}.
Next, we compare the network structures and training methods. Firstly, compared to the A3C network in \cite{a3c,sigcomm17}, the eA3C network has additional input that is lower-layer information $\mathbf{X}_{m},m=1,\ldots,M$. Specifically, one 1D CNN and $L$ fully connected layers are constructed to deal with each lower layer information $\mathbf{X}_{m}$. Secondly, unlike \cite{sigcomm17,a3c}, we jointly train the actor and critic networks of eA3C network (i.e., jointly optimize the policy and value parameters) which may reduce the training time and improve the QoE. Therefore, it is expected that our proposed eA3C method has higher QoE and shorter training time than the state-of-art A3C method\cite{sigcomm17,a3c}, which will be illustrated in Section~\ref{sec:sim}.
}
\begin{figure*}[t]
%\vspace*{-1.2cm}
\begin{center}
   %\subfloat[\small{Average QoE versus number of layers}]
   %{\resizebox{8cm}{!}{\includegraphics{Offline_QoE_layer.eps}}}
   %\subfloat[\small{Average QoE versus number of neurons}]
   %{\resizebox{8cm}{!}{\includegraphics{Offline_QoE_neuron.eps}}}
      \subfloat[eA3C-OTP]
   {\resizebox{8cm}{!}{\includegraphics{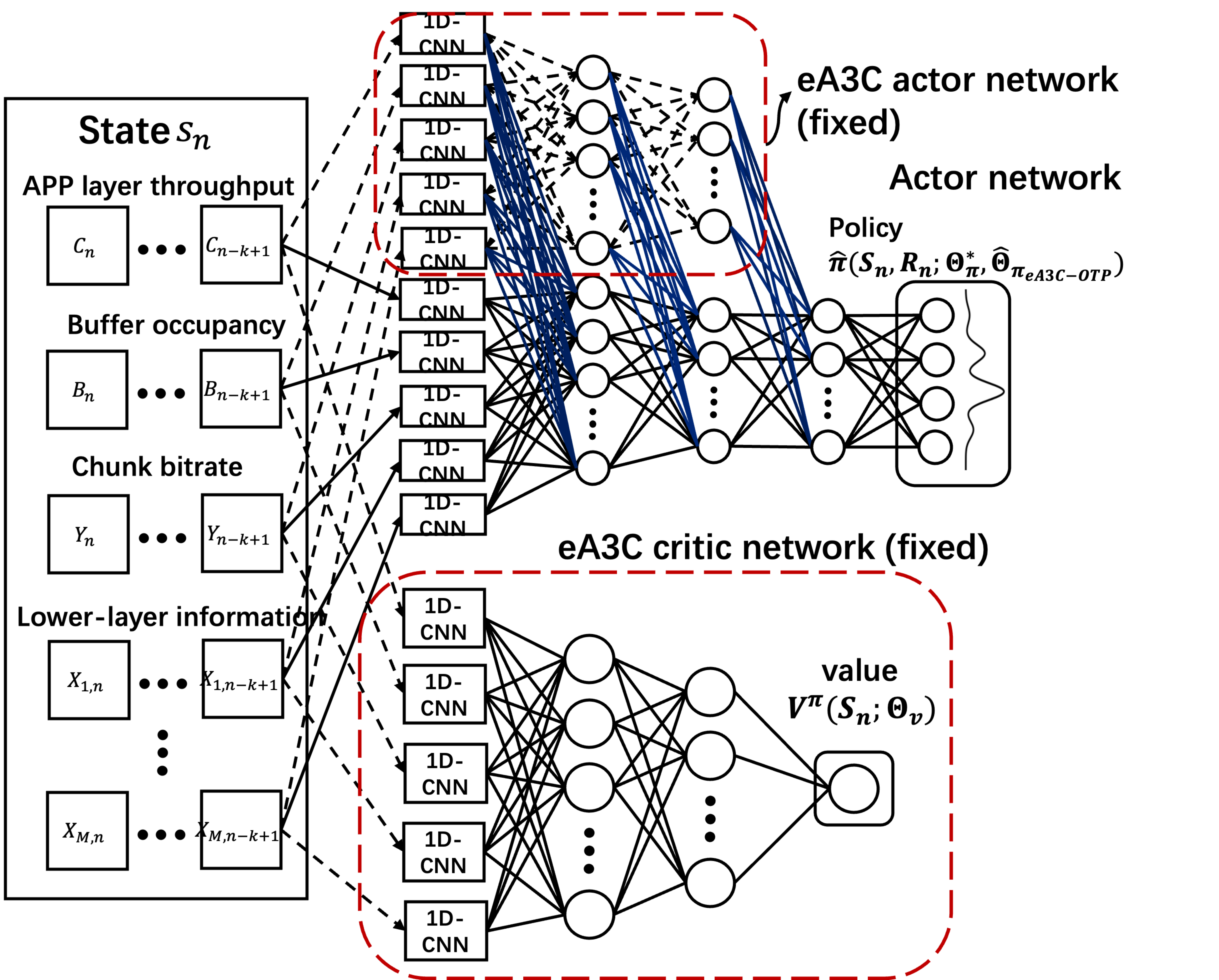}}}
   \subfloat[eA3C-OTPV]
   {\resizebox{8cm}{!}{\includegraphics{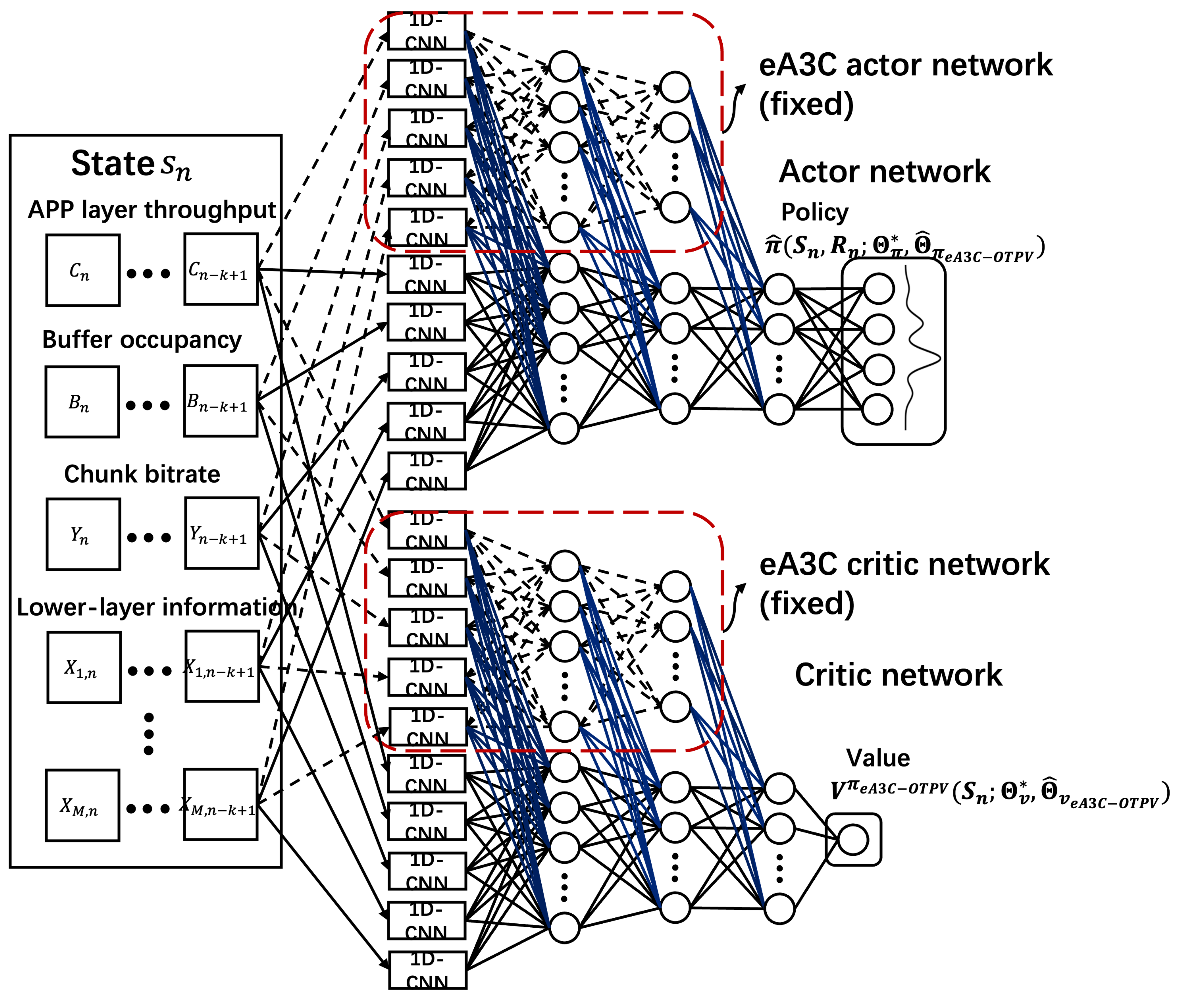}}}
 %\subfloat[\small{CDF of rebuffering time}]
 %{\resizebox{5.9cm}{!}{\includegraphics{pic/multi_rebuffering.eps}}}
 %\vspace*{-0.07cm}
 \end{center}
 %\vspace*{-0.37cm}
   \caption{Structures of the eA3C-OTP and eA3C-OTPV networks.}
   \label{fig:network_structure_online}
%\vspace*{-0.60cm}
\end{figure*}

\section{\textcolor{black}{Continual} Learning-based Online Tuning Methods in the Online Scenario}\label{sec_online_MDP}
The offline trained eA3C network based on the pre-collected data from a large number of users has good generalization ability. \textcolor{black}{However, different user may have different wireless network conditions as shown in Fig.~\ref{fig:cor}.} The offline trained eA3C network is probably suboptimal for the user due to model mismatches. Since video streaming is a long-lasting application, the real-time samples of a user watching the streaming video can be gradually collected during the video streaming process and used for tailoring the offline policy for this user\cite{JSAC20,mm22,mobicom20}.
In this section, we consider an online scenario where a specific user is watching a video, starting from stage $0$, and its samples, also denoted by $\{(\mathbf{S}_{n},R_{n}): n = 1,2,\ldots\}$ with a slight abuse of notation, are collected during the video streaming process. 
We \textcolor{black}{present} two \textcolor{black}{continual} learning-based online tuning methods to design better policies for the specific user, named online policies, leveraging the eA3C network presented in Section~\ref{sec:A3C} and gradually collected real-time samples. We employ a batch of the latest $q$ samples for each round of optimization process of the online tuning methods. At stage $n \in\{1,\ldots, q-1\}$, we use the offline policy \textcolor{black}{obtained from the eA3C network} for bitrate adaptation. From stage $n \in \{q, \ldots, q'-1\}$ for some $q' > q$ (collected real-time samples are enough to start the optimization process of online polices, but the optimization process is far from convergent), we still use the offline policy \textcolor{black}{derived from eA3C network} for bitrate adaptation but start applying the online tuning methods for designing better online polices. Starting from stage $n \in\{q',\ldots\}$, we keep on fine-tuning the online polices while simultaneously using them for bitrate adaptation.

\begin{table*}[t]
\caption{Notation of the weights and layer outputs of the eA3C-OTP and eA3C-OTPV networks ($j = \pi$ or $v$). $\mathbf{w}^{\star(i)}_{j,0},i = 1,\ldots, M+3, \mathbf{W}^{\star}_{j,l},l=1,\ldots,L_{j}$ are pre-trained weights from the eA3C network and are ``frozen" during the online tuning. For ease of presentation, we let $\pi$ network and $v$ network represent actor network and critic network, respectively. \textcolor{black}{Noting that the policy networks of the eA3C-OTP and eA3C-OTPV networks have the same structure, we do not distinguish the notation for their weights and layer outputs.} }
\resizebox{18cm}{!}{
\begin{tabular}{|c|c|c|c|} 
\hline 
\multicolumn{2}{|c|}{eA3C-OTP} &  \multicolumn{2}{|c|}{eA3C-OTPV}\\ \hline
Notation & Description & Notation& Description \\ \hline  
$\mathbf{w}^{\star(i)}_{j,0}\in\mathbb{R}^{n_{j},0}$ &  \thead{weight vector of $i$-th 1D CNN\\ in input layer of $j$ network, $i = 1,\ldots, M+3$ } & $\mathbf{w}^{\star(i)}_{j,0}\in\mathbb{R}^{n_{j},0}$ & \thead{  weight vector of $i$-th 1D CNN \\in input layer of $j$ network, $i = 1,\ldots, M+3$ }\\ \hline
$\hat{\mathbf{w}}^{(i)}_{\pi,0}\in\mathbb{R}^{\hat{n}_{\pi,0}}$ &  \thead{weight vector of $i$-th 1D CNN\\ in input layer of $j$ network, $i = M+4,\ldots,2(M+3)$} &$\hat{\mathbf{w}}^{(i)}_{j,0}\in\mathbb{R}^{\hat{n}_{j,0}}$ &  \thead{weight vector of $i$-th 1D CNN \\in input layer of $j$ network, $i = M+4,\ldots,2(M+3)$} \\ \hline
$\mathbf{W}^{\star}_{j,1}\in\mathbb{R}^{n_{j,1}\times (M+3)(k-n_{j,0}+1)}$ &  \thead{weight matrix from first $(M+3)$ 1D CNNs in input layer \\to first $n_{j,1}$ neurons in first hidden layer of $j$ network}  & $\mathbf{W}^{\star}_{j,1}\in\mathbb{R}^{n_{j,1}\times (M+3)(k-n_{j,0}+1)}$ &  \thead{weight matrix from first $(M+3)$ 1D CNNs in input layer \\to first $n_{j,1}$ neurons in first hidden layer of $j$ network }\\ \hline
$\hat{\mathbf{W}}_{\pi, 1} \in \mathbb{R}^{\hat{n}_{\pi,1}\times (M+3)(2k-\hat{n}_{\pi,0}-n_{\pi,0}+2)}$ &  \thead{weight matrix from input layer \\to remaining $\hat{n}_{\pi,1}$ neurons in first hidden layer of actor network} & $\hat{\mathbf{W}}_{j, 1} \in \mathbb{R}^{\hat{n}_{j,1}\times (M+3)(2k-\hat{n}_{j,0}-n_{j,0}+2)}$ &\thead{  weight matrix from input layer \\to remaining $\hat{n}_{j,1}$ neurons in first hidden layer of $j$ network}\\ \hline
\thead{$\begin{pNiceMatrix}
      \mathbf{W}^{\star}_{\pi,1} & \bigzero_{n_{\pi,1}\times (k-\hat{n}_{\pi, 0} + 1)} \\
      \Block{1-2}{\hat{\mathbf{W}}_{\pi,1}} & 
    \end{pNiceMatrix}$ \\$~~~~~~~~~~~~~~~~~~~~\in\mathbb{R}^{(n_{\pi,1} + \hat{n}_{\pi,1}) \times (M+3)(2k-\hat{n}_{\pi,0}-n_{\pi,0}+2)}$ } &  \thead{weight matrix from input layer \\to first hidden layer of actor network} & \thead{$\begin{pNiceMatrix}
      \mathbf{W}^{\star}_{j,1} & \bigzero_{n_{j,1}\times (k-\hat{n}_{j, 0} + 1)} \\
      \Block{1-2}{\hat{\mathbf{W}}_{j,1}} & 
    \end{pNiceMatrix} $\\$\in\mathbb{R}^{(n_{j,1} + \hat{n}_{j,1}) \times (M+3)(2k-\hat{n}_{j,0}-n_{j,0}+2)}$} &  \thead{weight matrix from input layer \\to first hidden layer of $j$ network}\\ \hline
$\mathbf{W}^{\star}_{j,l}\in\mathbb{R}^{n_{j,l} \times n_{j,l-1}}$ &  \thead{weight matrix from first $n_{j,l-1}$ neurons in $(l-1)$-th layer \\to first $n_{j,l}$ neurons in $l$-th layer of $j$ network, $l = 2,\ldots,L_{j} $ } &$\mathbf{W}^{\star}_{j,l}\in\mathbb{R}^{n_{j,l} \times n_{j,l-1}}$ &  \thead{weight matrix from first $n_{j,l-1}$ neurons in $(l-1)$-th layer \\to first $n_{j,l}$ neurons in $l$-th layer of $j$ network, $l = 2,\ldots,L_{j} $ }\\ \hline
$\hat{\mathbf{W}}_{\pi,l} \in\mathbb{R}^{\hat{n}_{\pi,l}\times (\hat{n}_{\pi,l-1} + n_{\pi,l-1})}$ &  \thead{weight matrix from $(l-1)$-th layer \\to remaining $\hat{n}_{\pi,l}$ neurons in $l$-th layer of actor network, $l = 2,\ldots,L_{\pi} $ } &$\hat{\mathbf{W}}_{j,l} \in\mathbb{R}^{\hat{n}_{j,l}\times (\hat{n}_{j,l-1} + n_{j,l-1})}$ &  \thead{weight matrix from $(l-1)$-th layer \\to remaining $\hat{n}_{j,l}$ neurons in $l$-th layer of $j$ network, $l = 2,\ldots,L_{j} $ }\\ \hline
\thead{$\begin{pNiceMatrix}
      \mathbf{W}^{\star}_{\pi,l} & \bigzero_{n_{\pi,l}\times \hat{n}_{\pi, l-1}} \\
      \Block{1-2}{\hat{\mathbf{W}}_{\pi,l}} & 
    \end{pNiceMatrix}$ \\$ \in\mathbb{R}^{(n_{\pi,l} + \hat{n}_{\pi,l}) \times (n_{\pi,l-1} + \hat{n}_{\pi,l-1})}$  }&  \thead{weight matrix from input layer \\to first hidden layer of actor network } &\thead{$\begin{pNiceMatrix}
      \mathbf{W}^{\star}_{j,l} & \bigzero_{n_{j,l}\times \hat{n}_{j, l-1}} \\
      \Block{1-2}{\hat{\mathbf{W}}_{j,l}} & 
    \end{pNiceMatrix} $\\$\in\mathbb{R}^{(n_{j,l} + \hat{n}_{j,l}) \times (n_{j,l-1} + \hat{n}_{j,l-1})}$} &  \thead{weight matrix from input layer\\ to first hidden layer }\\ \hline
$\hat{\mathbf{W}}_{\pi,l} \in\mathbb{R}^{\hat{n}_{\pi,l}\times \hat{n}_{\pi,l-1}}$  &  \thead{weight matrix from $(l-1)$-th layer \\to $l$-th layer of actor network, $l = L_{\pi} +1,\ldots, \hat{L}_{\pi} $ } & $\hat{\mathbf{W}}_{j,l} \in\mathbb{R}^{\hat{n}_{j,l}\times \hat{n}_{j,l-1}}$  &  \thead{weight matrix from $(l-1)$-th layer \\to $l$-th layer of $j$ network, $l = L_{j} +1,\ldots, \hat{L}_{j} - 1$ }\\ \hline
$\hat{\mathbf{h}}_{\pi,0} \in\mathbb{R}^{(M+3)(k-\hat{n}_{\pi,0}+1)}$ &  \thead{output of last $(M+3)$ 1D CNNs \\in input layer of actor network } & $\hat{\mathbf{W}}_{\pi,\hat{L}_{\pi}}\in\mathbb{R}^{\hat{n}_{\pi, \hat{L}_{\pi}} \times \hat{n}_{\pi,\hat{L}_{\pi}-1}  }$ &  \thead{weight vector from $(\hat{L}_{\pi}-1)$-th layer \\to output layer of actor network}\\ \hline
$\hat{\mathbf{h}}_{\pi,l} \in\mathbb{R}^{\hat{n}_{\pi,l}}$&  \thead{output of last $\hat{n}_{\pi,l}$ neurons \\in $l$-th layer of actor network, $l=1,\ldots,L_{\pi}$} & $\hat{\mathbf{w}}_{v,\hat{L}_{v}}\in\mathbb{R}^{\hat{n}_{v,\hat{L}_{v}-1}  }$ &  \thead{weight vector from $(\hat{L}_{v}-1)$-th layer \\to output layer of critic network }\\ \hline
$\hat{\mathbf{h}}_{\pi,l} \in\mathbb{R}^{\hat{n}_{\pi,l}}$ &  \thead{output of $l$-th layer of actor network, $l=L_{\pi}+1,\ldots,\hat{L}_{\pi} $}& $\hat{\mathbf{h}}_{j,0} \in\mathbb{R}^{(M+3)(k-\hat{n}_{j,0}+1)}$ &  \thead{output of last $(M+3)$ 1D CNNs \\in input layer of $j$ network} \\ \hline
& & $\hat{\mathbf{h}}_{j,l} \in\mathbb{R}^{\hat{n}_{j,l}}$&  \thead{output of last $\hat{n}_{j,l}$ neurons \\in $l$-th layer of $j$ network, $l=1,\ldots,L_{j}$ } \\ \hline
&& $\hat{\mathbf{h}}_{j,l} \in\mathbb{R}^{\hat{n}_{j,l}}$ &  \thead{output of $l$-th layer of $j$ network, $l=L_{j}+1,\ldots,\hat{L}_{j} -1 $ } \\ \hline
&&$\hat{h}_{v,\hat{L}_{v}}\in\mathbb{R}$ &  \thead{output of $\hat{L}_{v}$-th layer of critic network} \\ \hline
&&$\hat{\mathbf{h}}_{\pi,\hat{L}_{\pi}} \in\mathbb{R}^{\hat{n}_{\pi, \hat{L}_{\pi}}}$ & \thead{output of $\hat{L}_{\pi}$-th layer of actor network }\\ \hline
\end{tabular}
} \label{table_online}
\end{table*}

\subsection{Online Tuning for Policy Network}\label{sec:tlp}
In this part, we \textcolor{black}{present} a \textcolor{black}{continual} learning-based online tuning method which builds a new policy network based on the eA3C policy network, adopts the eA3C value network, and trains the new component of its policy network using gradually collected real-time samples and the value network. \textcolor{black}{The online tuning method is called the eA3C-OTP method.} The resulting network, namely the eA3C-OTP network, consists of an actor network and a critic network and has the same inputs as eA3C network, as shown in Fig.~\ref{fig:network_structure_online} (a). \textcolor{black}{The notation of the weights and layer outputs of the eA3C-OTP network is summarized in Table~\ref{table_online}.}
%we assume that the environment changes very fast, i.e., the statistics of download rate and network information change very fast. 
%Specifically, we extend Progressive Network\cite{tflearning} to a more general structure. 
%The example can be see from Fig.~\ref{fig:A3C_p}. While online tuning incorporates prior knowledge only at initialization, progressive networks retain a pool of pretrained models throughout training, and learn lateral connections from these to extract useful features for the new environment. 
%The details of the network are as follows.

\noindent\textbf{eA3C-OTP actor network:} The eA3C-OTP actor network is built based on the eA3C actor network. 
  %We add a new neural network with the parameters $\hat{\bm{\Theta}}_{\pi_{\text{TL}}}\in\mathbb{R}^{\hat{n}_{\pi_{\text{TL}}}}$ and the \textcolor{black}{continual} is enabled via lateral connections to the actor network of eA3C.
  It has one input layer, $\hat{L}_{\pi}- 1~(\hat{L}_{\pi} \geq L_{\pi})$ hidden layers, and one output layer, indexed by $0,1,\ldots,\hat{L}_{\pi}-1,\hat{L}_{\pi}$, respectively. The input layer has $2(M+3)$ 1D CNNs, with the first $M+3$ 1D CNNs coming from the eA3C network. For all $l=1,\ldots,L_{\pi}-1$, the $l$-th hidden layer is a partially connected layer containing $n_{\pi,l} + \hat{n}_{\pi,l}$ neurons, with the first $n_{\pi,l}$ neurons coming from the eA3C network and the remaining $\hat{n}_{\pi,l}$ neurons that also utilize the ReLU activation function. For all $l=L_{\pi},\ldots,\hat{L}_{\pi}-1$, the $l$-th hidden layer is a fully connected layer with $\hat{n}_{\pi,l}$ neurons that utilize the ReLU activation function. The output layer is a fully connected layer with $\hat{n}_{\pi,\hat{L}_{\pi}}$ neurons that utilize the softmax activation function. 
 We rearrange all the tunable weights of the eA3C-OTP actor network as a column vector, denoted by $\hat{\bm{\Theta}}_{\pi_{\text{OTP}}} = (\hat{\mathbf{w}}^{(1)}_{\pi,0},\hat{\mathbf{w}}^{(2)}_{\pi,0},\ldots, \hat{\mathbf{w}}^{(M+3)}_{\pi,0}, \text{vec}(\hat{\mathbf{W}}_{\pi,1}),\ldots, \text{vec}(\hat{\mathbf{W}}_{\pi,\hat{L}_{\pi}}))\in\mathbb{R}^{\hat{n}_{\pi}}$, where $\hat{n}_{\pi} = (M+3)\hat{n}_{\pi,0} + \hat{n}_{\pi,1}(M+3)(2k-\hat{n}_{\pi,0}-n_{\pi,0} +2) + \sum^{L_{\pi}}_{l=2}\hat{n}_{\pi,l}(\hat{n}_{\pi,l-1} + n_{\pi,l-1}) + \sum_{l=L_{\pi}+1}^{\hat{L}_{\pi}}\hat{n}_{\pi,l} \hat{n}_{\pi,l-1}$. 
  %(iii) We denote the output of the last $(M+3)$ 1D CNNs in the input layer by $\hat{\mathbf{h}}_{\pi_{\text{eA3C-OTP}},0}$. For all $l=1,\ldots,L_{a}$, we denote the output of the last $\hat{n}_{\pi_{\text{eA3C-OTP}},l}$ neurons in the $l$-th layer by $\hat{\mathbf{h}}_{\pi_{\text{eA3C-OTP}},l}$. For all $l=L_{a}+1,\ldots,\hat{L}_{a}$, we denote the output of the $l$-th layer by $\hat{\mathbf{h}}_{\pi_{\text{eA3C-OTP}},l}$. 
  Then, based on the structure of the eA3C-OTP actor network, \textcolor{black}{we can derive the layer outputs given by \eqref{eq:output_OTP}, as shown at the top of the next page.}
\begin{figure*}
\begin{align}
\label{eq:output_OTP}
\begin{split}
&\hat{\mathbf{h}}_{\pi,0} 
= (\hat{\mathbf{w}}^{(1)}_{\pi,0} * \mathbf{C}_{n},\hat{\mathbf{w}}^{(2)}_{\pi,0} *\mathbf{B}_{n}, \hat{\mathbf{w}}^{(3)}_{\pi,0} *\mathbf{Y}_{n}, \hat{\mathbf{w}}^{(4)}_{\pi,0} * \mathbf{X}_{1,n}, \ldots,  \hat{\mathbf{w}}^{(M+3)}_{\pi,0} *\mathbf{X}_{M,n} ) ,\\
&(\mathbf{h}_{\pi,1}, \hat{\mathbf{h}}_{\pi,1}) = \textbf{ReLU}\left(\begin{pNiceMatrix}
      \mathbf{W}^{\star}_{\pi,1} & \bigzero_{n_{\pi,1}\times (k - \hat{n}_{\pi, 0}+1 )} \\
      \Block{1-2}{\hat{\mathbf{W}}_{\pi,1}} & 
    \end{pNiceMatrix}  \left(\mathbf{h}_{\pi,0}, \hat{\mathbf{h}}_{\pi,0} \right) \right),\\
&(\mathbf{h}_{\pi,l},\hat{\mathbf{h}}_{\pi,l}) = \textbf{ReLU}\left(\begin{pNiceMatrix}
      \mathbf{W}^{\star}_{\pi,l} & \bigzero_{n_{\pi,l}\times \hat{n}_{\pi, l-1}} \\
      \Block{1-2}{\hat{\mathbf{W}}_{\pi,l}} & 
    \end{pNiceMatrix}  \left(\mathbf{h}_{\pi,l-1},\hat{\mathbf{h}}_{\pi,l-1}\right)\right),~l= 2,\ldots,L_{\pi},\\
&\hat{\mathbf{h}}_{\pi,l} = \textbf{ReLU}(\hat{\mathbf{W}}_{\pi,l}\hat{\mathbf{h}}_{\pi,l-1}),~l= L_{\pi} + 1,\ldots,\hat{L}_{\pi}-1, \quad \hat{\mathbf{h}}_{\pi,\hat{L}_{\pi}} = \textbf{Softmax}(\hat{\mathbf{W}}_{\pi,L_{\pi}}\hat{\mathbf{h}}_{\pi,\hat{L}_{\pi}-1} ).
\end{split}
\end{align}
\hrulefill
\end{figure*}
%where $\mathbf{h}_{\pi,0}$ and $\mathbf{h}_{\pi,l},l=1,\ldots,L_{a}$ are the output of the input layer and the output of the $l$-th layer of the eA3C actor network, respectively, and are also the output of the first $(M+3)$ 1D CNNs in the input layer and the output of the first $n_{\pi,l}$ neurons in the $l$-th layer of the eA3C-OTP actor network, respectively. 
Based on the composition structure in \eqref{eq:output_OTP}, the output $\hat{\mathbf{h}}_{\pi,\hat{L}_{\pi}}$ can be rewritten as $(\hat{\pi}(\mathbf{s},r;\hat{\bm{\Theta}}_{\pi_{\text{OTP}}},\bm{\Theta}^{\star}_{\pi}))_{r\in\mathcal{D}}$, representing the distribution of a parameterized randomized online policy $\pi_{\text{OTP}}$. The policy parameters include newly introduced parameters $\hat{\bm{\Theta}}_{\pi_{\text{OTP}}}$, which will be optimized, and the parameters of the offline policy $\pi$, $\bm{\Theta}^{\star}_{\pi}$, which are ``frozen" during the online tuning.
%\begin{align}
%\hat{\bm{\Theta}}_{\pi,n+1} = \hat{\bm{\Theta}}_{\pi,n} &+ \alpha_{\pi,n} \nabla_{\hat{\bm{\Theta}}_{\pi,n}}\ln \pi(\tilde{\mathbf{S}}_{n},\tilde{R}_{n};\hat{\bm{\Theta}}_{\pi,n},\bm{\Theta}_{\pi})(Q(\tilde{\mathbf{S}}_{n},\tilde{R}_{n}) + \gamma V^{\pi}(\tilde{\mathbf{S}}_{n+1};\hat{\bm{\Theta}}_{v,n},\bm{\Theta}_{v}) - V^{\pi}(\tilde{\mathbf{S}}_{n};\hat{\bm{\Theta}}_{v,n},\bm{\Theta}_{v})) \nonumber\\
%&+ \alpha_{\pi,n} \beta_{\pi,n} \nabla_{\hat{\bm{\Theta}}_{\pi,n}}\sum_{\tilde{R}_{n}\in\mathcal{D}}\pi(\tilde{\mathbf{S}}_{n},\tilde{R}_{n};\hat{\bm{\Theta}}_{\pi,n},\bm{\Theta}_{\pi})\ln \pi(\tilde{\mathbf{S}}_{n},\tilde{R}_{n};\hat{\bm{\Theta}}_{\pi,n},\bm{\Theta}_{\pi}).\nonumber
%\end{align}

\noindent\textbf{eA3C-OTP critic network:} We use the trained eA3C critic network as the eA3C-OTP critic network and do not modify its internal structure and weights during the online tuning. In particular, the eA3C-OTP critic network, whose output is $V^{\pi}(\cdot;\bm{\Theta}^{\star}_{v})$, serves as a facilitator for training the eA3C-OTP actor network.
%\begin{align}
%\hat{\bm{\Theta}}_{v,n+1} = \hat{\bm{\Theta}}_{v,n} - \alpha_{v,n}  \nabla_{\hat{\bm{\Theta}}_{v,n}}(Q(\tilde{\mathbf{S}}_{n},\tilde{R}_{n}) + \gamma V^{\pi}(\tilde{\mathbf{S}}_{n+1};\hat{\bm{\Theta}}_{v,n},\bm{\Theta}_{v}) - V^{\pi}(\tilde{\mathbf{S}}_{n};\hat{\bm{\Theta}}_{v,n},\bm{\Theta}_{v}) )^{2}.\nonumber
%\end{align}

%\textcolor{black}{Note that eA3C-OTP policy network is set flexible to capture the temporal features of the input.} 
We aim to optimize the newly added parameters $\hat{\bm{\Theta}}_{\pi_{\text{OTP}}}$ for the online policy \textcolor{black}{obtained from the eA3C-OTP network} to maximize a metric similar to the one defined in \eqref{eq:loss_function}. Specifically, at each stage $n\geq q$, we use a batch of the latest $q$ samples and choose 
\begin{figure*}
\begin{dmath}
 \sum^{n-1}_{i=n- q}\frac{1}{q}\left(\log\hat{\pi}(\mathbf{S}_{i},R_{i};\bm{\Theta}^{\star}_{\pi},\hat{\bm{\Theta}}_{\pi_{\text{OTP}}}) \left(g(\mathbf{S}_{i},R_{i}) + \gamma V^{\pi}(\mathbf{S}_{i+1};\bm{\Theta}^{\star}_{v}) \right) 
 - \xi  \left(g(\mathbf{S}_{i},R_{i}) + \gamma V^{\pi}(\mathbf{S}_{i+1};\bm{\Theta}^{\star}_{v}) - V^{\pi}(\mathbf{S}_{i};\bm{\Theta}^{\star}_{v})\right)^{2}
	 + \beta \sum_{r\in\mathcal{R}}\hat{\pi}(\mathbf{S}_{i},r;\bm{\Theta}^{\star}_{\pi},\hat{\bm{\Theta}}_{\pi_{\text{OTP}}}) 
	\log\hat{\pi}(\mathbf{S}_{i},r;\bm{\Theta}^{\star}_{\pi},\hat{\bm{\Theta}}_{\pi_{\text{OTP}}}) \right).\label{eq:loss_function_OTP}
	\end{dmath}
	\hrulefill
	\end{figure*}
the loss function given by \eqref{eq:loss_function_OTP}, \textcolor{black}{as shown at the top of the next page,} for training the tunable components of the eA3C-OTP actor network (i.e., optimizing $\hat{\bm{\Theta}}_{\pi_{\text{OTP}}}$). Since $\pi(\cdot; \bm{\Theta}^{\star}_{\pi})$, whose nested functions are the components of the nested functions of $\hat{\pi}(\cdot;\hat{\bm{\Theta}}_{\pi_{\text{OTP}}},\bm{\Theta}^{\star}_{\pi})$, (see \eqref{eq:output_OTP}), and $V^{\pi}(\cdot;\bm{\Theta}^{\star}_{v})$ appear in the loss function in \eqref{eq:loss_function_OTP}, the policy and value function \textcolor{black}{obtained from the eA3C network} affect the optimal choice of the parameters $\hat{\bm{\Theta}}_{\pi_{\text{OTP}}}$ for the online policy \textcolor{black}{obtained from the eA3C-OTP network}. Intuitively, the optimization of $\hat{\bm{\Theta}}_{\pi_{\text{OTP}}}$ can be viewed as a constrained\footnote{The online policy subjects to a predetermined parametric form.} policy improvement given the value function $V^{\pi}(\cdot;\bm{\Theta}^{\star}_{v})$ under the offline policy $\pi$ with distribution $\pi(\cdot; \bm{\Theta}^{\star}_{\pi})$. Therefore, it is expected that the eA3C-OTP method has better performance for the specific user than the eA3C method. 

We train the eA3C-OTP actor network using the RMSP algorithm \cite{rmsp}. 
%The per-stage computational complexity during the online training is $\mathcal{O}((3\hat{n}_{\pi_{\text{TL}},1} +\hat{n}_{\pi_{\text{TL}},0} + 3n_{\pi,1} + n_{\pi,0})(M+3)(2k-\hat{n}_{\pi_{\text{TL}},0} -n_{\pi,0} +2) + 4\sum^{\hat{L}_{a}}_{l=2}\hat{n}_{\pi_{\text{TL}},l}(\hat{n}_{\pi_{\text{TL}},l-1} + n_{\pi,l-1}) + (3\hat{n}_{v,1} +\hat{n}_{v,0} + 3n_{v,1} + n_{v,0})(M+3)(2k-\hat{n}_{v,0} -n_{v,0} +2) + 4\sum^{\hat{L}_{c}}_{l=2}\hat{n}_{v,l}(\hat{n}_{v,l-1} + n_{v,l-1}))$ [41]. The per-stage computational complexity after the online training is $\mathcal{O}((\hat{n}_{\pi_{\text{TL}},1} +\hat{n}_{\pi_{\text{TL}},0} + n_{\pi,1} + n_{\pi,0})(M+3)(2k-\hat{n}_{\pi_{\text{TL}},0} -n_{\pi,0} +2) + \sum^{\hat{L}_{a}}_{l=2}\hat{n}_{\pi_{\text{TL}},l}(\hat{n}_{\pi_{\text{TL}},l-1} + n_{\pi,l-1}))$ [41]. The per-stage space complexity is $\mathcal{O}(n_{v} + n_{\pi} + \hat{n}_{v} + \hat{n}_{\pi_{\text{TL}}})$. 
After training, we denote the values of the tunable weights of the eA3C-OTP actor network by $\hat{\bm{\Theta}}^{\star}_{\pi_{\text{OTP}}}$. The output of the eA3C-OTP actor network, $\hat{\pi}(\mathbf{s},r;\hat{\bm{\Theta}}^{\star}_{\pi_{\text{OTP}}},\bm{\Theta}^{\star}_{\pi})$, is the distribution of the resulting online policy \textcolor{black}{corresponding to eA3C-OTP}.

\subsection{Online Tuning for Policy and Value Networks}
In this part, we \textcolor{black}{present} a \textcolor{black}{continual} learning-based online tuning method which builds new policy and value networks based on the policy and value networks of the eA3C network, respectively, and trains the new components of its policy and value networks using gradually collected real-time samples. \textcolor{black}{The online tuning method is called the eA3C-OTPV method.} The resulting network, namely the eA3C-OTPV network, consists of an actor network and a critic network and has the same inputs with eA3C, as shown in Fig.~\ref{fig:network_structure_online} (b). \textcolor{black}{The notation of the weights and layer outputs of the eA3C-OTPV network is summarized in Table~\ref{table_online}.}

\noindent\textbf{eA3C-OTPV actor network:} The eA3C-OTPV actor network has the same structure as the one of eA3C-OTP. However, different from the eA3C-OTP actor network, the eA3C-OTPV actor network will be jointly trained with the eA3C-OTPV critic network. We omit the details of the network structure due to page limitation and  similarly rearrange all the tunable weights of the eA3C-OTPV actor network as a column vector, denoted by $\hat{\bm{\Theta}}_{\pi_{\text{OTPV}}} = (\hat{\mathbf{w}}^{(1)}_{\pi,0},\hat{\mathbf{w}}^{(2)}_{\pi,0},\ldots, \hat{\mathbf{w}}^{(M+3)}_{\pi,0}, \text{vec}(\hat{\mathbf{W}}_{\pi,1}),\ldots, \text{vec}(\hat{\mathbf{W}}_{\pi,\hat{L}_{\pi}}))\in\mathbb{R}^{\hat{n}_{\pi}}$, where $\hat{n}_{\pi} = (M+3)\hat{n}_{\pi,0} + \hat{n}_{\pi,1}(M+3)(2k-\hat{n}_{\pi,0}-n_{\pi,0} +2) + \sum^{L_{\pi}}_{l=2}\hat{n}_{\pi,l}(\hat{n}_{\pi,l-1} + n_{\pi,l-1}) + \sum_{l=L_{\pi}+1}^{\hat{L}_{\pi}}\hat{n}_{\pi,l} \hat{n}_{\pi,l-1}$. 
%\begin{align}
%\hat{\bm{\Theta}}_{\pi,n+1} = \hat{\bm{\Theta}}_{\pi,n} &+ \alpha_{\pi,n} \nabla_{\hat{\bm{\Theta}}_{\pi,n}}\ln \pi(\tilde{\mathbf{S}}_{n},\tilde{R}_{n};\hat{\bm{\Theta}}_{\pi,n},\bm{\Theta}_{\pi})(Q(\tilde{\mathbf{S}}_{n},\tilde{R}_{n}) + \gamma V^{\pi}(\tilde{\mathbf{S}}_{n+1};\hat{\bm{\Theta}}_{v,n},\bm{\Theta}_{v}) - V^{\pi}(\tilde{\mathbf{S}}_{n};\hat{\bm{\Theta}}_{v,n},\bm{\Theta}_{v})) \nonumber\\
%&+ \alpha_{\pi,n} \beta_{\pi,n} \nabla_{\hat{\bm{\Theta}}_{\pi,n}}\sum_{\tilde{R}_{n}\in\mathcal{D}}\pi(\tilde{\mathbf{S}}_{n},\tilde{R}_{n};\hat{\bm{\Theta}}_{\pi,n},\bm{\Theta}_{\pi})\ln \pi(\tilde{\mathbf{S}}_{n},\tilde{R}_{n};\hat{\bm{\Theta}}_{\pi,n},\bm{\Theta}_{\pi}).\nonumber
%\end{align}

\noindent\textbf{eA3C-OTPV critic network:} The eA3C-OTPV critic network is built based on the eA3C critic network. The new neural network has one input layer, $\hat{L}_{v}- 1  (\hat{L}_{v} \geq L_{v})$ hidden layers, and one output layer, indexed by $0,1,\ldots,\hat{L}_{v}-1,\hat{L}_{v}$, respectively. The input layer has $2(M+3)$ 1D CNNs, with the first $M+3$ 1D CNNs coming from the eA3C network. For all $l=1,\ldots,L_{v}-1$, the $l$-th hidden layer is a partially connected layer containing $n_{v,l} + \hat{n}_{v,l}$ neurons, with the first $n_{v,l}$ neurons coming from the eA3C network and the remaining $\hat{n}_{v,l}$ neurons that also utilize the ReLU activation function. For all $l=L_{v},\ldots,\hat{L}_{v}-1$, the $l$-th hidden layer is a fully connected layer with $\hat{n}_{v,l}$ neurons that utilize the ReLU activation function. The output layer is a fully connected layer without any activation function.
 Similarly, we rearrange all the tunable weights of the eA3C-OTPV critic network as a column vector, denoted by $\hat{\bm{\Theta}}_{v_{\text{OTPV}}} = (\hat{\mathbf{w}}^{(1)}_{v,0},\hat{\mathbf{w}}^{(2)}_{v,0},\ldots, \hat{\mathbf{w}}^{(M+3)}_{v,0}, \text{vec}(\hat{\mathbf{W}}_{v,1}),\ldots, \text{vec}(\hat{\mathbf{W}}_{v,\hat{L}_{v}-1}),$ $ \hat{\mathbf{w}}_{v,\hat{L}_{v}})\in\mathbb{R}^{\hat{n}_{v}}$, where $\hat{n}_{v} = (M+3)\hat{n}_{v,0} + \hat{n}_{v,1}(M+3)(2k-\hat{n}_{v,0} -n_{v,0} +2) + \sum^{L_{v}}_{l=2}\hat{n}_{v,l}(\hat{n}_{v,l-1}+n_{v,l-1}) + \sum_{L_{v}+1}^{\hat{L}_{v}}\hat{n}_{v,l}\hat{n}_{v,l-1}$. 
 %(iii) We denote the output of the last $(M+3)$ 1D CNNs in the input layer by $\hat{\mathbf{h}}_{v,0}$. For all $l=1,\ldots,L_{c}$, we denote the output of the last $\hat{n}_{v, l}$ neurons in the $l$-th layer by $\hat{\mathbf{h}}_{v,l}$. For all $l=L_{c}+1,\ldots,\hat{L}_{c}-1$, we denote the output of the $l$-th layer by $\hat{\mathbf{h}}_{v,l}$. We denote the output of the $\hat{L}_{c}$-th layer by $\hat{h}_{v,\hat{L}_{c}}$. 
 Then, based on the structure of the eA3C-OTPV critic network, \textcolor{black}{we can derive the layer outputs given by \eqref{eq:output_OTPV}, as shown at the top of the next page.}
\begin{figure*}
\begin{align}
\label{eq:output_OTPV}
\begin{split}
&\hat{\mathbf{h}}_{v,0} 
= (\hat{\mathbf{w}}^{(1)}_{v,0} * \mathbf{C}_{n},\hat{\mathbf{w}}^{(2)}_{v,0} *\mathbf{B}_{n}, \hat{\mathbf{w}}^{(3)}_{v,0} *\mathbf{Y}_{n}, \hat{\mathbf{w}}^{(4)}_{v,0} * \mathbf{X}_{1,n}, \ldots,  \hat{\mathbf{w}}^{(M+3)}_{v,0} *\mathbf{X}_{M,n} ) ,\\
&(\mathbf{h}_{v,1},\hat{\mathbf{h}}_{v,1}) = \textbf{ReLU}\left(\begin{pNiceMatrix}
      \mathbf{W}^{\star}_{v,1} & \bigzero_{n_{v,1}\times (k-\hat{n}_{v, 0} + 1)} \\
      \Block{1-2}{\hat{\mathbf{W}}_{v,1}} & 
    \end{pNiceMatrix}\left(\mathbf{h}_{v,0},\hat{\mathbf{h}}_{v,0}\right) \right),\\
&(\mathbf{h}_{v,l},\hat{\mathbf{h}}_{v,l}) = \textbf{ReLU}\left(\begin{pNiceMatrix}
      \mathbf{W}^{\star}_{v,l} & \bigzero_{n_{v,l}\times \hat{n}_{v, l-1} } \\
      \Block{1-2}{\hat{\mathbf{W}}_{v,l}} & 
    \end{pNiceMatrix}\left(\mathbf{h}_{v,l-1},\hat{\mathbf{h}}_{v,l-1} \right)\right),~l= 2,\ldots,L_{v},\\
&\hat{\mathbf{h}}_{v,l} = \textbf{ReLU}(\hat{\mathbf{W}}_{v,l}\hat{\mathbf{h}}_{v,l-1}) ,~l= L_{v} + 1,\ldots,\hat{L}_{v}-1, \quad \hat{h}_{v,\hat{L}_{v}} = \hat{\mathbf{w}}^{T}_{v,\hat{L}_{v}}\hat{\mathbf{h}}_{v,\hat{L}_{v}-1}.
%&\mathbf{h}_{\pi,L} = \pi(\mathbf{S}_{n},f_{i};\bm{\Theta}_{\pi,n}) =  \frac{\mathbf{w}^{T}_{\pi,L,i}\mathbf{h}_{L-1}}{\sum_{i=1}^{n_{\pi,L}}\mathbf{w}^{T}_{\pi,L,i}\mathbf{h}_{L-1}},~i=1,\ldots,n_{\pi,L}, \nonumber
\end{split}
\end{align}
\hrulefill
\end{figure*}
%where $\mathbf{h}_{v,0}$ and $\mathbf{h}_{v,l},l=1,\ldots,L_{c}$ are the output of the input layer and the output of the $l$-th layer of the eA3C critic network, respectively, and are also the output of the first $(M+3)$ 1D CNNs in the input layer and the output of the first $n_{v,l}$ neurons in the $l$-th layer of the eA3C-OTPV critic network of. 
\textcolor{black}{Based on the composition structure in \eqref{eq:output_OTPV},} the output $\hat{h}_{v,\hat{L}_{v}}$ can be rewritten as $V^{\pi_{\text{OTPV}}}(\cdot;\bm{\Theta}^{\star}_{v},\hat{\bm{\Theta}}_{v_{\text{OTPV}}})$, representing the approximated value function under policy $\pi_{\text{OTPV}}$. The value parameters include newly introduced parameters $\hat{\bm{\Theta}}_{v_{\text{OTPV}}}$ which will be optimized and the parameters of the value function from eA3C $\bm{\Theta}^{\star}_{v}$ which are ``frozen" during the online tuning. 
%\begin{align}
%\hat{\bm{\Theta}}_{v,n+1} = \hat{\bm{\Theta}}_{v,n} - \alpha_{v,n}  \nabla_{\hat{\bm{\Theta}}_{v,n}}(Q(\tilde{\mathbf{S}}_{n},\tilde{R}_{n}) + \gamma V^{\pi}(\tilde{\mathbf{S}}_{n+1};\hat{\bm{\Theta}}_{v,n},\bm{\Theta}_{v}) - V^{\pi}(\tilde{\mathbf{S}}_{n};\hat{\bm{\Theta}}_{v,n},\bm{\Theta}_{v}) )^{2}.\nonumber
%\end{align}

%\textcolor{black}{Note that both eA3C-OTPV policy and value networks are set flexible to capture the temporal features of the input.} 
We aim to optimize the newly added parameters $\hat{\bm{\Theta}}_{\pi_{\text{OTPV}}}$ and $\hat{\bm{\Theta}}_{v_{\text{OTPV}}}$ for the online policy and the value function \textcolor{black}{obtained from the eA3C-OTPV network}, respectively, to maximize a metric similar to the one defined in \eqref{eq:loss_function}. Specifically, at each stage $n\geq q$, we use a batch of the latest $q$ samples and choose the 
\begin{figure*}
\begin{dmath}
 \sum^{n-1}_{i=n- q}\frac{1}{q}\left(\log\hat{\pi}(\mathbf{S}_{i},R_{i};\bm{\Theta}^{\star}_{\pi},\hat{\bm{\Theta}}_{\pi_{\text{OTPV}}}) \left(g(\mathbf{S}_{i},R_{i}) + \gamma V^{\pi_{\text{OTPV}}}(\mathbf{S}_{i+1};\bm{\Theta}^{\star}_{v},\hat{\bm{\Theta}}_{v_{\text{OTPV}}}) \right) 
 - \xi  \left(g(\mathbf{S}_{i},R_{i}) + \gamma V^{\pi_{\text{OTPV}}}(\mathbf{S}_{i+1};\bm{\Theta}^{\star}_{v},\hat{\bm{\Theta}}_{v_{\text{OTPV}}}) - V^{\pi_{\text{OTPV}}}(\mathbf{S}_{i};\bm{\Theta}^{\star}_{v},\hat{\bm{\Theta}}_{v_{\text{OTPV}}})\right)^{2} 
	 + \beta \sum_{r\in\mathcal{R}}\hat{\pi}(\mathbf{S}_{i},r;\bm{\Theta}^{\star}_{\pi},\hat{\bm{\Theta}}_{\pi_{\text{OTPV}}}) 
	\log\hat{\pi}(\mathbf{S}_{i},r;\bm{\Theta}^{\star}_{\pi},\hat{\bm{\Theta}}_{\pi_{\text{OTPV}}}) \right).\label{eq:loss_function_OTPV}
	\end{dmath}
	\hrulefill
\end{figure*}
\textcolor{black}{loss function in given by \eqref{eq:loss_function_OTPV}, as shown at the top of the next page}, for jointly training the tunable components of the actor and critic networks of the eA3C-OTPV network (i.e., optimizing $\hat{\bm{\Theta}}_{\pi_{\text{OTPV}}}$ and $\hat{\bm{\Theta}}_{v_{\text{OTPV}}}$). Since $\pi(\cdot;\bm{\Theta}^{\star}_{\pi})$ and $V^{\pi}(\cdot;\bm{\Theta}^{\star}_{v})$, whose nested functions are components of the nested functions of $\hat{\pi}(\cdot;\bm{\Theta}^{\star}_{\pi},\hat{\bm{\Theta}}_{\pi_{\text{OTPV}}})$ and $V^{\pi_{\text{OTPV}}}(\cdot;\bm{\Theta}^{\star}_{v},\hat{\bm{\Theta}}_{v_{\text{OTPV}}})$, respectively, as shown in \eqref{eq:output_OTP} and \eqref{eq:output_OTPV}, appear in the loss function in \eqref{eq:loss_function_OTPV}, the policy and value function obtained from the eA3C network affect the optimal choices of the parameters $\hat{\bm{\Theta}}_{\pi_{\text{OTPV}}}$ and $\hat{\bm{\Theta}}_{v_{\text{OTPV}}}$ for the online policy and the value function \textcolor{black}{obtained from the eA3C-OTPV network}, respectively. Intuitively, the joint optimization of $\hat{\bm{\Theta}}_{\pi_{\text{OTPV}}}$ and $\hat{\bm{\Theta}}_{v_{\text{OTPV}}}$ resembles that in the proposed eA3C method in Section~\ref{sec:off_formulation} and can be viewed as its online improvement version. Therefore, it is expected that the eA3C-OTPV method has better performance for the specific user than the eA3C method. 

Similarly, we train the eA3C-OTPV network using the RMSP algorithm \cite{rmsp}. 
%The per-stage computational complexity during the online training is $\mathcal{O}((3\hat{n}_{\pi_{\text{TL}},1} +\hat{n}_{\pi_{\text{TL}},0} + 3n_{\pi,1} + n_{\pi,0})(M+3)(2k-\hat{n}_{\pi_{\text{TL}},0} -n_{\pi,0} +2) + 4\sum^{\hat{L}_{a}}_{l=2}\hat{n}_{\pi_{\text{TL}},l}(\hat{n}_{\pi_{\text{TL}},l-1} + n_{\pi,l-1}) + (3\hat{n}_{v,1} +\hat{n}_{v,0} + 3n_{v,1} + n_{v,0})(M+3)(2k-\hat{n}_{v,0} -n_{v,0} +2) + 4\sum^{\hat{L}_{c}}_{l=2}\hat{n}_{v,l}(\hat{n}_{v,l-1} + n_{v,l-1}))$ [41]. The per-stage computational complexity after the online training is $\mathcal{O}((\hat{n}_{\pi_{\text{TL}},1} +\hat{n}_{\pi_{\text{TL}},0} + n_{\pi,1} + n_{\pi,0})(M+3)(2k-\hat{n}_{\pi_{\text{TL}},0} -n_{\pi,0} +2) + \sum^{\hat{L}_{a}}_{l=2}\hat{n}_{\pi_{\text{TL}},l}(\hat{n}_{\pi_{\text{TL}},l-1} + n_{\pi,l-1}))$ [41]. The per-stage space complexity is $\mathcal{O}(n_{v} + n_{\pi} + \hat{n}_{v} + \hat{n}_{\pi_{\text{TL}}})$. 
After training, we denote the values of the tunable weights of the actor and critic networks of the eA3C-OTPV network by $\hat{\bm{\Theta}}^{\star}_{\pi_{\text{OTPV}}}$ and $\hat{\bm{\Theta}}^{\star}_{v_{\text{OTPV}}}$, respectively. The output of the eA3C-OTPV actor network, i.e., $\hat{\pi}(\mathbf{s},r;\hat{\bm{\Theta}}^{\star}_{\pi_{\text{OTPV}}},\bm{\Theta}^{\star}_{\pi})$, is the distribution of resulting online policy \textcolor{black}{corresponding to eA3C-OTPV}.

\subsection{Comparisons of the eA3C-OTP and eA3C-OTPV Methods}
We compare the eA3C-OTP and eA3C-OTPV methods respectively, in terms of QoE, training time, and inference time. Firstly, only the tunable components of the eA3C-OTP policy network are trained based on the policy and value networks of the eA3C network. By contrast, the tunable components of both policy and value networks of the eA3C-OTPV network are trained based on the policy and value networks of the eA3C network. Therefore, it is expected that the eA3C-OTPV method outperforms the eA3C-OTP method in QoE. Secondly, the training time for the eA3C-OTPV network is longer than the one for the eA3C-OTP network (and hence $q'$ for the eA3C-OTPV network is larger than the one for the eA3C-OTP network) since it has more tunable parameters and a more complex neural network structure. Thirdly, the inference times for the policies network of the eA3C-OTP and eA3C-OTPV networks to produce the actions (bitrate selections) are identical if their policy networks have the same structure, i.e., the parametric forms $\hat{\pi}(\cdot;\hat{\bm{\Theta}}^{\star}_{\pi_{\text{OTP}}},\bm{\Theta}^{\star}_{\pi})$ and $\hat{\pi}(\cdot;\hat{\bm{\Theta}}^{\star}_{\pi_{\text{OTPV}}},\bm{\Theta}^{\star}_{\pi})$ are the same. One can select the eA3C-OTP network or eA3C-OTPV network based on the different QoE, training time, and inference time requirements.

\begin{table}[t]
\caption{Bitrates (in Mbit/s).}
\centering
\resizebox{8.5cm}{!}{  
\begin{tabular}{|c|c|c|c|c|c|c|} 
\hline  
Quality level $d$& 1 & 2 & 3 & 4 & 5 & 6\\ \hline  
Encoding rate  $r_{d}$ (in Mbit/s)& $0.3$ &  $0.75 $ & $1.2$ & $1.85$ & $2.85$ & $4.3$\\ \hline 
\end{tabular}
} \label{table1}
\end{table}

\begin{figure*}[t]
%\vspace*{-1.2cm}
\begin{center}
   \subfloat[MAC rate]
   {\resizebox{4.4cm}{!}{\includegraphics{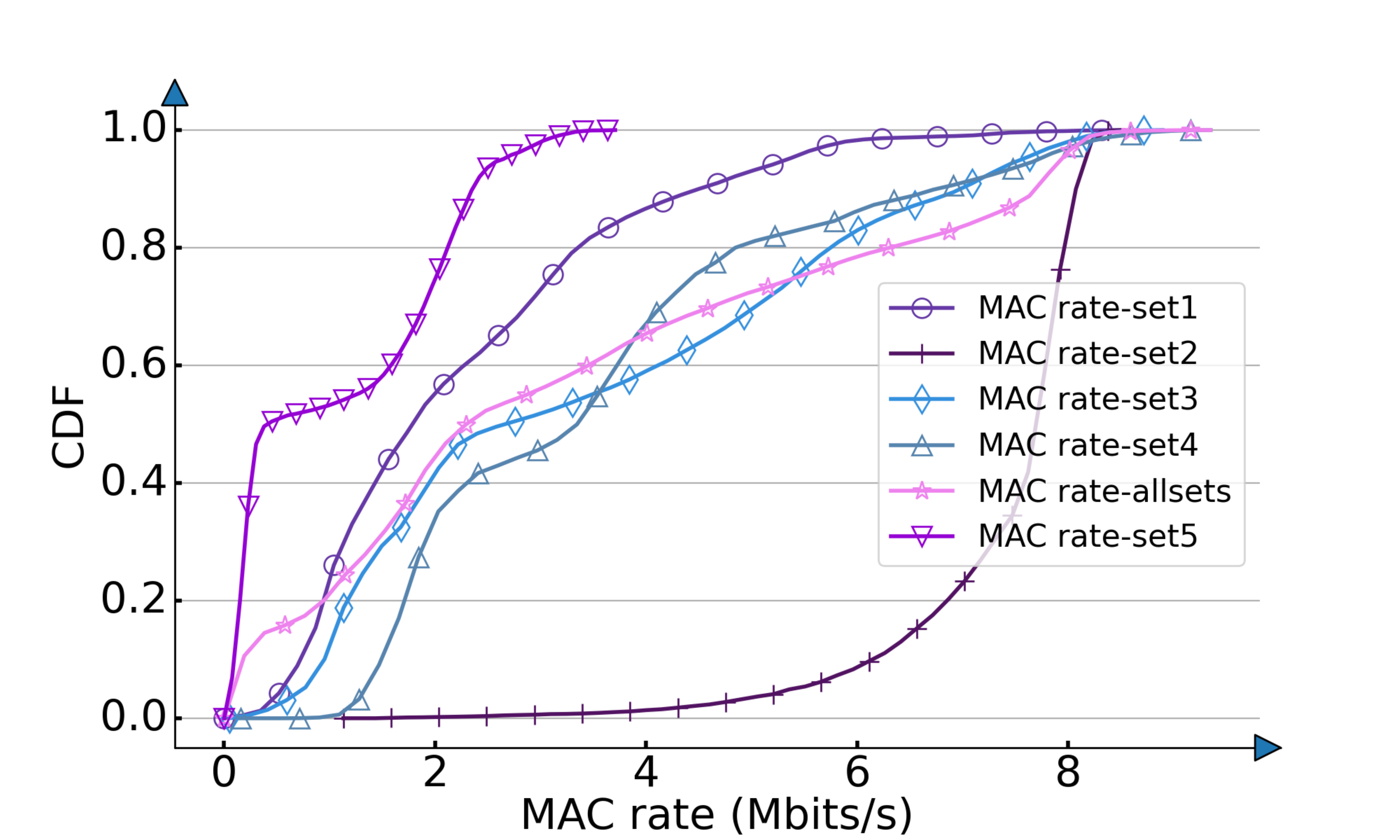}}}
   \subfloat[MCS index]
   {\resizebox{4.4cm}{!}{\includegraphics{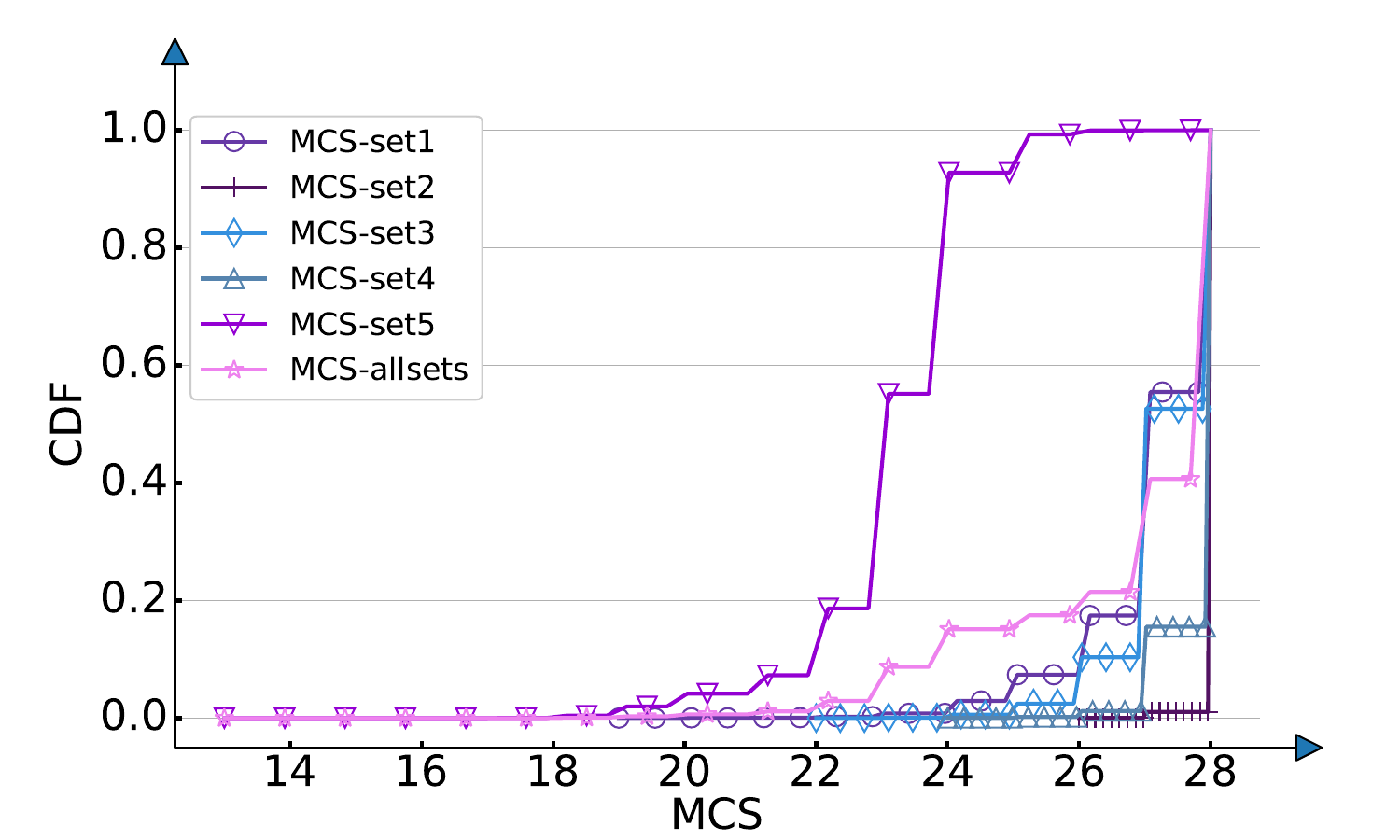}}}
   \subfloat[PRB number]
   {\resizebox{4.4cm}{!}{\includegraphics{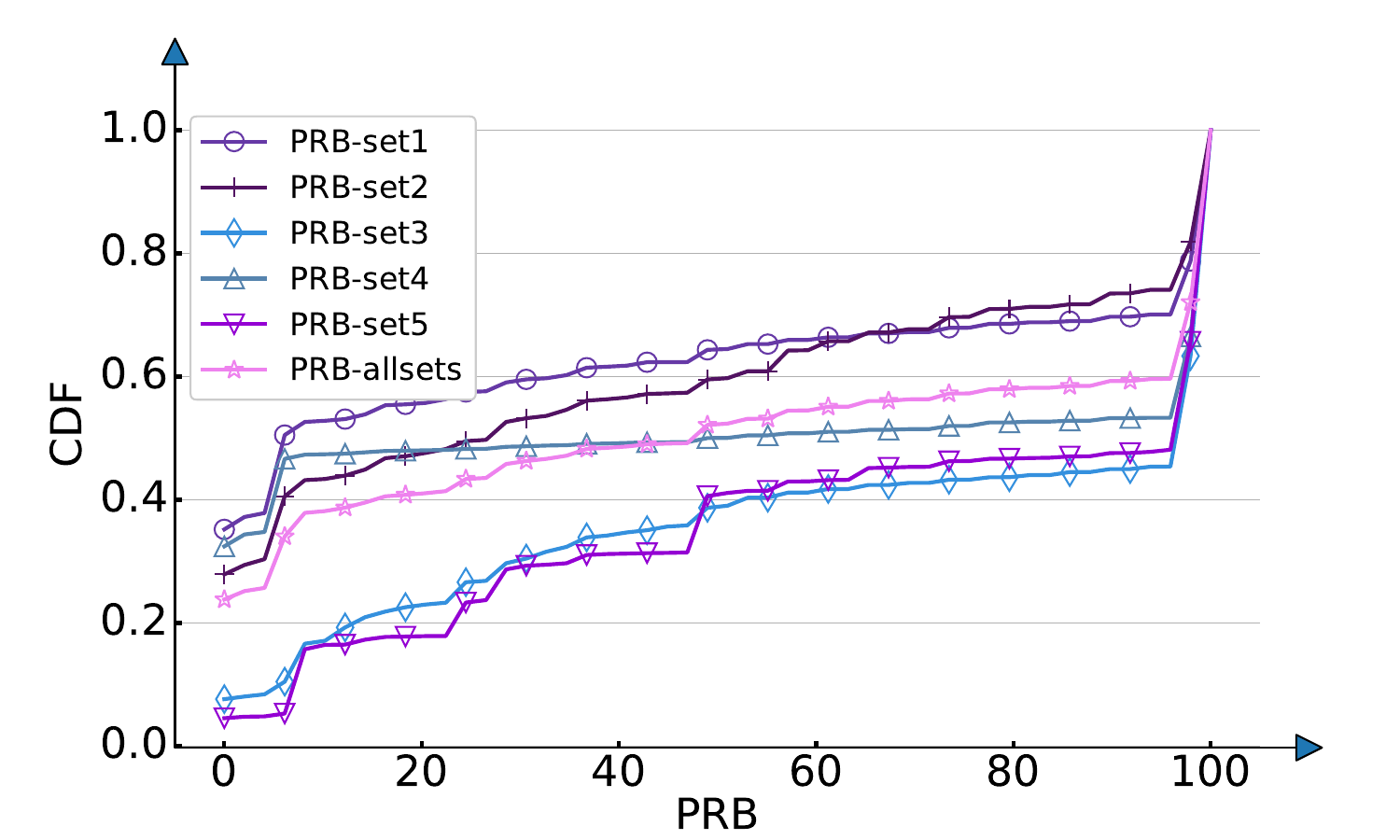}}}
   \subfloat[APP layer throughput]
   {\resizebox{4.4cm}{!}{\includegraphics{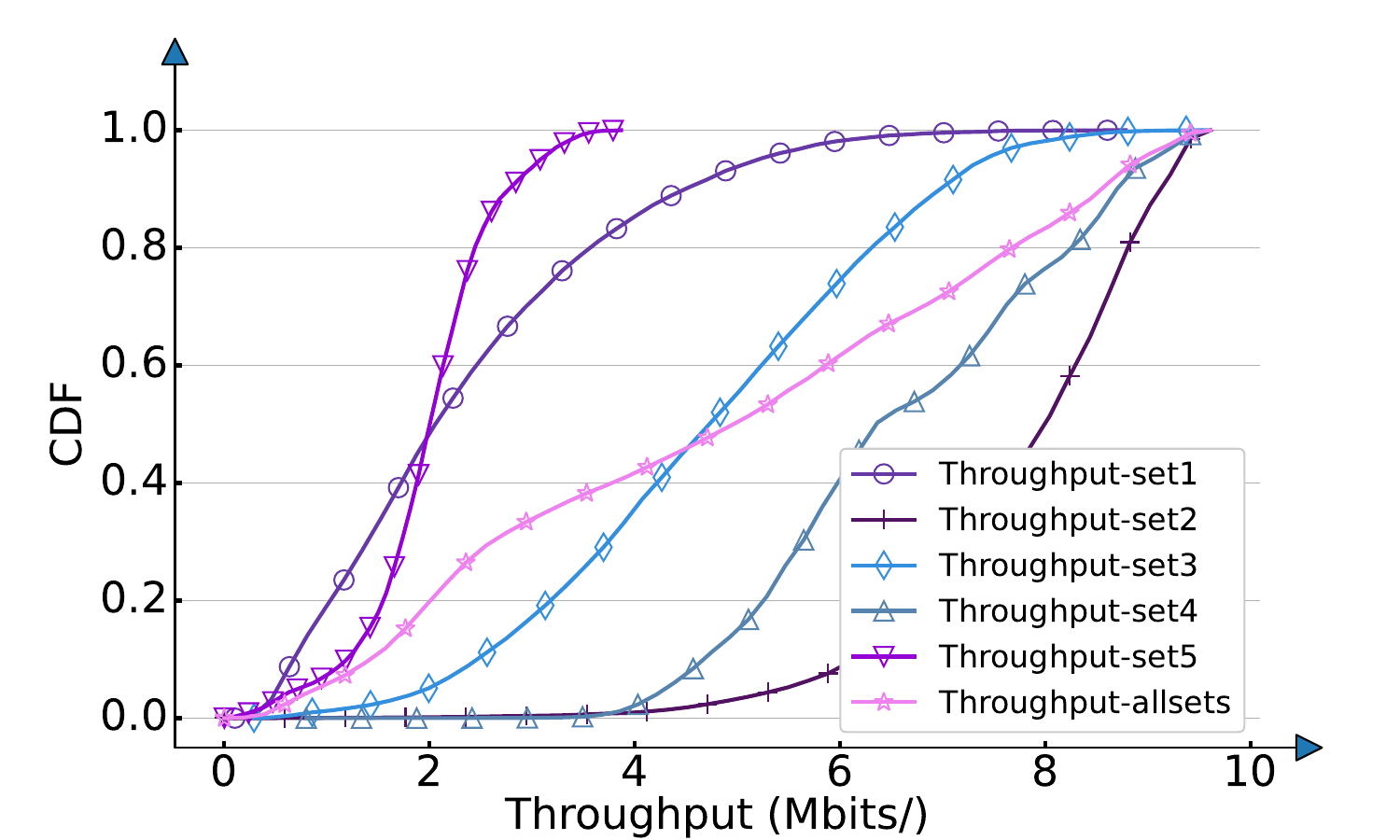}}}
 %\subfloat[\small{CDF of rebuffering time}]
 %{\resizebox{5.9cm}{!}{\includegraphics{pic/multi_rebuffering.eps}}}
 %\vspace*{-0.07cm}
 \end{center}
 %\vspace*{-0.37cm}
   \caption{APP layer throughput and lower-layer information.}
   \label{fig:dataset}
%\vspace*{-0.60cm}
\end{figure*}

\begin{table}[t] 
\caption{Training setup.}
\centering
\resizebox{8.5cm}{!}{  
\begin{tabular}{|c|c|c|} 
\hline  
{} & Offline training & Online training\\ \hline  
Training Parameter & \multicolumn{2}{c|}{Value}  \\\hline 
Number of training samples & $19200$ &  $3200\times 5 $  \\ \hline
Number of validation samples  & $6400$ &  \diagbox{}{} \\\hline 
Number of testing samples & $6400$ &  \diagbox{}{} \\\hline 
Maximization epochs & $100 $ &  $20 $ \\ \hline 
Discount factor $\gamma$ & $0.99 $ &  $0.9 $  \\ \hline
Entropy factor $\beta$ & Decay from $3$ to $0.1$ &  $0.5$ \\ \hline  
Learning rate & $10^{-4} $ &  $10^{-3} $ \\ \hline
Batch size & $8$ &  $8$\\ \hline
Weight $\xi$ & $1 $ &  $1 $ \\ \hline 
\end{tabular}
} \label{table3}
\end{table}

\section{Performance Evaluation}\label{sec:sim}

\subsection{Experimental Setup}
We consider adaptive streaming of one video, i.e., \textit{EnvivioDash3} provided by\cite{video}.\footnote{\textcolor{black}{The evaluation results for different videos are similar due to the assumption in footnote~1.}} The video lasts 192 seconds and is encoded into 48 chunks, each of 4 seconds. That is to say, we set $N=48$ and $T=4$ seconds. We set $D=6$ and set $r_{d},d\in\mathcal{D}$ according to Table~\ref{table1}\cite{sigcomm17}. We set $B=60$ seconds and adopt the utility function in \cite{sigcomm17}, i.e., $U(R) = \log(\frac{R}{r_{1}})$. We conduct five experiments under different network environments and collect 5 datasets. Each dataset has multiple traces, each containing 200 samples. Each sample consists of the APP layer throughput and $M=3$ lower-layer quantities including MAC rate, PRB number, and MCS index. 
%During data collection, a test user is downloading a 4K video from a BS. The lower-layer information from the BS and the application layer throughput from test user are reported to a data collection server. We collect the dataset under different channel conditions and number of background users. Each dataset contains the application layer throughput and three lower-layer information, i.e., . 
%The detailed setup can be seen in Table~\ref{table2}. 
The distributions of the APP layer throughput, MAC rate, PRB number, and MCS index across all datasets are shown in Fig.~\ref{fig:dataset}. For all $i=1,2,\ldots,5,$ we partition the $i$-th dataset into two subsets with $67\%$ and $33\%$ of samples, named Set-$i$-OFF and Set-$i$-ON, respectively. We further divide Set-$i$-OFF into three subsets with $60\%$, $20\%$, and $20\%$ of samples, named Set-$i$-OFF-Train, Set-$i$-OFF-Val, and Set-$i$-OFF-Test, respectively, and merge all samples in Set-$i$-OFF-Train/Set-$i$-OFF-Val/Set-$i$-OFF-Test, $i=1,2,\ldots, 5$ into one training/validation/testing set, namely Set-OFF-Train/Set-OFF-Val/Set-OFF-Test. The sizes of the datasets are shown in Table~\ref{table3}. Set-OFF-Train, Set-OFF-Val, and Set-OFF-Test are used in the offline scenario, whereas Set-$i$-ON, $i=1,2,\ldots,5$ are used in the online scenario. In the rest of the section, we evaluate the proposed eA3C, eA3C-OTP, and eA3C-OTPV methods in terms of the average QoE, inference time, and training time, respectively. \textcolor{black}{We implement eA3C, eA3C-OTP, and eA3C-OTPV on a desktop with the processor of AMD Ryzen 9 5900HX, 16GB RAM, and NVIDIA GeForce RTX 3080 graphics card.} Unless otherwise specified, for any DRL-based method, \textcolor{black}{the input of the neural network at stage $n$ includes the APP layer information of the most recent $k=8$ stages, i.e., APP layer throughputs $\mathbf{C}_{n}$, buffer occupancies $\mathbf{B}_{n},$ and bitrate selections $\mathbf{Y}_{n}$, and for each proposed method, the input of the neural network includes additional lower-layer information (at different amounts) of the most recent $k=8$ stages.} \textcolor{black}{The numbers of neurons and hidden layers of the policy and value networks are set as the same for simplicity,} and the average QoE and inference time are for the optimized neural network structure.

\begin{table*}[t]
\centering
\caption{Details of the baseline and proposed methods.}
\resizebox{17cm}{!}{ 
\begin{tabular}{c|c|c|ccccccc|cc|cc}
\toprule
\multicolumn{3}{c}{Methods} &  \multicolumn{7}{c}{Input state} & \multicolumn{2}{c}{Network }&\multicolumn{2}{c}{Datasets}\\
\midrule
{} &{} &{}   &APP layer throughput   & buffer occupancy & bitrate selection & MAC rate & PRB & MCS index & $M$  & A3C   & eA3C & training & evaluation \\
\midrule
\multirow{9}{*}{\thead{Offline \\scenario}} 
& \multirow{5}{*}{\thead{Proposed \\methods}} & eA3C-MAC   &  \checkmark & \checkmark   & \checkmark  & \checkmark & $\times$ & $\times$ & $1$ &$\times$ & \checkmark & \multirow{5}{*}{Set-OFF-Train} & \multirow{5}{*}{Set-OFF-Test} \\
& & eA3C-PRB  &  \checkmark & \checkmark   & \checkmark  & $\times$ & \checkmark & $\times$ & $1$ &$\times$ & \checkmark &  & \\
&  & eA3C-MCS   &  \checkmark  &  \checkmark   & \checkmark  & $\times$ & $\times$ & \checkmark & $1$ &$\times$ & \checkmark &  & \\
&  & eA3C-2   &  \checkmark  &  \checkmark   & \checkmark  & $\times$ & \checkmark   & \checkmark  & $2$&$\times$ & \checkmark &  &  \\
&  & eA3C-3   &  \checkmark  &  \checkmark   & \checkmark  & \checkmark & \checkmark & \checkmark & $3$ &$\times$ & \checkmark &  & \\ \cline{2-14}
&  \multirow{4}{*}{\thead{Baseline\\ methods}}& Pensieve\cite{sigcomm17}  &  \checkmark  &  \checkmark   & \checkmark   & $\times$ & $\times$ & $\times$ & $0$ & \checkmark  &$\times$ & \multirow{2}{*}{Set-OFF-Train} & \multirow{3}{*}{Set-OFF-Test}\\
 & & Pensieve-3\cite{a3c}   &  \checkmark  &  \checkmark   & \checkmark  & \checkmark & \checkmark & \checkmark & $3$& \checkmark  &$\times$ &  & \\
& & MPC\cite{sigcomm15} &  \checkmark  &  \checkmark   & $\times$  & $\times$ & $\times$ & $\times$ & $0$ & $\backslash$ & $\backslash$ & $\backslash$ & \\  \hline
 
\multirow{7}{*}{\thead{Online\\ scenario}} 
&  \multirow{2}{*}{\thead{Proposed\\ methods}} & eA3C-OTP-3   &  \checkmark & \checkmark   & \checkmark  & \checkmark & \checkmark & \checkmark & $3$ &$\times$ & \checkmark & \multirow{2}{*}{Set-$2$($5$)-ON} & \multirow{2}{*}{Set-$2$($5$)-ON} \\
& & eA3C-OTPV-3  &  \checkmark & \checkmark   & \checkmark  & \checkmark & \checkmark & \checkmark & $3$ &$\times$ & \checkmark &  & \\ \cline{2-14}
&  \multirow{5}{*}{\thead{Baseline\\ methods}}& eA3C-3-1set   &  \checkmark  &  \checkmark   & \checkmark  & \checkmark & \checkmark & \checkmark & $3$& $\times$ &\checkmark & Set-$2$($5$)-OFF-Train & \multirow{5}{*}{Set-$2$($5$)-ON}  \\
 & & eA3C-3  &  \checkmark  &  \checkmark   & \checkmark   & \checkmark & \checkmark & \checkmark & $3$ & $\times$  &\checkmark & \multirow{3}{*}{Set-OFF-Train} & \\
& & Pensieve\cite{sigcomm17}   &  \checkmark  &  \checkmark   & \checkmark  & $\times$ & $\times$ & $\times$ & $0$& \checkmark  &$\times$ & & \\
 & & MERINA\cite{mm22}   &  \checkmark  &  \checkmark   & \checkmark  & $\times$ & $\times$ & $\times$ & $0$& \checkmark  &$\times$ &  & \\
 & & MPC\cite{sigcomm15} &  \checkmark  &  \checkmark   & $\times$  & $\times$ & $\times$ & $\times$ & $0$ & $\backslash$ & $\backslash$ & $\backslash$ & \\ 
 
\bottomrule
\end{tabular}
}
\label{table4}
\end{table*}

\begin{figure*}[t]
\begin{center}
 {\resizebox{18cm}{!}{\includegraphics{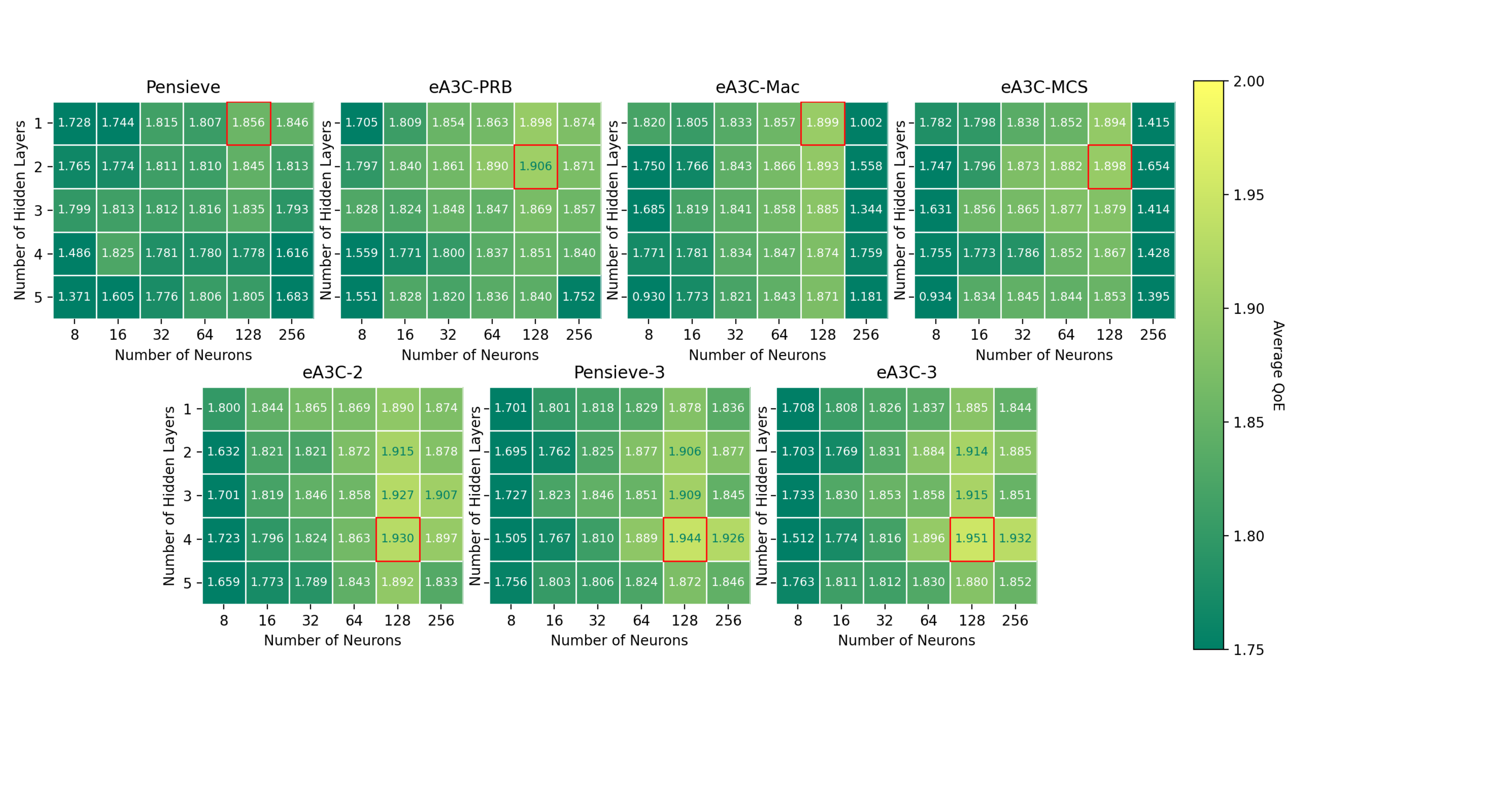}}}
 \vspace*{-1.57cm}
\end{center}
   \caption{Heatmaps of the average QoE at $k = 8$ in the offline scenario. The QoE of the optimized neural network structure is marked with a red rectangle.}
   \label{fig:heatmap}
%\vspace*{-1.60cm}
\end{figure*}

\begin{figure*}[t]
%\vspace*{-1.2cm}
\begin{center}
\subfloat[Average QoE versus weight.]
   {\resizebox{4.4cm}{!}{\includegraphics{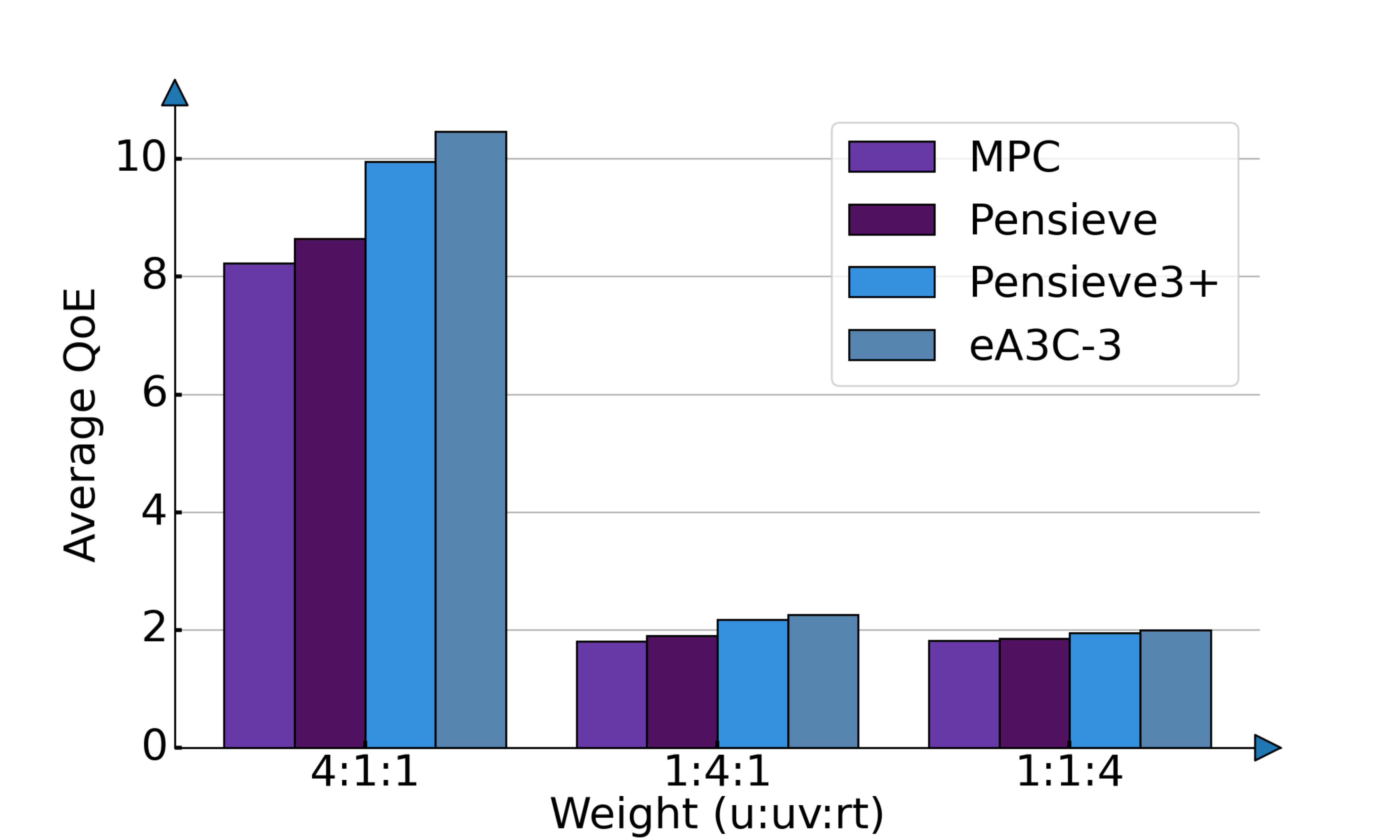}}}
   \subfloat[Average video quality versus weight.]
   {\resizebox{4.4cm}{!}{\includegraphics{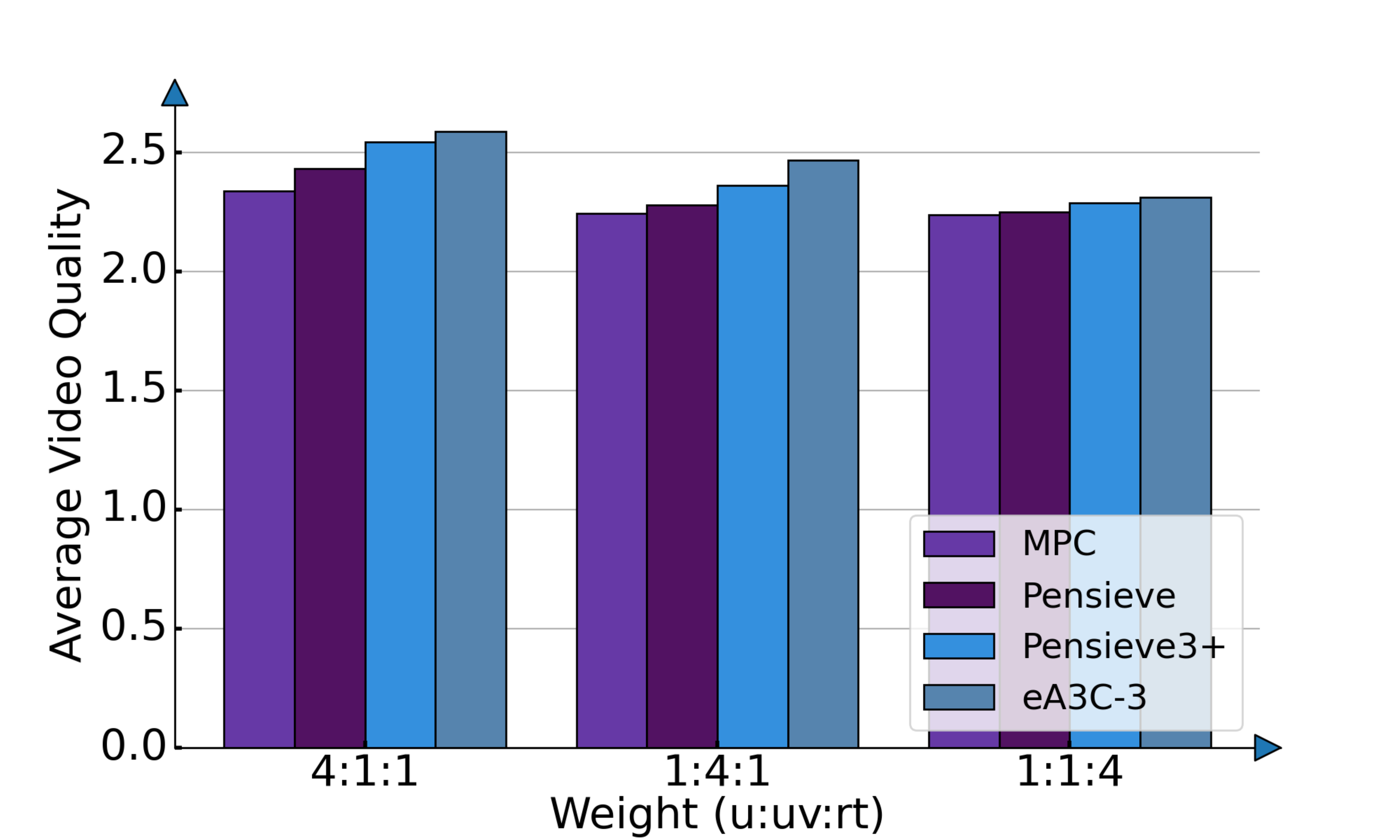}}}
   \subfloat[Average video quality variation versus weight.]
   {\resizebox{4.4cm}{!}{\includegraphics{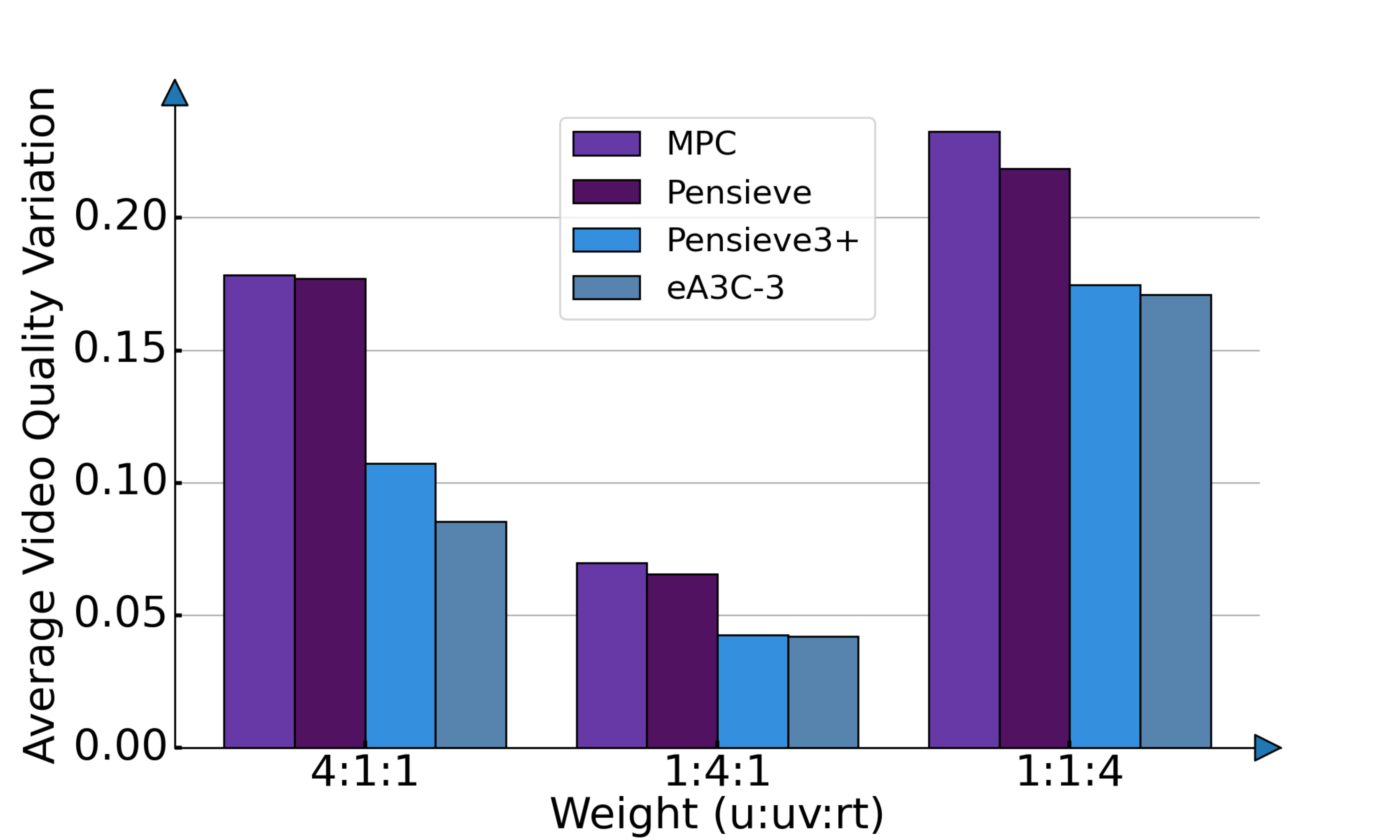}}}
    \subfloat[Average rebuffering versus weight.]
   {\resizebox{4.4cm}{!}{\includegraphics{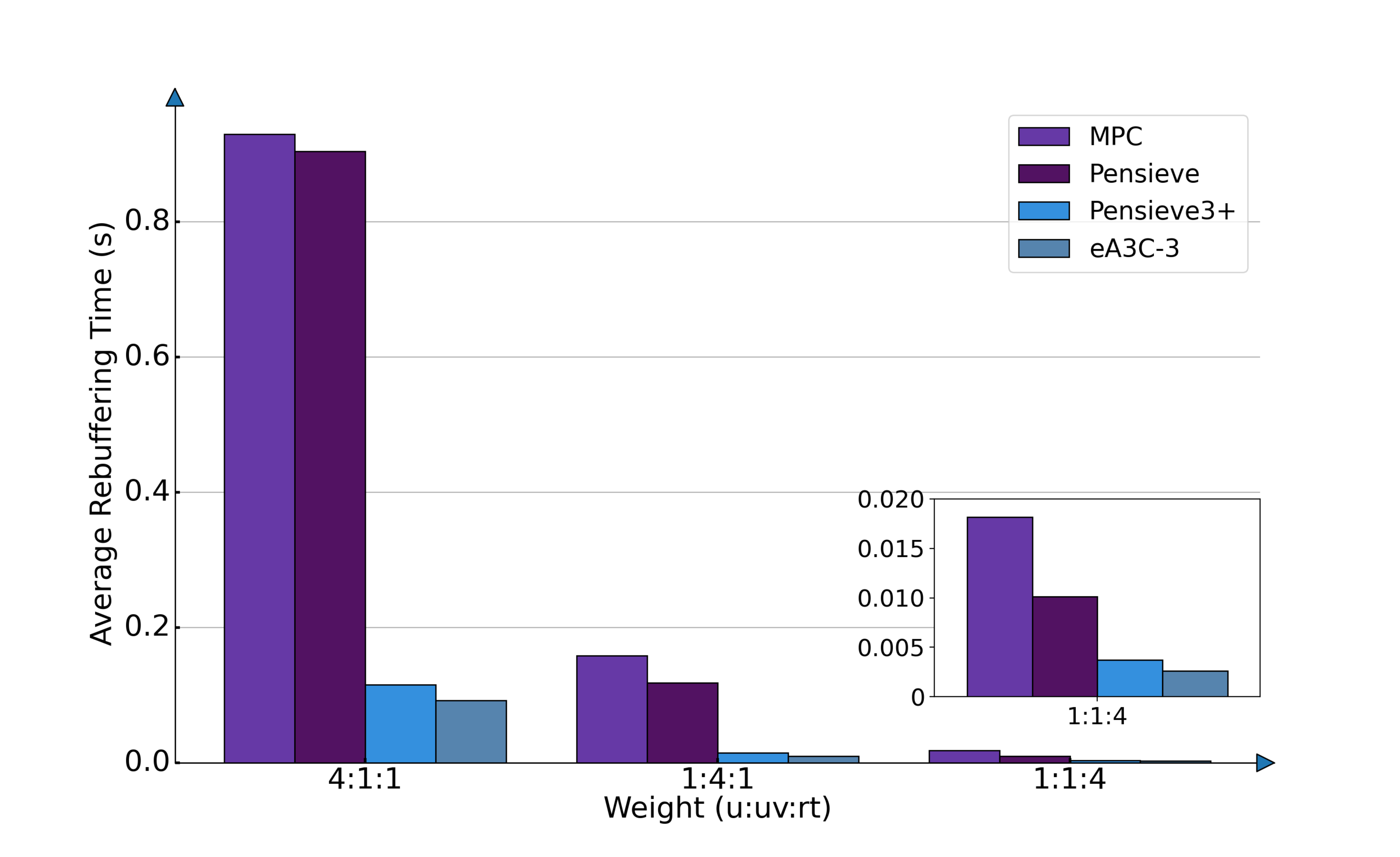}}}
 %\subfloat[\small{CDF of rebuffering time}]
 %{\resizebox{5.9cm}{!}{\includegraphics{pic/multi_rebuffering.eps}}}
 %\vspace*{-0.07cm}
 \end{center}
 %\vspace*{-0.37cm}
   \caption{Individual components in QoE of proposed and baseline schemes under different weights  at $k = 8$ in the offline scenario.}
   \label{fig:weight}
%\vspace*{-0.60cm}
\end{figure*}

\begin{figure*}[t]
%\vspace*{-1.2cm}
\begin{center}
      \subfloat[Average QoE]
   {\resizebox{6cm}{!}{\includegraphics{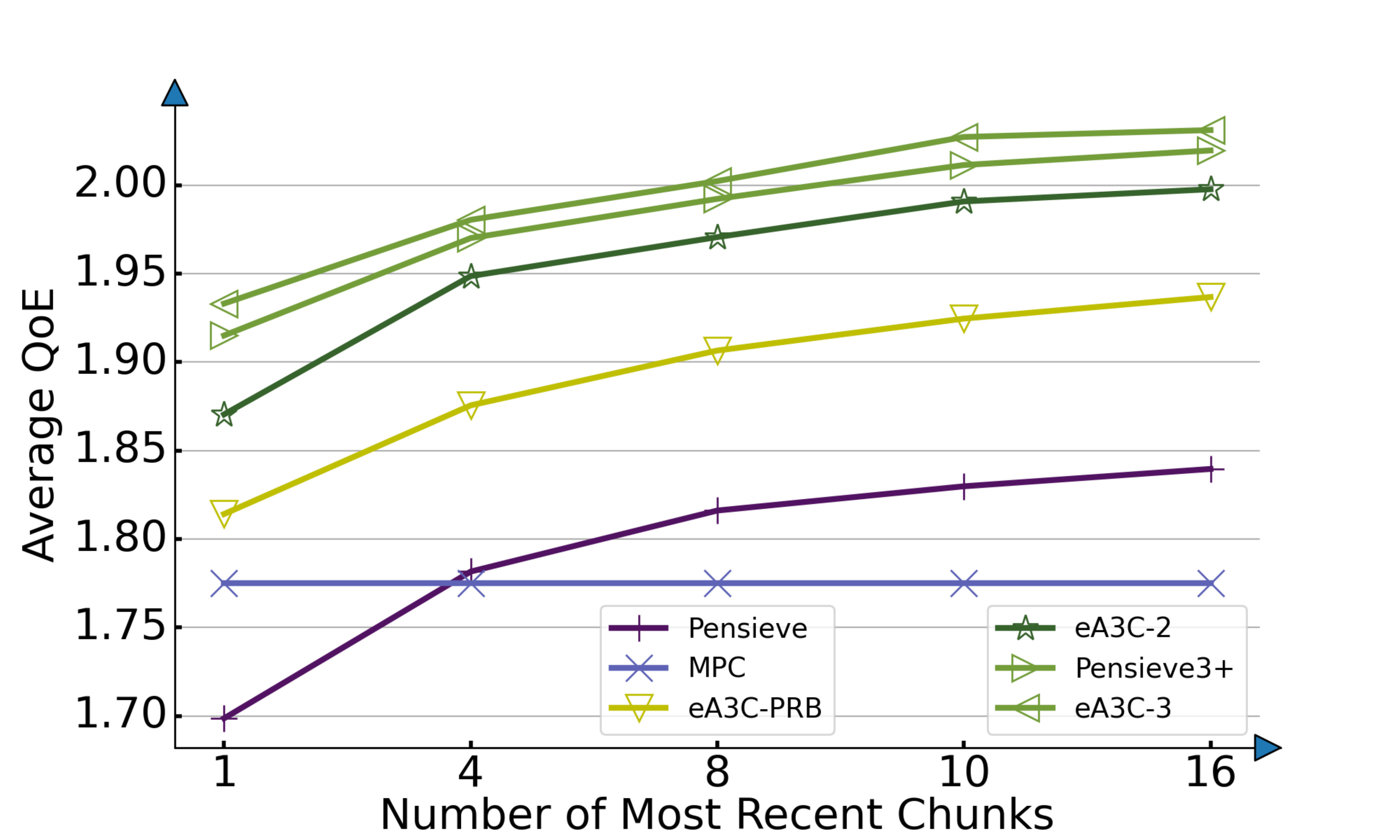}}}
   \subfloat[Average inference time]
   {\resizebox{6cm}{!}{\includegraphics{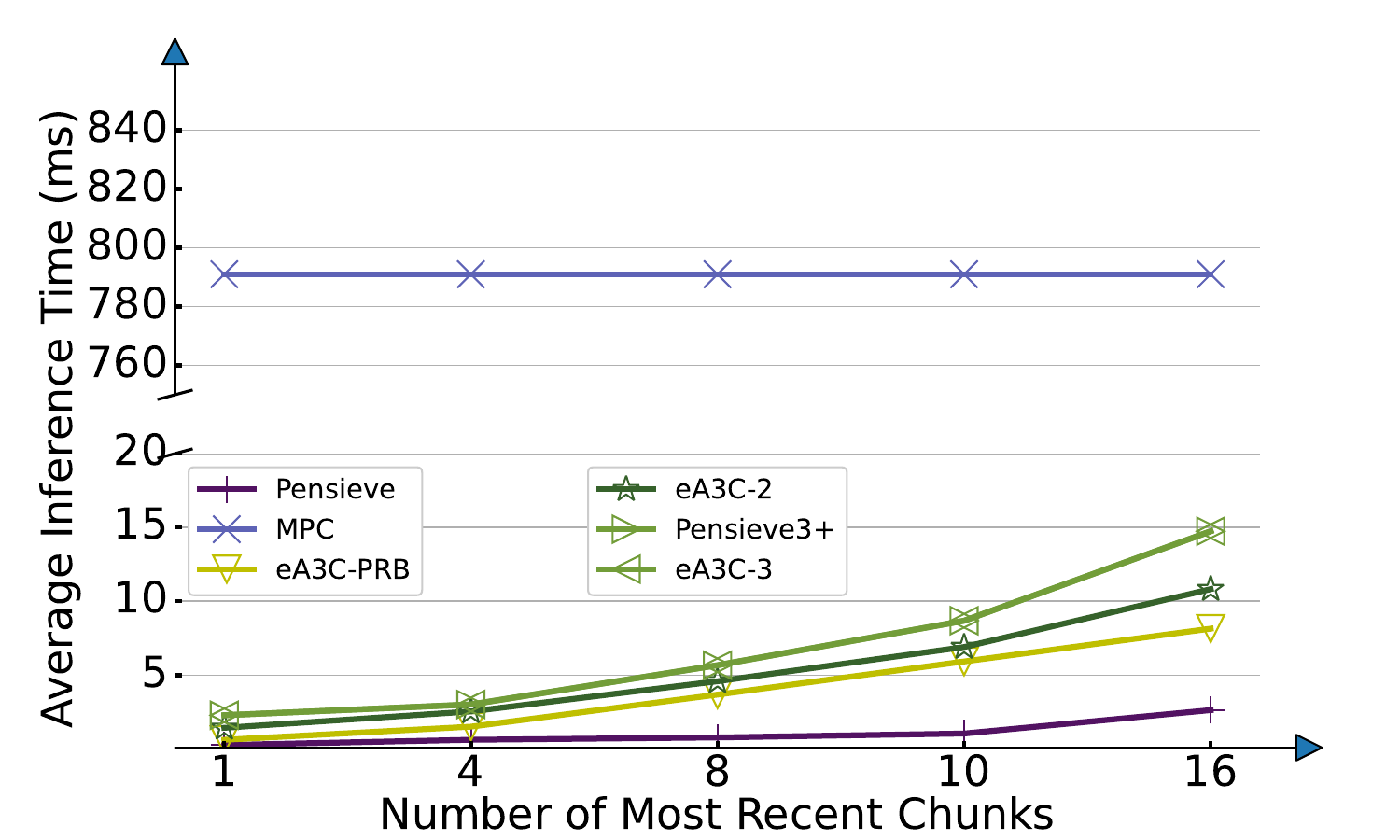}}}
 %\subfloat[\small{CDF of rebuffering time}]
 \subfloat[Average training time]
   {\resizebox{6cm}{!}{\includegraphics{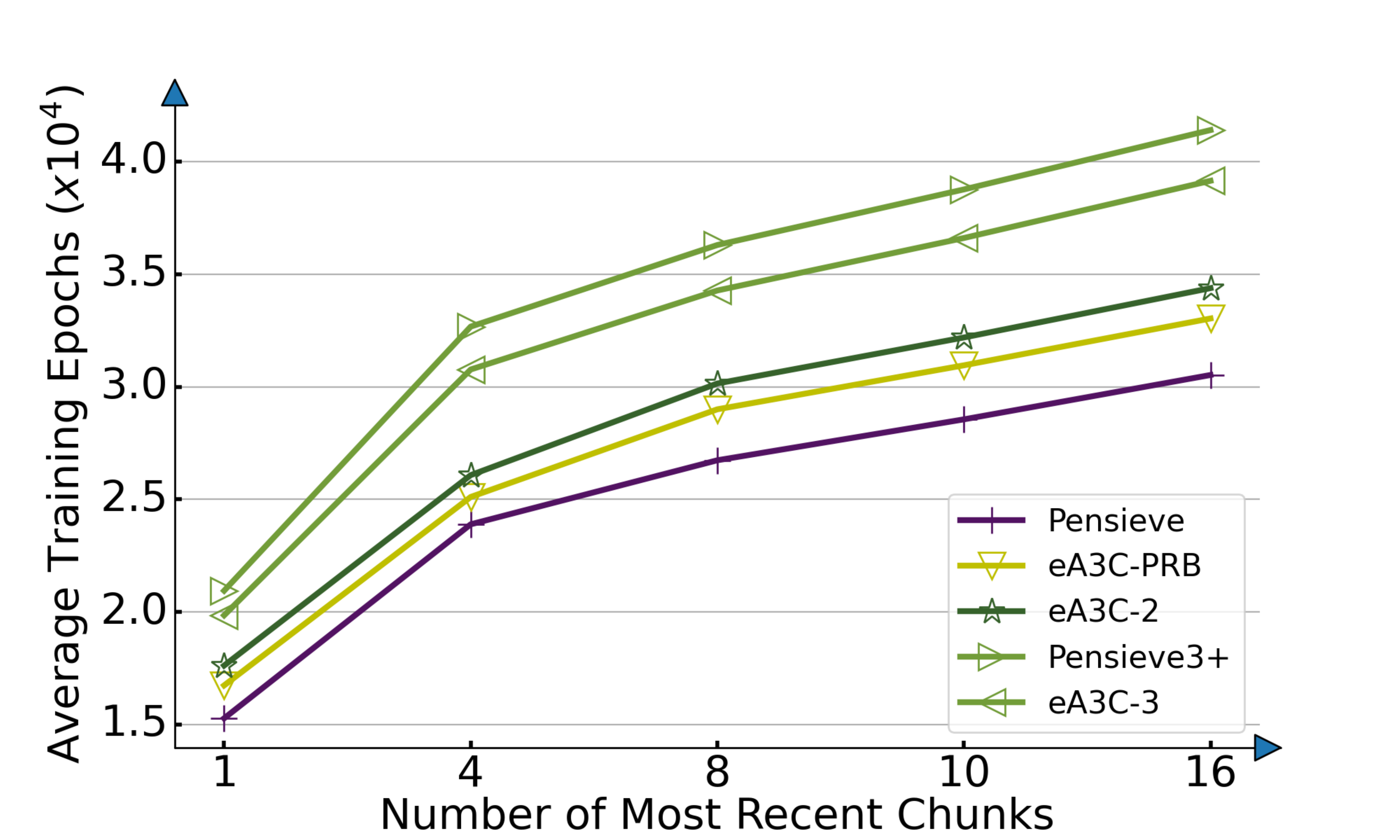}}}
 %\subfloat[\small{CDF of rebuffering time}]
 \end{center}
   \caption{Different performance metrics of proposed and baseline schemes under different numbers of most recent chunks $k$ in the offline scenario.}
   \label{fig:offline_pastchunk}
%\vspace*{-0.60cm}
\end{figure*}

\subsection{Offline Scenario}\label{sec:offline_sim} 
In the offline scenario, we consider five instances of the proposed eA3C method which use different lower-layer information. Specifically, the instances with one lower-layer quantity ($M=1$) being MAC rate, PRB number, and MCS index are referred to as eA3C-MAC, eA3C-PRB, eA3C-MCS, respectively; the instance with two lower-layer quantities ($M=2$) being MCS index and PRB number is referred to as eA3C-2; and the instance with all three lower-layer quantities ($M=3$) is referred to as eA3C-3. We train the five respective eA3C networks using the proposed eA3C method.
\textcolor{black}{We consider three baseline methods, including Pensieve\cite{sigcomm17}, Pensieve-3 (an enhanced version of Pensieve\cite{sigcomm17} which utilizes all three lower-layer quantities), and MPC\cite{sigcomm15}. \textcolor{black}{Specifically, Pensieve (Pensieve-3) adopts the A3C method\cite{a3c} and trains the policy network and value network of the A3C network alternatively. The training setups of all DRL-based methods are identical and are illustrated in Table~\ref{table3}.}
For each DRL-based method,} when the value of a loss function on the validation set does not change for five epochs, the training process is stopped, and the corresponding parameters are saved. \textcolor{black}{MPC adapts the video bitrate according to the current APP layer throughput and buffer occupancy based on model predictive control optimization method\cite{dp_2}. Specifically, MPC\cite{sigcomm15} applies $5$-step lookahead optimization\cite{dp_2} and adopts exhaustive search over $\mathbb{R}^{5}$ to obtain the optimal bitrate.} The details of the proposed and baseline methods are summarized in Table~\ref{table4}. The performance of all methods are evaluated on Set-OFF-Test in the offline scenario.

In the following, we evaluate the proposed eA3C method and the above mentioned baseline methods in the offline scenario. Fig.~\ref{fig:heatmap} shows the heatmap of the average QoE at different numbers of neurons and hidden layers. We have the following observations: the average QoE of each method increases and then decreases with the number of neurons; the average QoEs of Pensieve and eA3C-MAC always decrease with the number of hidden layers; and the average QoEs of Pensieve-3, eA3C-MCS, eA3C-PRB, eA3C-2, and eA3C-3 increase and then decrease with the number of hidden layers. These phenomena indicate that overfitting may happen when training a complex neural network, \textcolor{black}{and the structure (numbers of hidden layers and neurons) of the optimized neural network with the maximum QoE is more complex for a larger input size.}

\textcolor{black}{Fig.~\ref{fig:weight} shows all components in the QoE under different weight combinations. From Fig.~\ref{fig:weight}, we can see that the performance of the corresponding QoE component can be improved if more attention (larger weight) is paid (allocated) to it. This fact indicates that the tradeoff among video quality, video quality variation and rebuffering time can be achieved by choosing different weight combinations.}

\textcolor{black}{Fig.~\ref{fig:offline_pastchunk} shows the average QoE, inference time, and training time versus the number of most recent chunks $(k)$ whose information is input to each network. We can make the following observations. Firstly, the average QoE, inference time, and training time of the proposed eA3C method increase with $M$, and 
the average QoE, inference time, and training time of each DRL-based method increase with $k$. These phenomena imply that the tradeoff among average QoE, inference time, and training time can be achieved by choosing different $M$ and $k$. Secondly, eA3C-3 outperforms Pensieve-3 in the average QoE and training time and has the same inference time as Pensieve-3. The gains of eA3C-3 over Pensieve-3 in average QoE and training time come from the joint optimization of the policy and value parameters. Their identical inference time derives from the fact that their optimized networks have the same structure. Thirdly, the proposed eA3C method outperforms Pensieve in the average QoE at the cost of increased training time and inference time. The gains in QoE of the three instances of the proposed eA3C method over Pensieve ($6.8\% \sim 13.8\%$) mainly come from utilizing the lower-layer information (at different amounts). The increased training time and inference time are due to the more complex structures of the optimized eA3C networks for effectively utilizing lower-layer information.  
Fourthly, the proposed eA3C method outperforms MPC in the average QoE and inference time. The gains of the three instances of the proposed eA3C method over MPC in the average QoE ($9.1\% \sim 14.4\% $) come from wise utilization of past and lower-layer information. Their gains over MPC in the average inference time are due to lower computation time for obtaining the bitrate via neural network than exhaustive search.
Finally, eA3C-PRB and eA3C-2 at $k = 4$ outperform Pensieve at $k=16$ in the average QoE, inference time, and training time, implying that a small amount of lower-layer information can compensate for the lack of a large amount of past APP layer information including APP layer throughput, buffer occupancy, and bitrate selection. It becomes evident that judicious utilization of lower-layer information can reduce the memory requirement without compromising performance.\footnote{eA3C-PRB and eA3C-2 have 16 quantities and 20 quantities as inputs at $k = 4$ while Pensieve has 48 quantities as inputs at $k=16$.}
}

\begin{figure*}[t]
%\vspace*{-1.2cm}
\begin{center}
   %\subfloat[\small{Average QoE versus number of layers}]
   %{\resizebox{8cm}{!}{\includegraphics{Offline_QoE_layer.eps}}}
   %\subfloat[\small{Average QoE versus number of neurons}]
   %{\resizebox{8cm}{!}{\includegraphics{Offline_QoE_neuron.eps}}}

   {\resizebox{17cm}{!}{\includegraphics{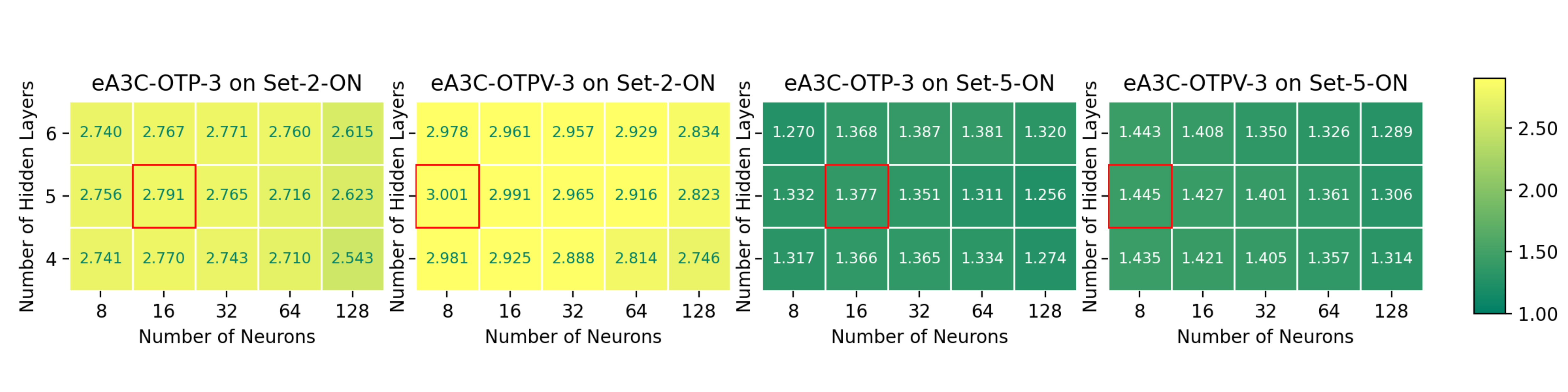}}}
   %\subfloat[\small{Average running time (ms)}]
   %{\resizebox{8cm}{!}{\includegraphics{run.eps}}}
 %\subfloat[\small{CDF of rebuffering time}]
 %{\resizebox{5.9cm}{!}{\includegraphics{pic/multi_rebuffering.eps}}}
 %\vspace*{-0.07cm}
 \end{center}
 %\vspace*{-0.37cm}
   \caption{Heat maps of the average QoE at $k=8$ in the online scenario. The QoE of the optimized neural network structure is marked with a red rectangle.}
   \label{fig:online}
%\vspace*{-0.60cm}
\end{figure*}

\begin{figure*}[t]
\begin{center}
      \subfloat[Average QoE]
   {\resizebox{6cm}{!}{\includegraphics{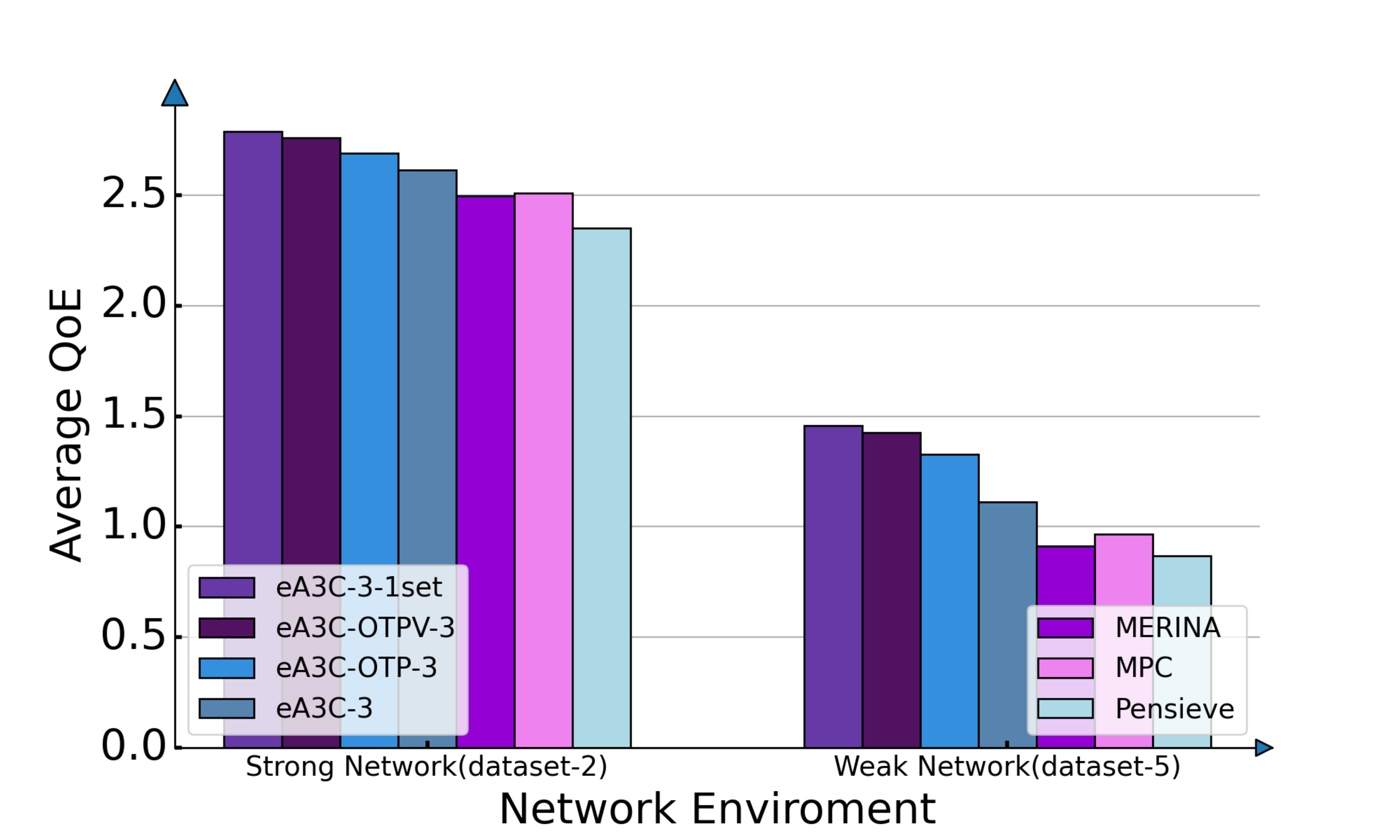}}}
         \subfloat[Average inference time]
   {\resizebox{6cm}{!}{\includegraphics{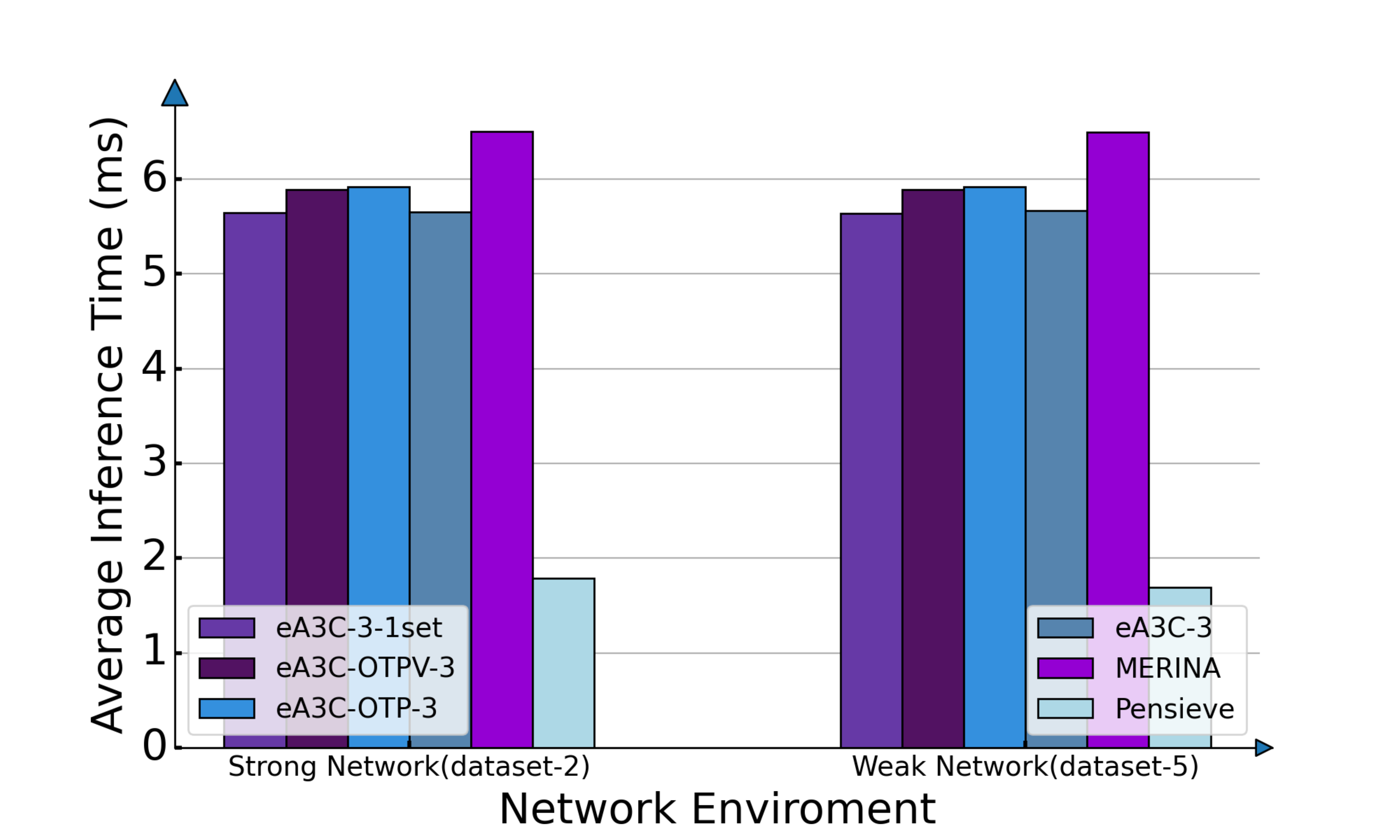}}}
         \subfloat[Average training time ($q'$)]
   {\resizebox{6cm}{!}{\includegraphics{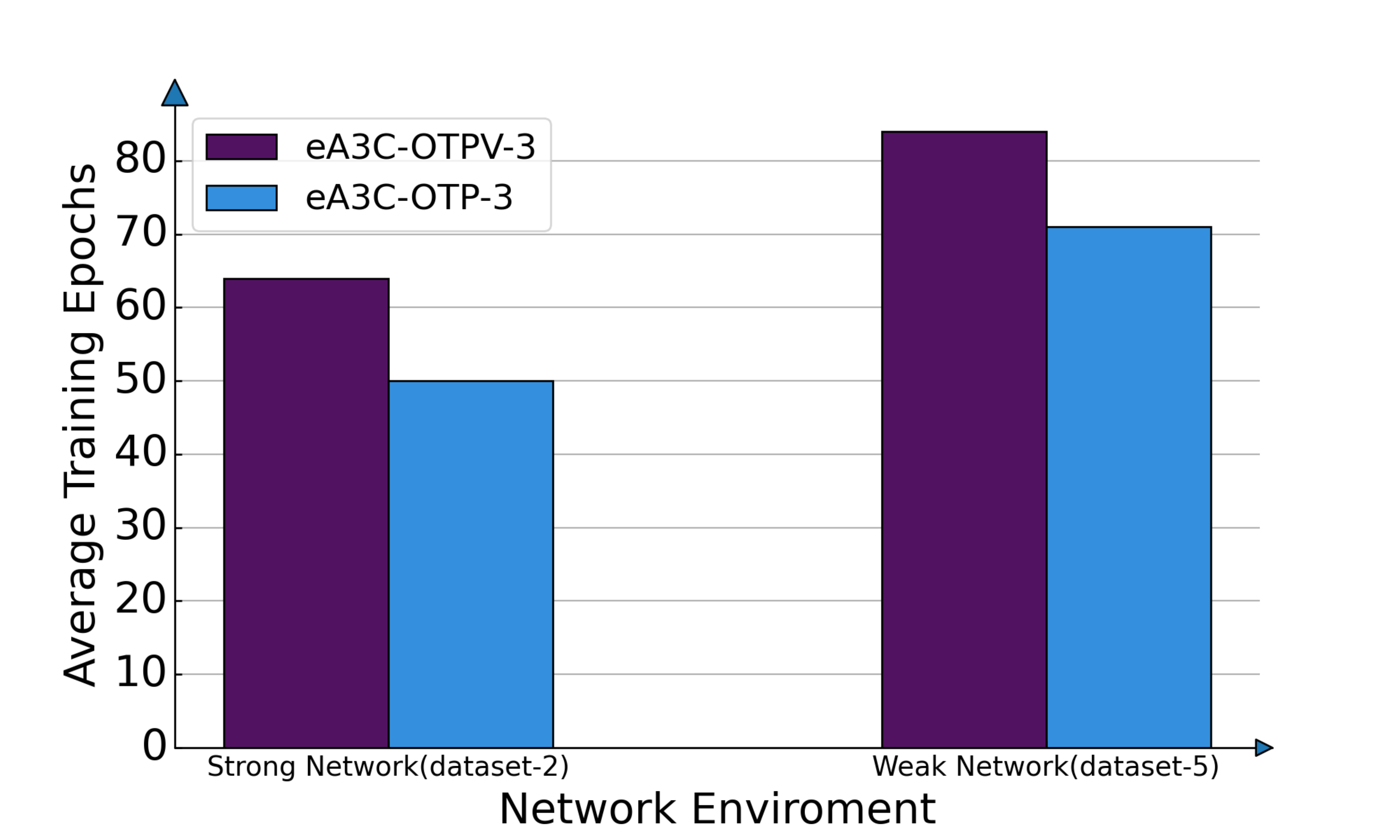}}}
 %\vspace*{-0.57cm}
\end{center}
   \caption{Different performance metrics of proposed and baseline schemes at $k=8$ in the online scenario.}
   \label{fig:propvsbaseline}
%\vspace*{-0.60cm}
\end{figure*}

\subsection{Online Scenario}\label{sec:online:sec}
In the online scenario, we consider two instances, one for each proposed online tuning method, with all three lower-layer quantities ($M=3$), namely eA3C-OTP-3 and eA3C-OTPV-3, respectively, to show the maximum achievable QoE of the proposed online tuning methods.\footnote{It has been shown in Section~\ref{sec:offline_sim} that the corresponding offline eA3C method, eA3C-3 achieves the highest QoE.} The training setups of the tunable components of the eA3C-OTP-3 and eA3C-OTPV-3 networks are identical and are shown in Table~\ref{table3}. The training process lasts over the entire video streaming process. We consider five baseline methods, including eA3C-3, Pensieve\cite{sigcomm17}, eA3C-3-1set, MERINA\cite{mm22}, and MPC\cite{sigcomm15}. Specifically, eA3C-3 and Pensieve apply the offline trained eA3C and A3C networks in Section~\ref{sec:offline_sim}, respectively. eA3C-3-1set uses the same network as eA3C-3 in Section~\ref{sec:offline_sim} but trains it in the offline scenario using Set-$i$-OFF-Train, $i=2,5$. MERINA\cite{mm22} consists of an APP layer throughput prediction network and an A3C network. The prediction network takes the APP layer throughputs of the most recent $k=8$ stages as input and outputs the predicted future APP layer throughput. The A3C network takes the output of the prediction network as additional input. The two networks are jointly trained in the offline scenario using Set-OFF-Train. eA3C-3-1set and MERINA adopt the same offline training setup as shown in Table~\ref{table3}. MPC is the same as the one in Section~\ref{sec:offline_sim}. The details of the proposed and baseline methods are summarized in Table~\ref{table4}. The performances of all methods are evaluated on Set-$i$-ON, $i=2,5$.

%We evaluate all schemes under dataset 2 (user 2) and dataset 5 (user 5), respectively. Note that the average throughput in dataset 2 is higher than that in dataset 5.

%In the POMDP scenario, the numbers of hidden layers and neurons of the progressive prediction network are set to $2$ and $8$, respectively.

In the following, we evaluate the two proposed online tuning methods and the above mentioned baseline methods in the online scenario.
Fig.~\ref{fig:online} shows the heatmaps of the average QoE at different numbers of neurons and hidden layers. We have the following observations: the average QoEs of eA3C-OTP-3 and eA3C-OTPV-3 increase and then decrease with the number of hidden layers; the average QoE of eA3C-OTP-3 increases and then decreases with the number of neurons; the average QoE of eA3C-OTPV-3 always decreases with the number of neurons. eA3C-OTPV-3 has a simpler optimized policy network structure than eA3C-OTP-3, as it can also capture input features via the tunable components of its value network. 
%From Fig.~\ref{fig:online} (b), we can see that the average training time of eA3C-OTP-3 and eA3C-OTPV-3 increase with the numbers of hidden layers and neurons. Besides, the average training time of eA3C-OTPV-3 is higher than the one of eA3C-OTP-3 and the additional training time comes from training the value network. Therefore, eA3C-OTP-3 and eA3C-OTPV-3 have different tradeoffs between the average QoE and training time. eA3C-OTP-3 has lower QoE and shorter training time while eA3C-OTPV-3 has higher QoE and longer the training time.

Fig.~\ref{fig:propvsbaseline} shows the average QoE, inference time, and training time under Set-$2$-ON and Set-$5$-ON, respectively. Note that Set-$2$-ON has a larger average APP layer throughput than that of Set-$5$-ON, as shown in Fig.~\ref{fig:dataset}. We can make the following observations from Fig.~\ref{fig:propvsbaseline}.
Firstly, eA3C-3-1set outperforms eA3C-3 in the average QoE ($6.6\% \sim 31.3\%$), indicating the model mismatch issue when applying the offline trained eA3C network to the online scenario.  
Secondly, eA3C-OTP-3 (eA3C-OTPV-3) outperforms eA3C-3 in the average QoE at the cost of increased inference time. Its gain in the average QoE ($2.9\% \sim 28.3\%$) arises from online tuning. Its increment in inference time is due to the more complex structure of the policy network. Furthermore, the gain in the average QoE of eA3C-OTP-3 (eA3C-OTPV-3) over eA3C-3 under Set-$5$-ON ($19.4\% \sim 28.3\%$) is higher than that under Set-$2$-ON ($2.9\% \sim 5.5\%$), implying that the gain from online tuning is larger in a weaker network environment.
Thirdly, eA3C-OTPV-3 outperforms eA3C-OTP-3 in the average QoE and inference time at the cost of longer training time. Its gain in the average QoE ($2.6\% \sim 7.4\%$) comes from adapting the value network to the online environment. Its gain in the average inference time is due to its simpler optimized policy network. The training time increasement arises from training more tunable components (including those in the policy and value networks).
Fourthly, eA3C-OTP-3 (eA3C-OTPV-3) exceeds MERINA in the average QoE and inference time. Its gain in the average QoE ($7.7\% \sim 55.9\%$) comes from better offline and online designs. Its gain in the average inference time is because MERINA has extra inference time from the prediction network.
Finally, eA3C-OTP-3 (eA3C-OTPV-3) outperforms Pensieve at the cost of increased inference time and outperforms MPC in the average QoE and inference time. The gains of eA3C-OTP-3 (eA3C-OTPV-3) over Pensieve and MPC in the average QoE mainly arise from better offline design and online tuning. The increased inference time compared to Pensieve is due to its more complex policy network structure. The reason for the gain over MPC in the average inference time is identical to that illustrated in Section~\ref{sec:offline_sim}. 
%We have the same observations from Fig.~\ref{fig:propvsbaseline} (b) and omit here due to the page limitation.

%Fig.~\ref{fig:MCTSvssamples} (a) and (c) show the average QoE versus the number of samples on the dataset 2 and dataset 6, respectively. Fig.~\ref{fig:MCTSvssamples} (b) and (d) show the average running time versus the number of sample on the dataset 2 and dataset 5, respectively. Fig.~\ref{fig:MCTSvssamples} demonstrates that the average QoE and the average running time of Prop-3-ML increase with the number of samples. The remaining observations are the same as those in Fig.~\ref{fig:MCTSvssteps}.
%Fig.~\ref{fig:MCTSvssamples_POMDP} shows the average QoE and average running time versus the number of samples on the dataset 2 and dataset 5, respectively. We have the same observations and omit here due to the page limitation.

\begin{figure*}
\begin{align}
\Pr[&\mathbf{S}_{n+1} = \mathbf{s}_{n+1}|\mathbf{S}_{n} = \mathbf{s}_{n},\ldots, \mathbf{S}_{1} = \mathbf{s}_{1}, R_{n} = r_{n}, \ldots R_{1} =r_{1}]  \nonumber\\
%&= \Pr[\mathbf{C}_{n+1} = \mathbf{c}_{n+1}, B_{n+1} = b_{n+1},\overline{\mathbf{X}}_{n+1} = \overline{\mathbf{x}}_{n+1},Y_{n+1} = y_{n+1}|\mathbf{C}_{n} = \mathbf{c}_{n},\ldots, \mathbf{C}_{1} =\mathbf{c}_{1}, B_{n} = b_{n},\ldots,B_{1} = b_{1}, \overline{\mathbf{X}}_{n} = \overline{\mathbf{x}}_{n}, \ldots, \overline{\mathbf{X}}_{1} = \overline{\mathbf{x}}_{1}, Y_{n} = y_{n}, \ldots, Y_{1} = y_{1}, R_{n} = r_{n},\ldots, R_{1} = r_{1}] \nonumber\\
 &\overset{(a)}{=} \Pr[\mathbf{C}_{n+1} = \mathbf{c}_{n+1} | \overline{\mathbf{X}}_{n+1}= \overline{\mathbf{x}}_{n+1},\mathbf{C}_{n} = \mathbf{c}_{n},\overline{\mathbf{X}}_{n} = \overline{\mathbf{x}}_{n}] \times \Pr[\overline{\mathbf{X}}_{n+1}=\overline{\mathbf{x}}_{n+1} | \overline{\mathbf{X}}_{n} = \overline{\mathbf{x}}_{n}] \nonumber\\
 &\times  \Pr[B_{n+1} = b_{n+1},Y_{n+1} = y_{n+1}|\mathbf{C}_{n} = \mathbf{c}_{n},\ldots, \mathbf{C}_{1} =\mathbf{c}_{1},  B_{n} = b_{n},\ldots,B_{1} = b_{1}\nonumber\\
 & ~~~~~~~~~~,\overline{\mathbf{X}}_{n} = \overline{\mathbf{x}}_{n}, \ldots, \overline{\mathbf{X}}_{1} = \overline{\mathbf{x}}_{1}, Y_{n} = y_{n}, \ldots, Y_{1} = y_{1}, R_{n} = r_{n},\ldots, R_{1} = r_{1}]  \nonumber\\
 &\overset{(b)}{=} \Pr[\mathbf{C}_{n+1} = \mathbf{c}_{n+1} | \overline{\mathbf{X}}_{n+1}= \overline{\mathbf{x}}_{n+1}, \mathbf{C}_{n} = \mathbf{c}_{n},\overline{\mathbf{X}}_{n} = \overline{\mathbf{x}}_{n}] \times \Pr[\overline{\mathbf{X}}_{n+1}=\overline{\mathbf{x}}_{n+1} | \overline{\mathbf{X}}_{n} = \overline{\mathbf{x}}_{n}]  \nonumber\\
 &\times \Pr[B_{n+1} = b_{n+1},Y_{n+1} = y_{n+1}|C_{n} = c_{n}, B_{n} = b_{n}, Y_{n} = y_{n}, R_{n} = r_{n}]\nonumber\\ 
 &\overset{(c)}{=}\Pr[\mathbf{C}_{n+1} = \mathbf{c}_{n+1} | \overline{\mathbf{X}}_{n+1}= \overline{\mathbf{x}}_{n+1}, \mathbf{C}_{n} = \mathbf{c}_{n},\overline{\mathbf{X}}_{n} = \overline{\mathbf{x}}_{n}] \times \Pr[\overline{\mathbf{X}}_{n+1}=\overline{\mathbf{x}}_{n+1} | \overline{\mathbf{X}}_{n} = \overline{\mathbf{x}}_{n}]  \nonumber\\
& \times \mathbb{I}\left[b_{n+1} = \left(\left(b_{n}-\frac{r_{n}T}{c_{n}}\right)^{+} +T,B\right)^{+} \right] \times \mathbb{I}\left[y_{n+1} = r_{n}\right]. \label{eq:proof_lemma_1}
\end{align}
\hrulefill
\end{figure*}

%\subsubsection{Discussions}
%In this part, we would discuss the advantages and disadvantages of Prop-MDP-3-TL and Prop-MDP-3-ML. Prop-MDP-3-TL and Prop-MDP-3-ML have two stages. Prop-MDP-3-TL (with 16 neurons and 5 hidden layers) takes 1257s (962s) first to train a new neural network on dataset-2 (dataset-5), and then the neural network is used for adaptive video streaming. The running time of Prop-MDP-3-TL is 0.144ms (0.491ms) on dataset-2 (dataset-5). Prop-MDP-3-LA (with 5 steps and 100 samples) needs to collect at least 100 samples first and then applies the MCTS for adaptive video streaming. The running time of Prop-MDP-3-LA is 992ms (954ms) on dataset-2 (dataset-5). Besides, the average QoE of Prop-MDP-3-TL is higher than that of Prop-MDP-3-LA. That is to say, Prop-MDP-3-TL can achieve higher QoE and has less running time comparing to Prop-MDP-3-LA while Prop-MDP-3-TL needs more time to train the neural network.

\section{Conclusion}
This paper focused on enhancing DRL-based adaptive wireless video streaming by incorporating lower-layer information, deriving a rigorous training method, and adopting online tuning with real-time data. We formulated a more comprehensive and accurate infinite stage discounted MDP problem for adaptive wireless video streaming. \textcolor{black}{In the offline scenario (only with pre-collected data), we \textcolor{black}{presented} an enhanced A3C method, eA3C, which improves the state-of-art DRL method, A3C, based on lower-layer information and a rigorous training method (based on the joint optimization of the policy and value parameters). In the online scenario (with additional real-time data), we \textcolor{black}{presented} two \textcolor{black}{continual} learning-based online tuning methods, i.e., eA3C-OTP and eA3C-OTPV, for designing better policies for a specific user, which enhance the proposed eA3C method leveraging the offline trained eA3C network and online tuning based on real-time samples. The two online tuning methods compensate each other, providing flexibility to meet diverse QoE and training time requirements. Experimental results show that the proposed eA3C method outperforms the state-of-arts in the offline scenario, and the proposed eA3C-OTP and eA3C-OTPV methods achieve further gains over the proposed eA3C method and surpass the state-of-arts in the online scenario.}
%There are still some key aspects that we leave for future investigations. One direction is to Another interesting perspective is 

\section*{Appendix A: Proof of Lemma \ref{theorem}}\label{proof:theorem_1}
For all $n\in\mathbb{N}$, \textcolor{black}{we can obtain \eqref{eq:proof_lemma_1}, as shown at the top of the current page,}
where (a) is due to Assumption~\ref{asmp:1}, (b) is due to \eqref{eq:buffer_evolve} and \eqref{eq:state_augmentation}, and (c) is due to the fact that $B_{n+1}$ is conditionally independent of $Y_{n+1}$ given $C_{n},B_{n},R_{n}$, and $Y_{n+1}$ is conditionally independent of $B_{n+1}$ given $R_{n}$. Thus, we can show Lemma~\ref{theorem}.

\section*{Appendix B: Proof of Lemma \ref{lemma_2}}\label{proof:theorem_2}
Firstly, we equivalently convert the problem in \eqref{prob:offline_policy_gradient_formulation} to a constrained problem by including $V^{\pi}(\mathbf{s}),\mathbf{s}\in\boldsymbol{\mathcal{S}}$ and equations in \eqref{eq:bellman_equ_policy} as optimization variables and constraints:
\begin{align}
\max_{\bm{\Theta}_{\pi},(V^{\pi}(\mathbf{s}))_{\mathbf{s}\in\boldsymbol{\mathcal{S}}}}&\quad \sum_{\mathbf{s}\in\boldsymbol{\mathcal{S}}}
\eta^{\pi}(\mathbf{s}) V^{\pi}(\mathbf{s}),\label{prob:offline_policy_gradient_formulation_proof}\\
	&\mathrm{s.t.}\quad \eqref{eq:bellman_equ_policy}.\nonumber
\end{align}
Next, we transform the constrained problem in \eqref{prob:offline_policy_gradient_formulation_proof} into the unconstrained problem in \eqref{prob:policy_gradient_equal} \textcolor{black}{whose objective function is the weighted sum of the objective of the problem in \eqref{eq:bellman_equ_policy} and the penalty for violating the constraints of the problem in \eqref{eq:bellman_equ_policy} by the penalty method\cite[pp. 388]{cvx}. According to \cite[Proposition 5.2.1]{cvx}, we can show Lemma~\ref{lemma_2}.}

%\section*{Appendix A: Proof of Lemma~\ref{lemma_DC_transform}}\label{proof:approximation_slow_fading}

%Note that $\mathbf{e}(\mathbf{h})$ and $\mathbf{u}(\mathbf{h})$ are auxiliary variables, and \eqref{eq:DC_R<e_perslot}, \eqref{eq:DC_e<u_perslot}, and \eqref{eq:equivalentDCfunction_perslot} are extra constraints. 

%Then, a KKT point of Problem~\ref{prob:UMwT_LMsg_1slot_decouple_DC} can be obtained by CCCP \cite{TSP17}. The main idea is to solve a sequence of successively refined approximate convex problems, each of which is obtained by linearizing \textcolor{black}{the noncovex term of each constraint function in \eqref{eq:equivalentDCfunction_perslot}, i.e.,} $\frac{\sum\nolimits_{\mathcal{G}\in\boldsymbol{\mathcal{X}}}|
%\mathbf{h}^{H}_{k,n}\mathbf{w}_{\mathcal{G},n}(\mathbf{h})|^{2}+
%\sum\nolimits_{\mathcal{G}'\in\boldsymbol{\mathcal{G}}\backslash\boldsymbol{\mathcal{G}}^{(k)}}|\mathbf{h}^{H}_{k,n}\mathbf{w}_{\mathcal{G}^{'},n}(\mathbf{h})|^{2}
%+\sigma^2}{u_{k,n,\boldsymbol{\mathcal{X}}}(\mathbf{h})}$, and preserving the remaining \textcolor{black}{convex terms of each constraint function in \eqref{eq:equivalentDCfunction_perslot}.} 

\end{document}